\documentclass{aa}  

\usepackage{graphicx}
\usepackage{txfonts,xcolor}
\usepackage{hyperref}  
\usepackage{rotating}
\usepackage{pdflscape}
\def\ts     {\thinspace} 
\usepackage{amsmath}

\def\kms  {\ifmmode{{\rm \ts km\ts s}^{-1}}\else{\ts km\ts s$^{-1}$\ts}\fi}
\def\msol {\ifmmode{{\rm M}_{\odot}}\else{M$_{\odot}$\ts}\fi}
\def\lsun {\ifmmode{{\rm L}_{\odot}}\else{L$_{\odot}$\ts}\fi}
\def\cii  {\ifmmode{{\rm [C}{\rm \small II}]}\else{[C\ts {\scriptsize II}]\ts}\fi}
\def\ci   {\ifmmode{{\rm C}{\rm \small I}}\else{C\ts {\scriptsize I}\ts}\fi}
\def\m    {\ifmmode{\mu {\rm m}}\else{$\mu$m}\fi}
\def\hi   {\ifmmode{{\rm H}{\rm \small I}}\else{H\ts {\scriptsize I}\ts}\fi}
\def\hii  {\ifmmode{{\rm H}{\rm \small II}}\else{H\ts {\scriptsize II}\ts}\fi}
\def\nii  {\ifmmode{{\rm [N}{\rm \small II}]}\else{[N\ts {\scriptsize II}]\ts}\fi}
\def\oiii {\ifmmode{{\rm [O}{\rm \small III}]}\else{[O\ts {\scriptsize III}]\ts}\fi}
\def\hh   {\ifmmode{{\rm H}_2}\else{H$_2$\ts}\fi}
\def\nhh  {\ifmmode{N({\rm H}_2)}\else{$N$(H$_2$)\ts}\fi}
\def\microns {\ifmmode{\mu{\rm m}}\else{$\mu$m\ts}\fi}

\begin{document} 

\title{The ALMA Frontier Fields Survey}

\subtitle{III: 1.1\,mm Emission Line Identifications in Abell\,2744, MACSJ0416.1-2403, MACSJ1149.5+2223, Abell\,370, and Abell S1063}

\author{J. Gonz{\'{a}}lez-L{\'{o}}pez\inst{1}
\and
{F. E. Bauer}\inst{1,2,3}
\and
{M. Aravena}\inst{4}
\and
{N. Laporte}\inst{5}
\and
{L. Bradley}\inst{6}
\and
{M. Carrasco}\inst{7}
\and
{R. Carvajal}\inst{1}
\and
{R. Demarco}\inst{8}
\and
{L. Infante}\inst{1,9}
\and
{R. Kneissl}\inst{10,11}
\and
{A. M. Koekemoer}\inst{6}
\and
{A. M. Mu{\~{n}}oz Arancibia}\inst{12}
\and
{P. Troncoso}\inst{1,13}
\and
{E. Villard}\inst{10,11}
\and
{A. Zitrin}\inst{14}
}
\institute{Instituto de Astrof\'{\i}sica and Centro de Astroingenier{\'{\i}}a, Facultad de F\'{i}sica, Pontificia Universidad Cat\'{o}lica de Chile, Casilla 306, Santiago 22, Chile.\\
              \email{jgonzal@astro.puc.cl}
         \and
{Millennium Institute of Astrophysics, MAS, Nuncio Monse\~{n}or S\'{o}tero Sanz 100, Providencia, Santiago de Chile.} 
         \and
{Space Science Institute, 4750 Walnut Street, Suite 205, Boulder, Colorado 80301.} 
         \and
{N\'{u}cleo de Astronom\'{i}a, Facultad de Ingenier\'{i}a y Ciencias, Universidad Diego Portales, Av. Ej\'{e}rcito 441, Santiago, Chile.}
		\and
{Department of Physics and Astronomy, University College London, Gower Street, London WC1E 6BT, UK.}
		 \and
{Space Telescope Science Institute, 3700 San Martin Dr., Baltimore, MD 21218 USA.}
		 \and
{Zentrum f\"ur Astronomie, Institut f\"ur Theoretische
Astrophysik, Philosophenweg 12, 69120 Heidelberg, Germany.}
		 \and
{Department of Astronomy, Universidad de Concepcion, Casilla 160-C, Concepci\'{o}n, Chile.}
        \and
{Carnegie Institution for Science, Las Campanas Observatory, Casilla 601, Colina El Pino S/N, La Serena, Chile.}
         \and
{Joint ALMA Observatory, Alonso de C\'{o}rdova 3107, Vitacura, Santiago, Chile.}
         \and
{European Southern Observatory, Alonso de C\'{o}rdova 3107, Vitacura, Casilla 19001, Santiago, Chile.}
		 \and
{Instituto de F\'isica y Astronom\'ia, Universidad de Valpara\'iso, Avda. Gran Breta\~na 1111, Valpara\'iso, Chile.}
		\and
{Universidad Aut\'onoma de Chile, Chile. Av. Pedro de Valdivia 425, Santiago, Chile.}
		\and
{Physics Department, Ben-Gurion University of the Negev, P.O. Box 653, Be’er-Sheva 8410501, Israel.}
}
%    \date{Received September 15, 1996; accepted March 16, 1997}
%    \date{v.11.03.10, to be submitted...}

% \abstract{}{}{}{}{} 
% 5 {} token are mandatory
 
  \abstract
   {Most sub-mm emission line studies of galaxies to date have targeted sources with known redshifts where the frequencies of the lines are well constrained. Recent blind line scans circumvent the spectroscopic redshift requirement, which could represent a selection bias.}
   {Our aim is to detect emission lines present in continuum oriented observations. The detection of such lines provides spectroscopic redshift information and yields important properties of the galaxies.}
   {We perform a search for emission lines in the ALMA observations of five clusters which are part of the Frontier Fields and assess the reliability of our detection. We additionally investigate plausibility by associating line candidates with detected galaxies in deep near-infrared imaging.}
   {We find 26 significant emission lines candidates, with observed line fluxes between \hbox{0.2--4.6\,Jy\,\kms} and velocity dispersions (FWHM) of 25--600 \kms. Nine of these candidates lie in close proximity to near-infrared sources, boosting their reliability; in six cases the observed line frequency and strength are consistent with expectations given the photometric redshift and properties of the galaxy counterparts. We present redshift identifications, magnifications and molecular gas estimates for the galaxies with identified lines. We show that two of these candidates likely originate from starburst galaxies, one of which is a so-called jellyfish galaxy that is strongly affected by ram pressure stripping, while another two are consistent with being main sequence galaxies based in their depletion times.}
   {This work highlights the degree to which serendipitous emission lines can be discovered in large mosaic continuum observations when deep ancillary data are available. The low number of high-significance line detections, however, confirms that such surveys are not as optimal as blind line scans. We stress that Monte Carlo simulations should be used to assess the line detections significances, since using the negative noise suffers from stochasticity and incurs significantly larger uncertainties.
}


   \maketitle
%
%________________________________________________________________

\section{Introduction}

One of the goals of modern astrophysics is to characterize galaxies at high redshift. Measuring the properties of galaxies at different redshifts can help to understand how they grew and what physical factors dominated their evolution. 
Observations of high redshift galaxies at the far infrared (FIR) and millimeter (mm) wavelengths are crucial in this respect, since they can reveal the state of the interstellar medium (ISM). Continuum observations at these wavelengths trace cold dust emission in galaxies, allowing constraints on dust masses and temperatures, as well as total star-formation rates (SFRs), which may be obscured from optical and near-infrared (NIR) observations. Emission line observations, on the other hand, allow measurements of gas densities, temperatures, metallicities, and kinematics, as well as alternative measures of SFRs and galaxy dynamics.

Emission lines are produced by several mechanisms in different phases of the ISM \citep[for a recent review, see][]{Carilli_Walter2013}. The relatively short wavelength coverage of FIR and mm instruments has limited the extent of line observations to focus mainly on individual galaxies with known redshifts, whereby specific lines are targeted at their corresponding rest-frame wavelengths.

Recently, blind line-scan surveys have been undertaken to search for emission lines over relatively large frequency coverage and areas on the sky \citep{Aravena2012,Lentati2015}. \cite{Decarli2014a} presented the first blind molecular line-scan over the Hubble Deep Field North \citep{Williams1996} using the IRAM Plateau de Bure Interferometer (PdBI). The survey was designed to cover the 3\,mm atmospheric band to search for CO emission lines spanning a wide range of redshifts. Such observations eliminate biases associated with targeted emission line observations in galaxies with high SFRs and/or stellar masses. However, it can be more difficult to associate emission line candidates to galaxies, particularly if no deep imaging is available. 

The Atacama Large Millimeter/submillimeter Array (ALMA) has recently opened up  the possibility of performing blind line-scan campaigns over larger areas and larger frequency ranges in a reasonable amount of time. For instance, the ALMA Spectroscopic Survey \citep[ASPECS;][]{Walter2016} observed a 1 arcmin$^2$ region in the Hubble Ultra Deep Field \citep[HUDF;][]{Beckwith2006} over the full 1\,mm (band 6) and 3\,mm (band 3) atmospheric bands available to ALMA. ASPECS enabled the detection of CO emission lines over a wide redshift range, as well as the detection of the \cii 158 \m\ts\ts emission line in $z\sim6$--8 galaxies \citep{Aravena2016c}. The key factor allowing these kind of surveys is the large spectral coverage achieved at good sensitivity, which strongly improves the chances of blindly detecting multiple lines. 

A less complete method for searching for emission lines is to use large area continuum surveys. ALMA offers 7.5 GHz of frequency coverage per setup, allowing line searches over limited spectral windows. A large area deep survey may have a sufficiently high probability of having a bright line-emitting galaxy at the correct redshift to be detectable. This method has been used already in some ALMA observations with mixed success \citep{Matsuda2015,Kohno2016}.

In this paper we present a search for emission lines in five cluster fields of the ALMA Frontier Fields survey \citep[ALMA-FFs; ][hereafter GL17]{Gonzalez-Lopez2017}. The main goal of the ALMA-FFs is to detect lensed high-redshift galaxies at 1.1\,mm continuum, but the data were acquired in such a way as to facilitate useful lines searches as well. The ALMA coverage builds on the Frontier Fields (FFs) Survey \citep{Lotz2017},\footnote{http://www.stsci.edu/hst/campaigns/frontier-fields/}, a legacy project that combines the power of gravitational lensing by massive clusters (with magnifications of $\mu$$>$5--10 over up to several arcmin$^{2}$ regions and 100's of multiple images) with extremely deep multi-band {\it HST} and {\it Spitzer} imaging of six lensing clusters and adjacent parallel fields \citep{Coe2015}. Of particular relevance here is the fact that the deep ancillary data and spectroscopic redshift coverage of these fields allow immediate assessment of any line candidates.

This paper is organized as follows: in $\S$2 we outline the observations and imaging procedures; in $\S$3 we discuss our line-search method; in $\S$4 we present the emission lines candidates; in $\S$5 we discuss the properties of the counterparts to the emission line candidates and in $\S$6 we summarize our results.
Throughout this paper, we adopt a cosmology with $H_{0} = 67.8$ \kms Mpc$^{-1}$, $\Omega_{\rm m}=0.307$ and $\Omega_{\rm \Lambda}=0.693$ \citep{Planck2014}. Errors are given at 1$\sigma$ confidence unless stated otherwise.

%__________________________________________________________________
\section{Data}
\subsection{Observations}
We use the data taken as part of the band 6 ALMA-FF survey (programs \#2013.1.00999.S in cycle 2 and \#2015.1.1425.S in cycle 3). In GL17, we presented the configuration, reduction and continuum imaging of the cycle 2 observations. Full details of the cycle 3 observations will be presented in \citet{Gonzalez-Lopez2017b}. Here we briefly summarize the observations.

\begin{figure}[!htbp]
\centering
\includegraphics[width=\hsize]{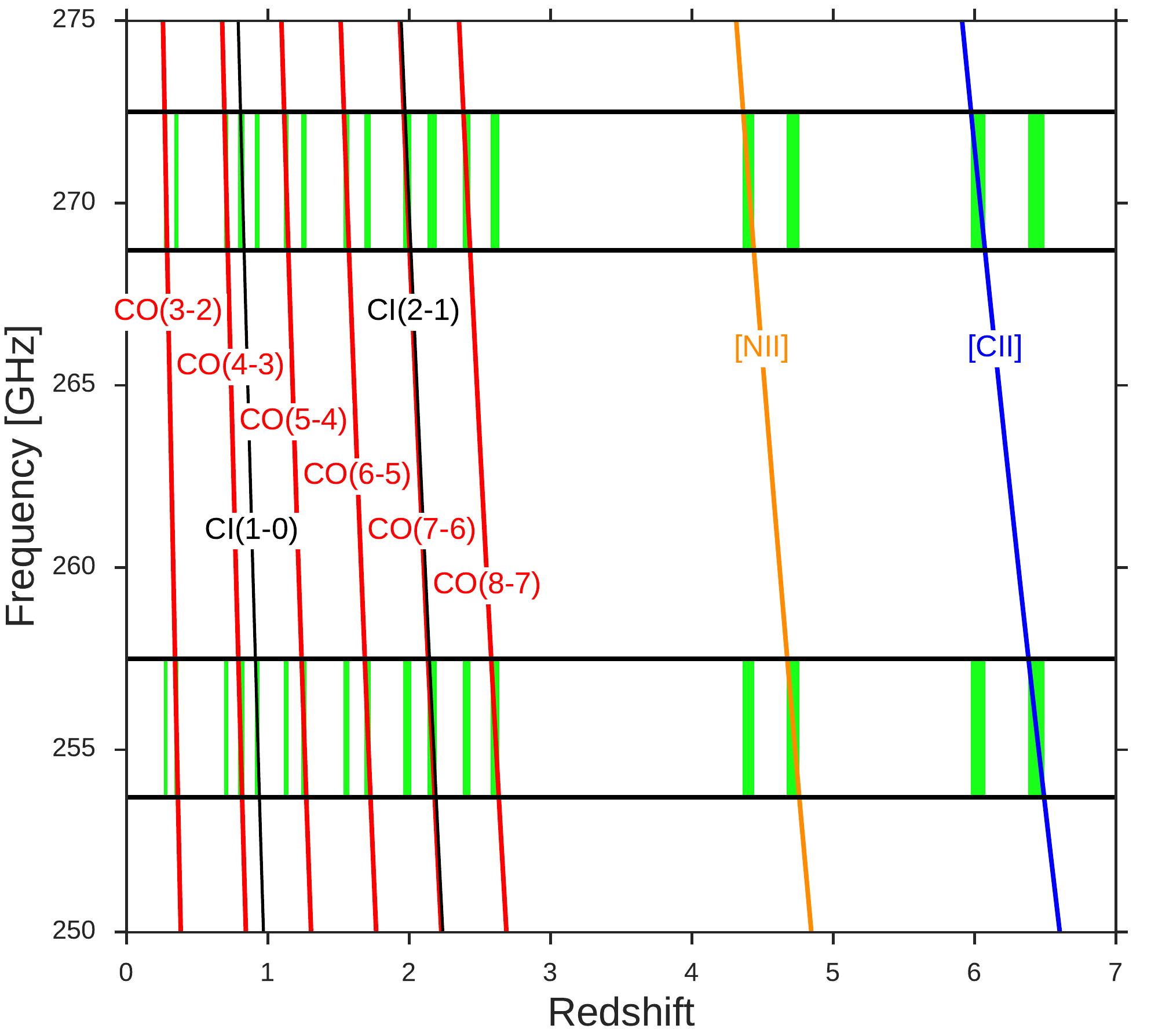}
\caption{Redshift coverage in the ALMA-FF observations for several bright emission lines expected from high redshift galaxies. The horizontal black lines denote the edges of the upper and lower sidebands for the ALMA-FF frequency setup, with each sideband comprised of two adjacent SPWs. The slanted red (CO), black (C{\sc i}), orange (\nii) and blue (\cii) curves show the frequencies expected for each emission line as a function of redshift. The green regions represent the redshift segments over which an emission line could be detected in either sideband. The ALMA-FFs should be sensitive to galaxies with strong star-formation activity at $z=0.3$--3 for some CO lines, as well as normal star-forming galaxies at $z\sim4.5$ for \nii and $z\sim6$ for \cii. 
\label{fig:redshift_coverage}}
\end{figure}

Five galaxy clusters belonging to the FFs sample were observed with ALMA using the 12\,m array: Abell 2744 ($z$$=$0.308); MACSJ0416.1-2403 ($z$$=$0.396); MACSJ1149.5+2223 ($z$$=$0.543); Abell 370 ($z$$=$0.375), Abell S1063 ($z$$=$0.348); hereafter A2744, MACSJ0416, MACSJ1149, A370, and AS1063, respectively.\footnote{The sixth FF cluster, MACSJ0717.5+3745, was only partially observed in cycle 3 to much worse sensitivity and thus is not useful for the present study.} Each field was covered with 126 pointings to create mosaics spanning an area of $\approx4.6$ square arcminutes within the half power ``region'' formed by the individual beams at the edge (i.e. their Half Power Beam Width (HPBW)).
The spectral configuration setup was the same for the five fields. The Local Oscillator frequency was set to 263.14 GHz ($\approx$1.1\,mm), with two spectral windows (SPWs) placed in each sideband. The correlator was set to Frequency Division Mode (FDM) with a bandwidth of 1875\,MHz and a channel spacing of 0.488\,MHz. This setup yielded a total frequency coverage of 7.45\,GHz after factoring in SPW overlaps. Figure \ref{fig:redshift_coverage} shows the redshift coverage for a variety of bright emission lines expected from high redshift galaxies, given our observational setup. Table \ref{tab:redshift_coverage_description} shows the specific redshift ranges where bright lines can be detected. While the frequency coverage is limited, each field contains $\sim4000$ galaxies coupled with strong lensing.

\begin{table}
\caption[]{Redshift coverage in the ALMA-FF observations for several bright emission lines expected from high redshift galaxies.
\label{tab:redshift_coverage_description}}
\centering
\begin{tabular}{lcc}
\hline     
\noalign{\smallskip}
Line Name & Upper sideband & Lower sideband\\
 & Redshift range & Redshift range\\
\hline
\noalign{\smallskip}
CO(3-2) & 0.269 -- 0.287 & 0.343 - 0.363 \\
CO(4-3) & 0.692 - 0.716 & 0.790 - 0.817 \\
\ci(1-0) & 0.806 - 0.832 & 0.911 - 0.94 \\
CO(5-4) & 1.115 - 1.145 & 1.238 - 1.271 \\
CO(6-5) & 1.538 - 1.573 & 1.685 - 1.726 \\
CO(7-6) & 1.960 - 2.002 & 2.133 - 2.180 \\
\ci(2-1) & 1.970 - 2.012 & 2.143 - 2.190 \\
CO(8-7) & 2.383 - 2.431 & 2.580 - 2.633 \\
\nii & 4.362 - 4.438 & 4.674 - 4.759 \\
\cii & 5.974 - 6.073 & 6.381 - 6.491 \\
\hline
\end{tabular}
%\tablefoot{}
\end{table}
We requested channel averaging inside the correlator with an averaging factor of $N=16$, resulting in a final spectral resolution of 7.813 MHz ($\sim9\kms$). The final mosaics have fairly homogeneous sensitivity except for a small region of MACSJ1149 (see Fig. 4 of GL17). 

\subsection{Imaging}

\begin{figure*}[!htbp]
\centering
\includegraphics[width=\hsize]{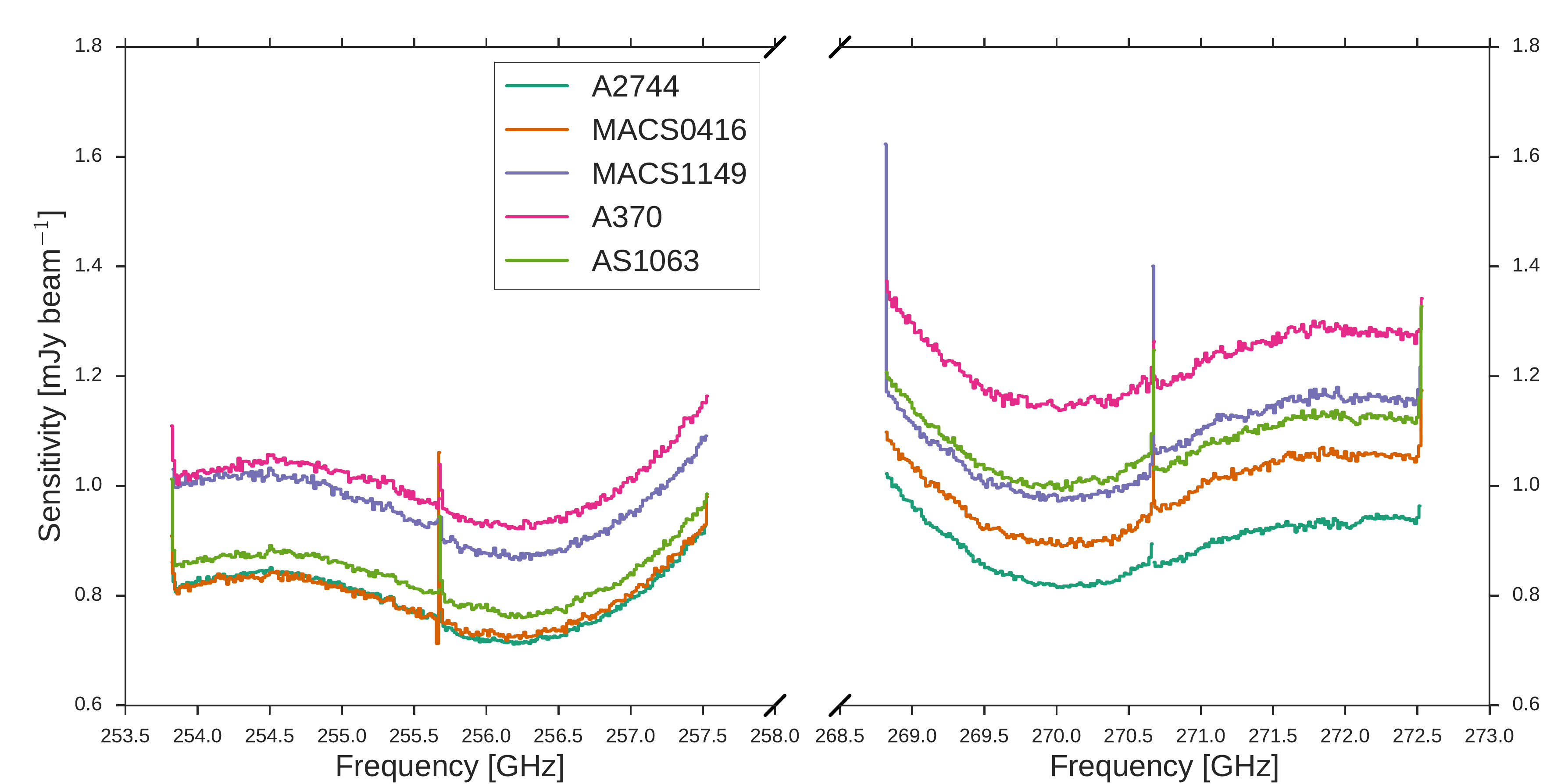}
\caption{Solid lines show $1\sigma$ sensitivity per channel (15.626\,MHz, or $\sim18\kms$ resolution) for the five ALMA-FF fields. The sensitivity is relatively homogeneous as a function of frequency, with variations up to $\sim30\%$ due largely to the transmission of the atmosphere, as the same shape is observed in all five clusters. The atmospheric transmission at these frequencies is fairly flat, with no strong features present in any of SPWs. Higher $1\sigma$ level values are observed near the edges of some SPWs (MACSJ1149 in particular), due to the loss of sensitivity at the edges and/or the combination of executions after transforming to the barycenter frequency reference.
\label{fig:sensitivity_frequency}}
\end{figure*}

We created data cubes of the observations using natural weighting, which provides the best sensitivity since it optimally weights all baselines according to their scatter. These cubes have synthesized beam sizes of 0\farcs63$\times$0\farcs49 and position angle of 86\fdg16 for A2744; 1\farcs52$\times$0\farcs85 and $-$85\fdg13 for MACSJ0416; 1\farcs20$\times$1\farcs08 and $-$43\fdg46 for MACSJ1149; 1\farcs25$\times$1\farcs00 and $-$89\fdg58 for A370; and 0\farcs96$\times$0\farcs79 and $-$80\fdg33 for AS1063. 

A set of spectral dirty cubes were created for each of the SPWs with spectral resolutions of 2, 4, 8 and 16 times the native resolution. The sensitivity as a function of frequency is presented in Fig.~\ref{fig:sensitivity_frequency}. In the lower and upper sidebands, we obtain sensitivity ranges of $\approx$0.8--1.0 and $\approx$0.9--1.3 mJy beam$^{-1}$ over a 15.626 MHz channel, respectively, for the five clusters. The sensitivity is fairly flat, with similar shapes for the five fields. Worse sensitivity is observed in the edge channels for several clusters.

\section{Emission line search}

\subsection{Method}
We need to find emission lines with different flux densities and line widths in the data cubes generated for each SPW. The measured full width half maximum (FWHM) for CO emission lines commonly lies between $\sim100\kms$ to $\sim1000\kms$ \citep{Carilli_Walter2013}. This wide span in line width produces challenges for search methods. For a given line width, a spectral channel of a similar or smaller width will return the highest sensitivity. Searching for very wide lines in high resolution spectral channel data or looking for narrow lines in low resolution spectral channel data will result in sub-optimal searches. The optimal case is when the FWHM of the emission line is resolved by three to five spectral channels.

To optimize the line search for different line widths and to limit the number of spectral convolutions required over a large number of channels, we generated four data cubes per SPW, each with a different spectral resolution. The highest spectral resolution cube uses a binning of two native channels of width 7.813 MHz ($\sim9\kms$), resulting in a channel resolution of 15.626 MHz ($\sim18\kms$); this should provide good sensitivity for emission lines with FWHM $\sim100\,\,\kms$ or narrower. The next three cubes adopt consecutively coarser spectral channels by factors of two, resulting in channel resolutions of 31.252 MHz ($\sim36\kms$), 62.504 MHz ($\sim72\kms$) and 125.008 MHz ($\sim144\kms$), respectively. The spectral resolution of $\sim144\kms$ should be wide enough to detect lines with widths up to $\sim750\kms$, which approaches the total size of an individual SPW ($\sim2000\kms$). 

This set of spectral resolutions should return good sensitivity for a wide range of line widths. Given the channel resolutions of the cubes, we expect that bright emission lines should appear in more than one cube, since adjacent channel resolutions are not so different.

The most common assumption made to find and fit emission lines is that they can be described by a Gaussian profile. This assumption is usually sufficient to describe emission lines at high redshift and low signal-to-noise ($S/N$). For bright emission lines, however, the kinematic structure can result in non-Gaussian profiles, e.g., rotating disks or mergers. Since we do not expect high $S/N$ detections here, because of the blind nature of this survey and the low probability of having a bright line-emitting galaxy at the correct redshift, we do not expect to resolve the kinematic features of the emission lines. Therefore assuming a Gaussian shape for emission line search is justified. 

The first step in the search is to take each data cube and convolve it with a Gaussian in the spectral axis. The $\sigma_{\rm convolution}$ values used for the Gaussian convolution range from 2 to 48 times the native resolution, giving a maximum Gaussian convolution with FWHM$\sim500\kms$ for the lowest resolution cube. After the convolution using the different $\sigma_{\rm convolution}$ values, we take the rms in each spectral channel and select those cells with $S/N\geq5$. The spaxels with high $S/N$ should represent the positions in the sky and frequency where a line candidate is located. 

Because of the multiple binning of the data cubes, and the multiple line searches performed on the same cubes assuming different line widths, estimating the significance of an observed $S/N$ is not necessarily straightforward. We need to estimate the expected number of cells that will have a given $S/N$ by chance assuming our search procedure in the data cubes.

\subsection{False rate detection through simulations}

\begin{figure}[!htbp]
\centering
\includegraphics[width=\hsize]{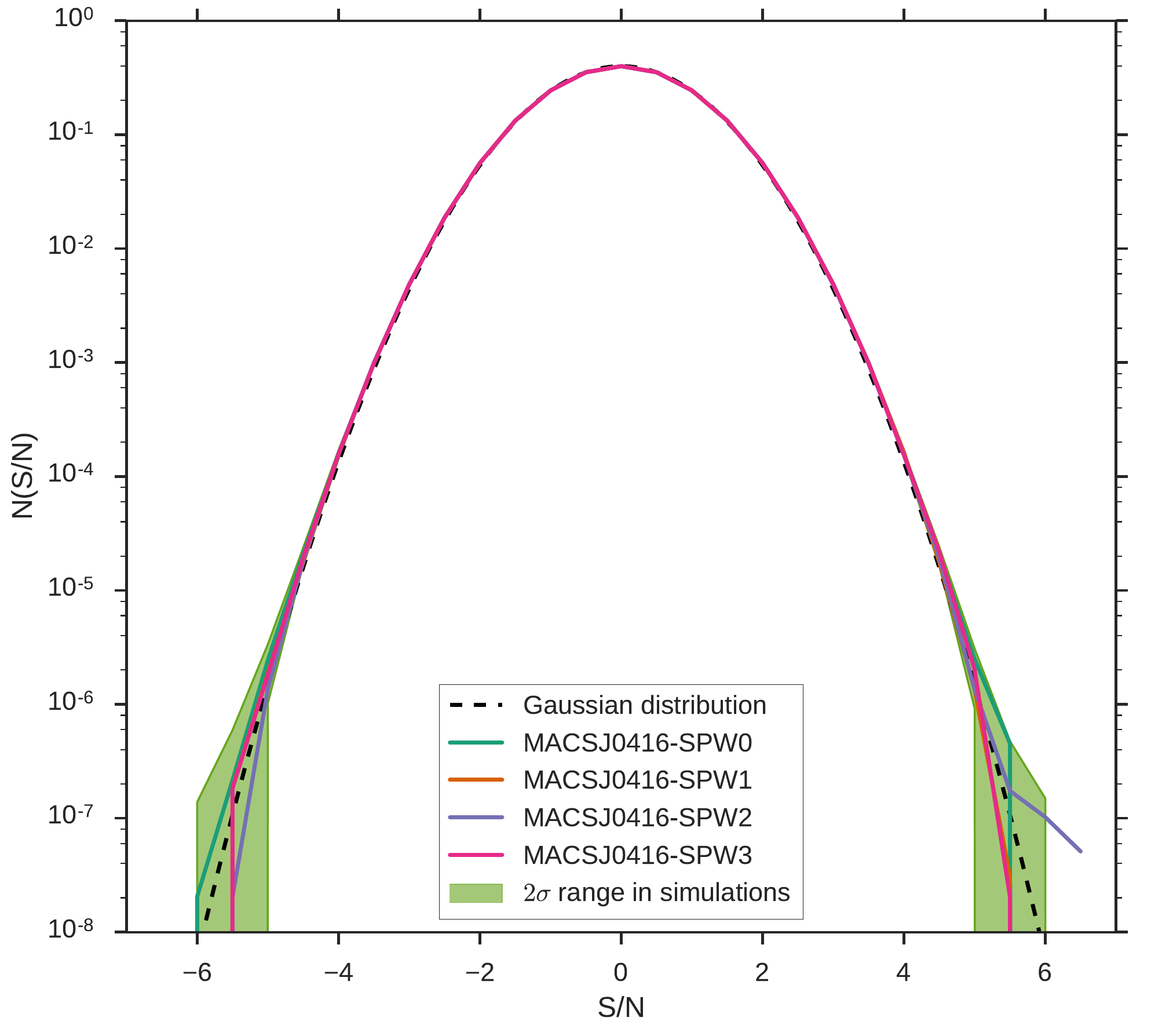}
\caption{The solid lines show the normalized histogram of the $S/N$ per cell for each of the four SPW cubes in the cluster MACSJ0416. Similar noise distributions are observed for SPWs associated with the other four clusters. The black dashed curve shows the $S/N$ histogram expected from a pure Gaussian noise distribution, while the green shaded regions correspond to the $95\%$ range based on over 100 simulated SPWs with the same characteristics as the real data. Importantly, while deviations from the perfect Gaussian distribution can be seen above/below roughly $\pm$5$\sigma$ in the real SPWs, most of these can be explained by low-number statistics, as the same behavior is observed in the simulations up to $S/N=6$. Thus one should be cautious about claims of detections based on positive excesses when reflecting the negative side of the noise distribution, as these methods have systematic uncertainties which can over- or underestimate the true significance.
\label{fig:sn_distribution}}

\end{figure}

Here we estimate the false detection rate (FDR) of an emission line candidate with a given $S/N$. FDR is the probability that a detected signal with $S/N$ is produced by noise in the data. To estimate the FDR, we perform the same search for emission lines as above but in simulated data cubes. To simulate comparable data cubes, we need to understand what the noise distribution in the data cubes is. We assume here that the noise in the ALMA-FFs data can be approximated by a Gaussian distribution;\footnote{In theory, it can deviate from Gaussianity for a number of reasons (amplification of non-Gaussian noise in the $uv$-plane due to atmosphere, antenna systematics, etc.), that leaves the noise in the image plane correlated and non-Gaussian. However, in practice, we do not see strong deviations in our data. We can only speculate that this is because we are dealing with low fluxes and no bright sources, detected with many antennas, which allow the data to obey the central limit theorem.} we test this hypothesis by comparing the noise distribution of our data with simulated cubes created assuming a Gaussian noise distribution. 
The simulated data cubes are created as follows:

\begin{enumerate}
\item We select the cells in the real data cubes where the Primary Beam correction (PBC) is higher than or equal to 0.5. The selected cells are replaced randomly adopting a Gaussian noise distribution, with each cell being independent of the rest, both in spatial and spectral space. 
\item The simulated spectral channels are then convolved with the exact synthesized beam of the original data cubes to obtain roughly the same independent angular elements. At this point the data cubes resemble those generated from real interferometric data.
\end{enumerate}

Posterior spectrally binned data cubes are then created from the simulated data cubes in the same manner as the original cubes. A comparison between the real and simulated noise distribution is shown in Fig. \ref{fig:sn_distribution}. We plot the $S/N$ distribution of all the cells in all the channels for each SPW in MACSJ0416, together with the $95\%$ confidence region assessed by simulating over 100 data cubes. We find that the noise distribution is well described by a Gaussian distribution for $|S/N|\leq5.0$. The strong deviations seen beyond $S/N<-5$ and $S/N>5$ can be explained by small-number statistics; given the large number of pixels, we expect only a few $5\sigma$ fluctuations, which can themselves suffer from stochasticity. It is important to note that because of this behavior in the edges of the distribution, extra care must be taken when using the negative pixels as a noise reference since the tails of the noise distribution need not be symmetrical, even when the overall noise is Gaussian. The latter could introduce additional uncertainty in the number of false positive detections when using negative pixels as a reference to estimate the FDR for emission line and continuum searches (hereafter the "negative image FDR"). For example, \citet{Dunlop2016} found that the FDR number derived from the negative continuum image for their ALMA observations over the Hubble Ultra Deep Field (HUDF) was $\simeq29$ sources, a factor of roughly three higher than the $\simeq10$ sources expected from Gaussian statistics based on the number of independent beam elements in the image. Most critically, even after applying a correction factor of two to the number of independent elements in interferometric observations, as recommended by \citet{Condon1997, Condon1998}, the HUDF negative image FDR remains a factor of $\approx1.5$ higher than expectations from Gaussian statistics. This highlights that caution should be exercised when using the negative counts to estimate the noise distribution. 
We propose that Monte Carlo simulations can yield a more realistic understanding of the true number of independent elements in interferometric observations and provide a better estimate of the true uncertainties. 

In each simulated SPW, a search for emission lines is performed. The total number of emission lines detected in the simulated cubes is then used to estimate the probability of a particular observed emission line with a given $S/N$ to appear in a given simulated observation (combination of the four individual SPWs). We define this probability as $P_{S/N}=N_{>S/N}/N_{total}$, with $N_{S/N}$ being the number of simulated observations with a line with signal-to-noise equal or higher than $S/N$ and $N_{total}$ being the total number of simulated observations. 

\begin{figure}[!htbp]
\centering
\includegraphics[width=\hsize]{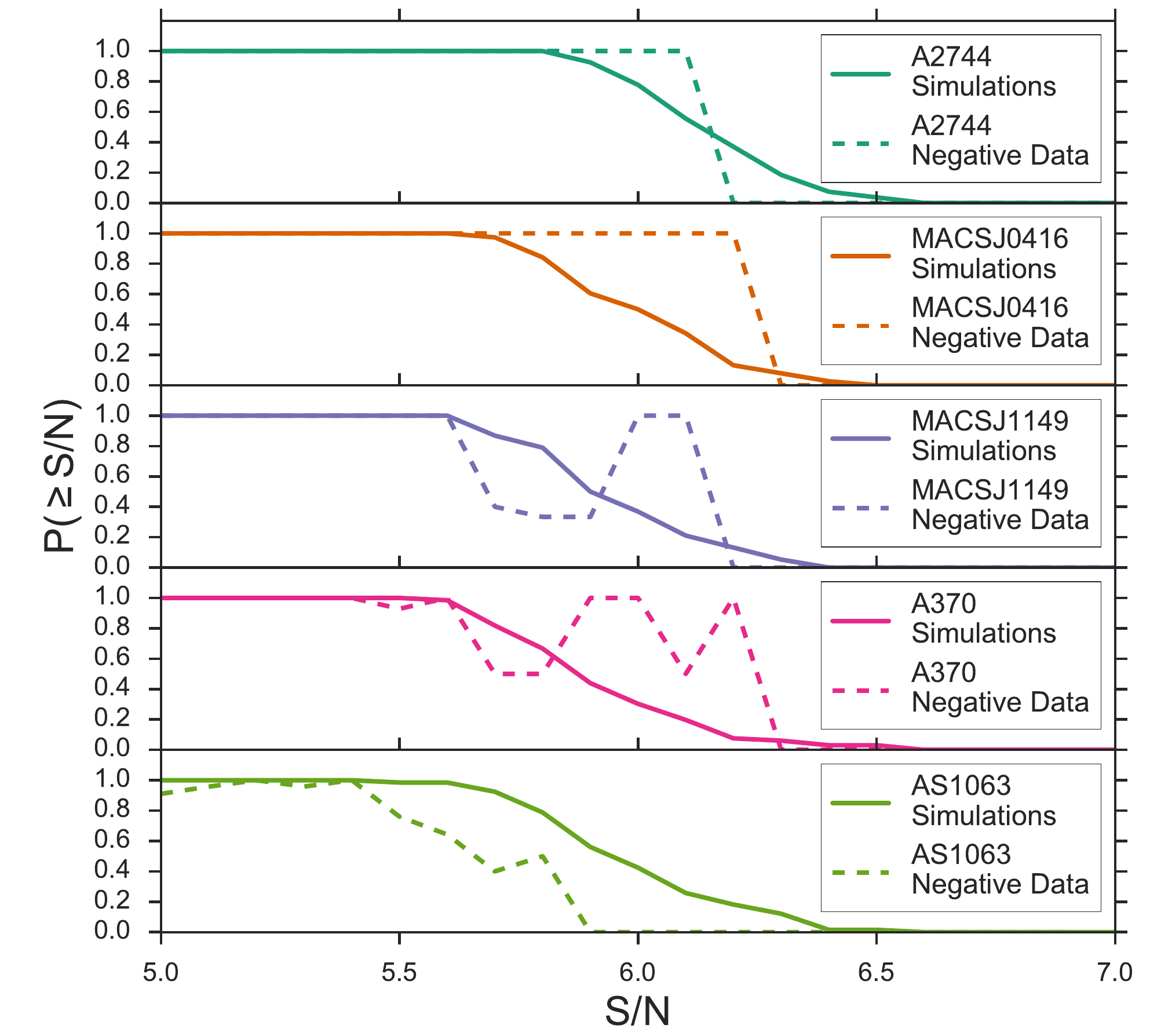}
\caption{Probability of an emission line with a given $S/N$ of being produced by noise for the five clusters. Solid curves correspond to the probabilities given by the simulations, while dashed curves correspond to the probabilities estimated from the negative data.
\label{fig:FDR_comparison}}
\end{figure}

To understand uncertainties in the sample, we initially apply a liberal threshold to include all emission lines candidates with $P_{S/N}\le0.99$. This choice of $P_{S/N}$ effectively includes only those lines with $S/N$$\ge$5.5--5.9,
depending on the cluster and the particular SPW; lower $S/N$ values will have $P_{S/N=2.0}=P_{S/N=5.0}=1.0$, meaning there is a 100\% chance that at least one line is produced by noise (see Figs.~\ref{fig:sn_distribution} and \ref{fig:FDR_comparison}). This cut yields a total of 26 candidates, as shown in Table~\ref{tab:ELC_detection_properties}. However, only two of these line candidates, MACSJ0416-EL01 and A370-EL01, appear to be detected at high significance (i.e, with $P_{S/N}<0.05$) based on their line properties alone. 

We compare the results from the simulations to what is expected when using the negative image FDR as reference. For the latter, we estimate the probability of an emission line with $S/N$ of being false by taking $P(S/N)=N_{neg}(\geq S/N)/N_{pos}(\geq S/N)$, where $N_{neg}$ is the number of emission lines with $\geq S/N$ found in the negative data while $N_{pos}$ is the number found in the positive ("real") data. This means that for equal numbers, there is a 100\% probability that the line is produced by noise. Alternatively, if there there are twice as many positive lines as negative ones, we assume there is a 50\% probability of each positive line being real.  A comparison between simulations and negative image FDR probabilities for each cluster is shown in  Fig.~\ref{fig:FDR_comparison}.

We see that using the negative image FDR as a reference for estimating the probability of a line of being false returns a much noisier and sharper probability distribution for all of the clusters. At a given S/N value, it can strongly under- or overestimate the probability, since it is dominated by small number statistics. Using simulations allow us to more accurately assign probabilities and understand the true detection rate at low $S/N$.

To confirm weaker real emission lines, we investigate the NIR and 1.1\,mm counterparts of all of our candidates.

\subsection{Association with optical/NIR counterparts}

Clearly galaxies with high SFRs are likely to be detectable both with {\it HST} and/or ALMA in continuum and in molecular emission lines. Because of this, we expect that an emission line candidate co-spatial to a NIR detected galaxy should increase the probability of this candidate being real and thus use the NIR images for this purpose.

However, the lack of a NIR or 1.1\,mm counterpart does not necessarily imply that the emission line is noise. A galaxy which remains undetected both in {\it HST}/F160W and ALMA continuum at $5\sigma$ could still have a $SFR\sim40\,\,\msol\,{\rm yr}^{-1}$ at $z>2$ and $SFR\sim30\,\,\msol\,{\rm yr}^{-1}$ at $z\sim1$, based on spectral energy distribution fitting. In such cases, it should still be possible to detect CO, \nii or \cii lines. For instance, we find that CO could be detected at $z\sim2$ from $J>5$ CO lines when a high CO excitation level ($r_{54}\sim0.39$) is present in combination with strong dust extinction ($A_{V}>7$). Detections of \nii and \cii lines can be made to lower SFRs and extinctions ($A_{V}>3$) at $z\sim6$.

Thus, some weak line candidates which lack NIR counterparts could still be real.
Unfortunately, in such cases, confirmation and identification of  line candidates will only be possible with deeper mm observations, which is beyond the scope of this study. Thus we restrict our analysis of weak emission line candidates to those with nearby optical/NIR detected sources. 

Based on the number and spatial distribution of NIR detections in the FFs, a given random position within any of the clusters should have a $35\%$ probability of lying within 1\farcs0 of a NIR galaxy, a $10\%$ probability of being within 0\farcs5 and a $2\%$ probability of being as close as 0\farcs2. Thus, out of the 26 initial emission lines candidates, we expect to find just by pure chance $\approx$9 at $<$1\farcs0 from a NIR source, $\approx$3 with $<$0\farcs5 and $\la$1 as close as 0\farcs2. By comparison, we have 15, 10, and 6 line candidates within these respective separations, implying $\approx$6--7 real detections based on spatial coincidence alone.  

For a given emission line candidate, we estimate the probability $P_{sep}$ of an emission line candidate being randomly associated to the nearest galaxy with angular separation $sep$. We estimate $P_{sep}$ using two approaches. In the first method, we calculate the corrected-Poissonian probability \citep{Downes1986} of $p= 1 - \exp(n\ \pi\ sep^2)$, where $n$ is the source density ($n\sim0.1-0.2$ arcsec$^{-2}$ for the clusters). In the second method, we compare the measured separations against the separation distribution expected for random simulated positions (this method should account for any source clustering which might be present in the HST images). We confirmed that both estimators return similar probabilities and are a function of the separation for the line candidates. For the final $P_{sep}$ we use the corrected-Poissonian estimator. 
We then calculate the overall emission line candidate probability as $P = P_{S/N}\times P_{sep}$. We select as reliable emission line candidates those candidates with $P\leq0.05$ and $P_{S/N}\le0.99$. This implies that if we replace the data with simulated cubes, in 5\% of the simulations we will find at least one significant line candidate with the same $S/N$ and separation to an optical galaxy. Using line candidates with $P_{S/N}\le0.99$ means that for weaker lines, the overall probability $P$ will be dominated by the separation between the line candidate and the NIR nearest source. 
The typical positional uncertainties of the {\it HST} sources are $\leq$0\farcs1, while they are $\approx$10\% of the beam size ($\sim$0\farcs05--0\farcs15) for the ALMA observations (GL17). Despite such small positional errors, separations larger than 0\farcs2 might be legitimately expected between FIR emission lines and optical/NIR continuum emission, since FIR lines could be produced inside of completely obscured star forming regions while the optical/NIR emission could arise from surrounding unobscured regions. At the same time, emission associated with outflowing or infalling material might also lead to larger physical offsets between emission lines and the continuum.

\section{Results}

\begin{table*}
\caption[]{Emission line candidates in five ALMA-FF galaxy clusters. 
\label{tab:ELC_detection_properties}}
\centering
\begin{tabular}{lcccccccc}
\hline     
\noalign{\smallskip}
ID\tablefootmark{a} & {R.A.} & {Dec} & {$S/N$} & {$P_{S/N}$} & {Sep.} & {$P_{sep}$}& {$P$} & {Line identification}\\
& &  &  & & {[\arcsec]} & &  & \\
\hline
\noalign{\smallskip}
A2744-EL01$\dagger$     & 00:14:18.43 & -30:24:47.50 & 6.0 & 0.777 & 0.5 & 0.120 & 0.093 & CO(8-7) at $z=2.62$, tentative\\
A2744-EL02      & 00:14:16.29 & -30:24:22.44 & 5.9 & 0.925 & 0.5 & 0.120 & 0.111 & \ldots\\
A2744-EL03      & 00:14:18.88 & -30:23:45.60 & 5.9 & 0.925 & 1.1 & 0.464 & 0.429 & \ldots\\
MACSJ0416-EL01* & 04:16:10.17 & -24:04:37.82 & 6.4 & 0.026 & 1.6 & 0.783 & 0.020 & No identification\\
MACSJ0416-EL02  & 04:16:09.24 & -24:05:21.67 & 6.1 & 0.342 & 1.3 & 0.635 & 0.217 & \ldots\\
MACSJ0416-EL03*$\dagger$ & 04:16:10.50 & -24:05:05.47 & 6.0 & 0.500 & 0.4 & 0.091 & 0.045 & CO(7-6) at $z=1.96$, tentative\\
MACSJ0416-EL04  & 04:16:10.56 & -24:04:02.67 & 5.9 & 0.605 & 0.8 & 0.317 & 0.051 & \ldots\\
MACSJ0416-EL05  & 04:16:12.32 & -24:05:02.47 & 5.9 & 0.605 & 1.2 & 0.576 & 0.348 & \ldots\\
MACSJ0416-EL06  & 04:16:08.68 & -24:03:42.37 & 5.7 & 0.973 & 1.0 & 0.449 & 0.437 & \ldots\\
MACSJ1149-EL01* & 11:49:36.09 & +22:23:47.90 & 6.0 & 0.368 & 0.1 & 0.006 & 0.002 & No identification\\
MACSJ1149-EL02  & 11:49:34.67 & +22:23:44.40 & 5.9 & 0.500 & 1.9 & 0.892 & 0.446 & \ldots\\
MACSJ1149-EL03* & 11:49:38.37 & +22:24:30.60 & 5.9 & 0.500 & 0.4 & 0.094 & 0.047 & No identification\\
A370-EL01*      & 02:39:51.02 & -01:33:45.39 & 7.6 & 0.000 & 0.0\tablefootmark{b} & 0.000 & 0.000 & CO(3-2) at $z=0.36$, secure  \\
A370-EL02       & 02:39:50.42 & -01:34:40.86 & 5.8 & 0.667 & 1.7 & 0.613 & 0.409 &  \ldots\\
A370-EL03       & 02:39:52.46 & -01:35:35.33 & 5.8 & 0.667 & 3.6 & 0.986 & 0.658 &  \ldots\\
A370-EL04*      & 02:39:53.25 & -01:33:49.13 & 5.6 & 0.985 & 0.2 & 0.013 & 0.013 &  CO(3-2) at $z=0.28$, tentative\\
A370-EL05       & 02:39:55.26 & -01:34:43.13 & 5.6 & 0.985 & 1.7 & 0.613 & 0.604 &  \ldots\\
AS1063-EL01*    & 22:48:40.14 & -44:30:50.56 & 5.9 & 0.561 & 0.0\tablefootmark{b} & 0.000 & 0.000 &  CO(3-2) at $z=0.35$, secure\\
AS1063-EL02     & 22:48:48.03 & -44:31:11.55 & 5.8 & 0.788 & 1.5 & 0.599 & 0.472 &  \ldots\\
AS1063-EL03*    & 22:48:47.45 & -44:32:59.00 & 5.8 & 0.788 & 0.1 & 0.004 & 0.003 & CO(9-8) at $z=3.07$, tentative\\
AS1063-EL04     & 22:48:40.37 & -44:31:58.00 & 5.7 & 0.924 & 0.9 & 0.281 & 0.260 &  \ldots\\
AS1063-EL05     & 22:48:43.65 & -44:31:35.60 & 5.7 & 0.924 & 2.0 & 0.803 & 0.742 &  \ldots\\
AS1063-EL06     & 22:48:47.26 & -44:31:49.20 & 5.7 & 0.924 & 1.0 & 0.334 & 0.309 &  \ldots\\
AS1063-EL07     & 22:48:48.42 & -44:31:50.05 & 5.7 & 0.924 & 2.4 & 0.904 & 0.835 &  \ldots\\
AS1063-EL08*    & 22:48:42.98 & -44:31:55.70 & 5.6 & 0.985 & 0.2 & 0.016 & 0.016 &  No identification\\
AS1063-EL09     & 22:48:43.35 & -44:32:31.40 & 5.6 & 0.985 & 0.9 & 0.281 & 0.277 &  \ldots\\
\hline
\end{tabular}
%\tablefoot{}
\tablefoottext{a}{Sources with $P<0.05$ are denoted by '*'. Sources which are spatially coincident with significant 1.1\,mm continuum detections are denoted by '$\dagger$'; given the low source density of 1.1\,mm continuum detections (GL17) and high likelihood of a physical connection, this implies $P<0.05$.}
\tablefoottext{b}{Sources that lie within the extents of large star-forming galaxies, and thus are considered to have separations of 0\farcs0.}
\end{table*}

Table~\ref{tab:ELC_detection_properties} presents the 26 emission line candidates with $P_{S/N}\le0.99$. Among these, we find nine candidates with $P\leq0.05$; one in A2744 and MACSJ1149, two in MACSJ0416 and A370, and three in AS1063. However, we do not consider MACSJ1149-EL01 further since the NIR counterpart identified from the F160W catalog is not confirmed upon visual inspection. The detection $S/N$ of the remaining eight ranges from 5.6 to 7.6, with the lowest having an associated probability of $P_{S/N}=0.985$ and the highest a $P_{S/N}=0.000$.
Note that the different beam sizes in each field mean that equivalent $S/N$ values will have somewhat different $P_{S/N}$ values.

The angular separations between the reliable emission line candidate centroids and the nearest galaxies range from 0\farcs0 to 1\farcs6, with the largest separation observed in MACSJ0416-EL01, where the high $S/N$ of the detection allows it to be selected even though it does not lie close to any NIR galaxy. In Figures ~\ref{fig:line_reliable} and ~\ref{fig:line_candidate} we present {\it HST} color images of all 26 candidates overlaid with the line and continuum emission $S/N$ contours. In two cases, A2744-EL01 and MACSJ0416-EL03, the emission line candidates are associated with 1.1\,mm continuum selected galaxies \citep[GL17;][]{Laporte2017}, effectively boosting their detection to $P<0.01$.

\begin{table*}
\caption[]{Properties of the emission line candidates obtained by fitting a Gaussian function. 
\label{tab:ELC_properties}}
\centering
\begin{tabular}{lccccc}
\hline     
\noalign{\smallskip}
ID & {Amplitude} & {Central frequency} & {FWHM} & {$F_{Gaussian}$} & Line identification\\
 & {[mJy beam$^{-1}$]} & {[GHz]} & {[km s$^{-1}$]} & {[Jy km s$^{-1}$]} & \\
\hline
\noalign{\smallskip}
A2744-EL01$\dagger$        &  $4.1 \pm 0.5$ & $254.659 \pm 0.019$ & $388 \pm 47$ & $1.70 \pm 0.29$ & CO(8-7) at $z=2.62$, tentative\\
A2744-EL02        &  $8.1 \pm 1.1$ & $271.967 \pm 0.004$ &  $67 \pm 10$ & $0.58 \pm 0.12$ & \ldots\\
A2744-EL03        &  $4.6 \pm 0.6$ & $270.363 \pm 0.004$ &  $62 \pm 10$ & $0.30 \pm 0.06$ & \ldots\\
MACSJ0416-EL01*    &  $8.4 \pm 1.1$ & $257.252 \pm 0.002$ &  $25 \pm  5$ & $0.22 \pm 0.05$ & No identification\\
MACSJ0416-EL02    &  $3.3 \pm 0.5$ & $270.004 \pm 0.008$ & $119 \pm 14$ & $0.42 \pm 0.08$ & \ldots\\
MACSJ0416-EL03*$\dagger$    &  $9.0 \pm 1.2$ & $272.501 \pm 0.002$ &  $32 \pm  5$ & $0.31 \pm 0.06$ & CO(7-6) at $z=1.96$, tentative\\
MACSJ0416-EL04    &  $4.1 \pm 0.6$ & $271.395 \pm 0.008$ & $125 \pm 19$ & $0.55 \pm 0.12$ & \ldots\\
MACSJ0416-EL05    &  $9.5 \pm 1.5$ & $272.318 \pm 0.002$ &  $32 \pm  7$ & $0.32 \pm 0.09$ & \ldots\\
MACSJ0416-EL06    &  $3.2 \pm 0.6$ & $254.733 \pm 0.009$ &  $99 \pm 21$ & $0.34 \pm 0.10$ & \ldots\\
MACSJ1149-EL01*    &  $5.5 \pm 1.0$ & $254.628 \pm 0.009$ &  $87 \pm 32$ & $0.51 \pm 0.21$ & No identification\\
MACSJ1149-EL02    &  $5.1 \pm 0.6$ & $272.251 \pm 0.007$ & $142 \pm 13$ & $0.77 \pm 0.12$ & \ldots\\
MACSJ1149-EL03*    &  $5.5 \pm 0.7$ & $254.206 \pm 0.005$ &  $82 \pm 12$ & $0.48 \pm 0.09$ & No identification\\
A370-EL01*         &  $4.5 \pm 0.5$ & $254.332 \pm 0.009$ & $176 \pm 23$ & $0.84 \pm 0.14$ & CO(3-2) at $z=0.36$, secure  \\
A370-EL02         &  $1.8 \pm 0.2$ & $255.809 \pm 0.055$ & $536 \pm 135$ & $1.03 \pm 0.28$ & \ldots\\
A370-EL03         &  $8.2 \pm 1.4$ & $272.201 \pm 0.004$ &  $43 \pm 11$ & $0.38 \pm 0.12$ & \ldots\\
A370-EL04*         &  $5.2 \pm 0.9$ & $270.915 \pm 0.006$ &  $77 \pm 16$ & $0.43 \pm 0.12$ &  CO(3-2) at $z=0.28$, tentative\\
A370-EL05         &  $2.6 \pm 0.4$ & $254.731 \pm 0.018$ & $253 \pm 45$ & $0.70 \pm 0.17$ & \ldots\\
AS1063-EL01*       &  $4.8 \pm 0.8$ & $255.994 \pm 0.007$ &  $81 \pm 21$ & $0.41 \pm 0.13$ &  CO(3-2) at $z=0.35$, secure\\
AS1063-EL02       &  $5.8 \pm 0.8$ & $254.832 \pm 0.003$ &  $46 \pm  6$ & $0.28 \pm 0.05$ & \ldots\\
AS1063-EL03*       &  $5.3 \pm 0.7$ & $254.802 \pm 0.014$ & $216 \pm 32$ & $1.22 \pm 0.24$ & CO(9-8) at $z=3.07$, tentative\\
AS1063-EL04       &  $5.3 \pm 0.7$ & $271.314 \pm 0.006$ &  $87 \pm 15$ & $0.49 \pm 0.11$ & \ldots\\
AS1063-EL05       &  $5.0 \pm 0.7$ & $270.531 \pm 0.006$ &  $86 \pm 15$ & $0.46 \pm 0.10$ & \ldots\\
AS1063-EL06       &  $6.2 \pm 1.0$ & $269.917 \pm 0.004$ &  $52 \pm 11$ & $0.34 \pm 0.09$ & \ldots\\
AS1063-EL07       &  $4.9 \pm 0.8$ & $254.118 \pm 0.005$ &  $61 \pm 10$ & $0.32 \pm 0.07$ & \ldots\\
AS1063-EL08*       &  $6.2 \pm 0.9$ & $271.415 \pm 0.004$ &  $64 \pm 13$ & $0.42 \pm 0.11$ &  No identification\\
AS1063-EL09       &  $5.3 \pm 0.7$ & $269.396 \pm 0.005$ &  $76 \pm 12$ & $0.43 \pm 0.09$ & \ldots\\
\hline
\end{tabular}
%\tablefoot{}
\end{table*}

In Table~\ref{tab:ELC_properties}, we list the line central frequencies, amplitudes, widths, and integrated fluxes obtained from fitting a Gaussian function to the spectra extracted at the centroid positions of the line candidates, assuming point sources. Three of the reliable line candidates (A2744-EL01, A370-EL01 and AS1063-EL03) have line widths $>100\kms$, which are comparable to similar level CO emission lines reported in the literature. On the other hand, one line candidate has a width of only $25\pm5$ \kms, which is quite small compared to published emission lines widths at high redshift. The spectra of the emission line candidates together with the Gaussian fits are shown in Figures \ref{fig:line_reliable} and \ref{fig:line_candidate}.

\begin{figure*}[!htbp]
\includegraphics[width=\hsize/2]{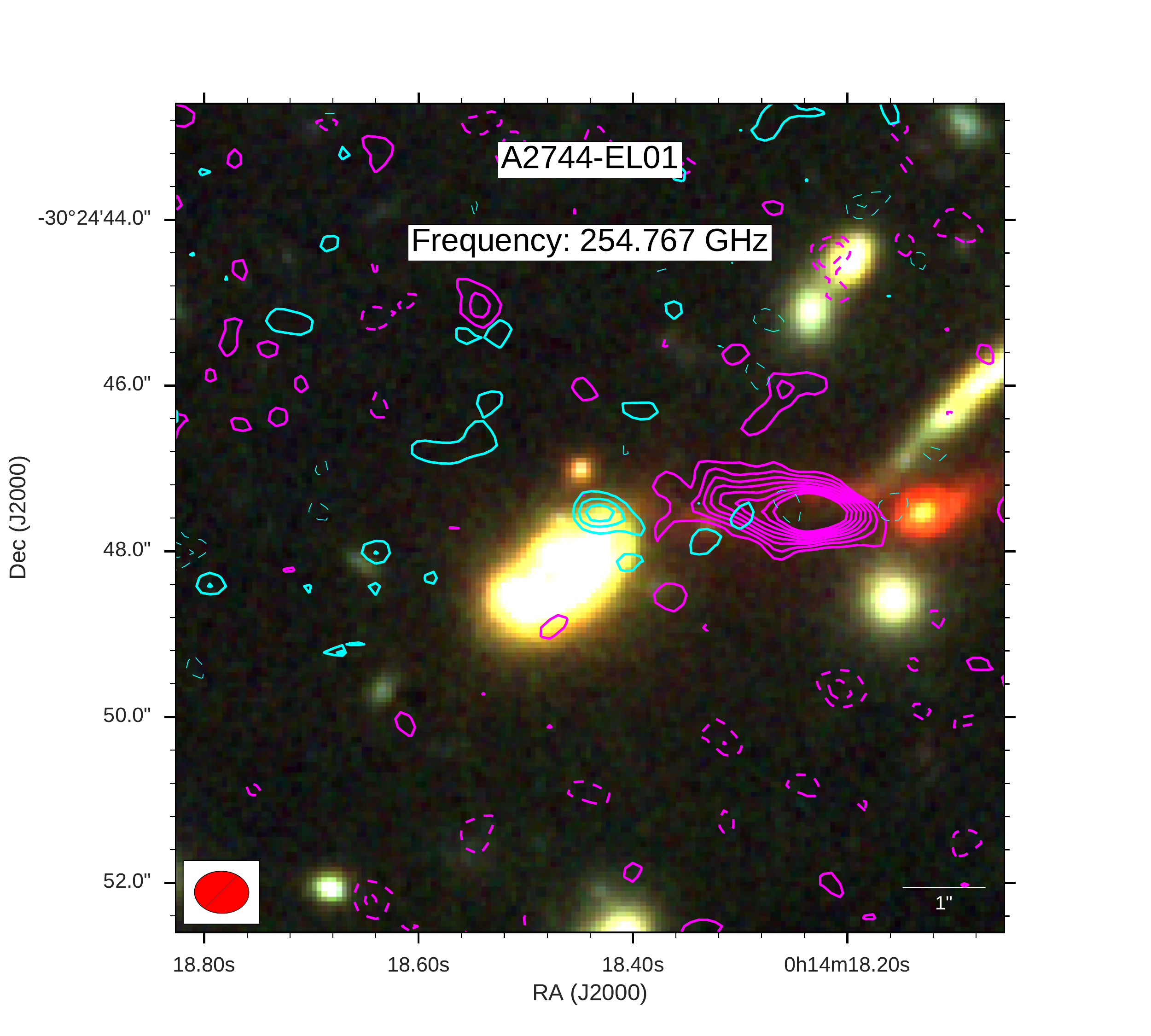}
\includegraphics[width=\hsize/2]{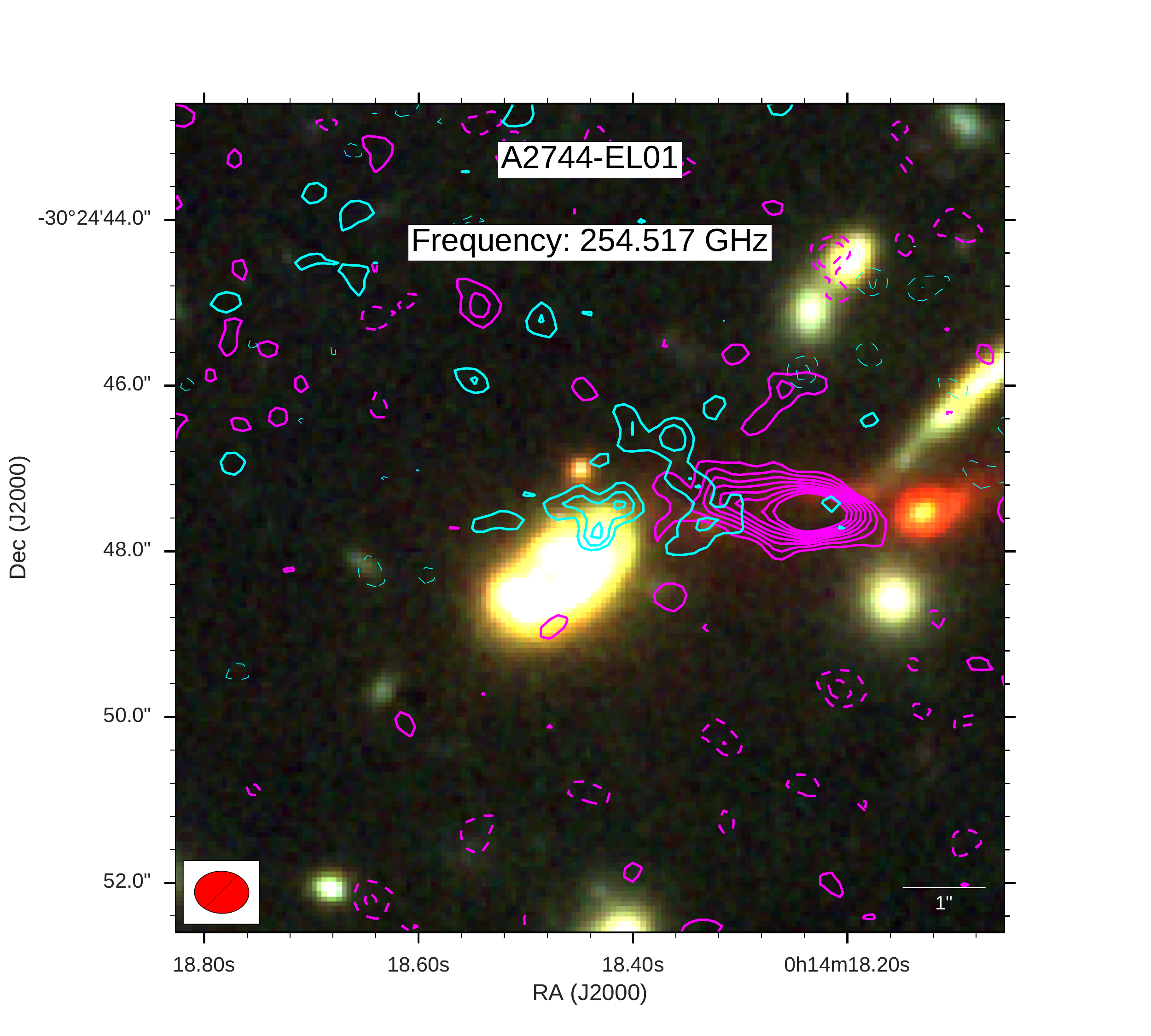}
\caption{{\it HST} color image cutouts for the ALMA-FFs emission line candidate A2744-EL01. Here bands $F435W+F060W$ correspond to blue, $F810W+F105W$ to green and $F125W+F140W+F160W$ to red, respectively. The cyan contours correspond to $S/N$ levels of the ALMA emission line candidate in $1\sigma$ steps starting at $\pm2$ times the primary beam corrected noise level at the position of the source. The blue portion of the line is presented in the {\it left} panel while the red portion in the {\it right} panel. The continuum 1.1\,mm emission is plotted in magenta in steps starting at $\pm2\sigma$ up to $9\sigma$. The blue portion of the line ({\it left}) appears to be unresolved, while the red portion ({\it right}) appears to be extended towards the continuum source. In the bottom left corner, we show the ALMA synthesized beam of 0\farcs63$\times$0\farcs49 and position angle of 86\fdg16.
\label{fig:channel_maps}}
\end{figure*}

We can compare the flux estimated from assuming a point source with that obtained from integrating the collapsed line. The fluxes agree for most of the reliable emission line candidates. However, for A2744-EL01 we find an integrated flux of $4.3 \pm 0.6$, at least twice as large as obtained when assuming a point source (Table \ref{tab:ELC_detection_properties}), indicating that the line could be resolved; this adds additional confidence to its detection. In Fig.~\ref{fig:channel_maps} we additionally present velocity channel contours from the blue and red wings of the corresponding emission line for A2744-EL01.  There appears to be a central unresolved component in both component maps, but significantly more extended emission in the red wing. The extended structure has a signal-to-noise of $\approx$$4.3$ over the 32 native channels ($\approx290\kms$) included in the red wing. Some of the extended emission appears to spatially overlap with the offset bright, extended continuum source, A2744-ID01.

\section{Discussion}

\subsection{Bright emission lines at high redshift}
With several reliable line candidates in hand, the next step is to identify which molecular species and transition might be detected. We adopt the brightest emission lines observed at high redshift as possible identifications.
For galaxies at $z<3$, the brightest emission lines correspond to lines from CO with upper level transitions in the range $J$$=$1--8 or atomic carbon (CI) $J$$=$1--2; other CI lines are generally considered too weak to be relevant \citep{Walter2011}. At $z$$\sim$4.5, a possible candidate is the 205 \m\ts\ts transition of \nii, which \cite{Decarli2014b} demonstrated as viable based on detections in a submillimeter galaxy (SMG) and two Lyman$-\alpha$ emitters (LAEs) belonging to the system BRI1202--0725 at $z=4.7$. Lastly, at $z\sim6.2$ the \cii emission line at the 158 \m\ts\ts transition is expected to be detectable at our observed frequencies \citep{Riechers2013,Capak2015,Willott2015}. 

We also estimate the emission line integrated flux expected for the sources detected in continuum at 263.14 GHz and presented in GL17. We use the relation presented by \citet{Scoville2016} to estimate the molecular gas mass and CO line luminosity based on the continuum flux density. As shown in \citet{Laporte2017}, all the continuum detections are expected to be in the redshift range of $z$$=$1--3, for which we could potentially detect CO lines with $5\leq J\leq8$. For the faintest source detected in GL17, with $S_{\nu}\approx0.4$ mJy, we obtain a molecular gas mass of $\sim4\times10^{10}\,{\rm M}_{\odot}$ in that redshift range. Assuming a CO-SLED appropriate for SMGs with ${\rm L'}_{CO(8-7)}/{\rm L'}_{CO(1-0)}\sim{\rm L'}_{CO(5-4)}/{\rm L'}_{CO(1-0)}\sim0.39$  and $\alpha_{\rm CO}= 0.8\ts\ts\msol / ({\rm K}\kms{\rm s}^{-1}{\rm pc}^2)$ \citep{Daddi2010b,Carilli_Walter2013} we obtain an integrated flux of $>4\,{\rm Jy}\kms$ for CO lines with $5\leq J\leq8$, $\sim2\times$ higher than the brightest line candidate detected. Even adopting a lower excitation CO-SLED with ${\rm L'}_{CO(8-7)}/{\rm L'}_{CO(1-0)}\sim{\rm L'}_{CO(5-4)}/{\rm L'}_{CO(1-0)}\sim0.02$, we should still expect to obtain CO lines integrated fluxes of $>0.2\,{\rm Jy}\kms$, which are of the same order as the faintest line candidates found in the ALMA-FF data cubes and therefore detectable by our observation (Table \ref{tab:ELC_properties}). None of the continuum sources show emission lines in their spectra, implying that all of them have redshifts outside the ranges presented in Table \ref{tab:redshift_coverage_description}.

\subsection{Redshift and line identification}
We describe below the process by which we identify the most likely molecular transition(s) for each of the eight most reliable emission line candidates. We make the implicit assumption here that the photometric or spectroscopic redshift of the nearest counterpart applies also to the emission line, in order to limit the possible line identifications we might detect. We use the NIR photometry of the nearest counterpart to estimate the SFR by fitting the spectral energy distribution using the high-redshift MAGPHYS code  \citep{daCunha2008,daCunha2015}.

\textbf{A2744-EL01:} This line candidate is found in between three potential counterpart galaxies. The closest galaxy corresponds to the red circle marked in Fig.~\ref{fig:line_reliable}. The emission line also lies at the edge of a bright galaxy 1\arcsec\ts\ts to the south of the line peak. However, despite lying relatively close to both galaxies, the emission line clearly extends toward the morphologically disturbed galaxy associated with the 1.1\,mm continuum source A2744-ID02 reported in GL17 and \citet{Laporte2017}. Given the low source density of ALMA 1.1\,mm continuum detections in FF clusters, and the low number of reliable emission line candidates, the probability of a chance match within 2\arcsec\ts\ts is low ($\lesssim$1\%). We thus favor the scenario where the emission line is somehow physically associated with A2744-ID02. The NIR counterpart of A2744-ID02 appears to be extended along an East-West axis, with 2--3 faint NIR objects ("clumps") lying on either side of the extended 1.1\,mm continuum emission and fainter "bridge" emission extending both in between these clumps as well as toward a bright compact peak of NIR emission $\sim1\farcs5$ to the west of the ALMA continuum source. The resolved continuum emission appears to be elongated roughly coincident with a suppression in the $F160W$ emission, suggesting that the ALMA source may arise from a dusty region that divides the faint $F160W$ clumps, which may represent less-obscured regions from a single extended object with strong and variable extinction. 

The redshift for A2744-ID02 is reported in \citet{Laporte2017} as $z\approx2.482$ for the brightest red clump. However, it should be noted that this redshift is estimated from the 4000\AA\ts\ts break and not from any clear emission lines in the NIR. Based on this, the best identification for A2744-EL01 would be CO(8-7) at $z=2.61975\pm0.0003$, which we consider as tentative.

\textbf{MACSJ0416-EL01:} With a line detection of $S/N=6.4$, this is one of only two candidates that are likely to be real based on the line detection alone. Surprisingly, the nearest galaxy lies $\sim1.6\arcsec$ from the emission line candidate. Such large distance implies that the line has a low probability of being physically related to this galaxy, and there is no obvious line identification near the counterpart redshift. The other nearby galaxies at larger separations also do not appear to be related to the emission line. Therefore we cannot find a reliable counterpart to the emission line candidate.
We consider the possibility of the counterpart galaxy being strongly obscured. As already discussed above, an obscured galaxy with SFR$<30-40 \,\,\msol\,{\rm yr}^{-1}$ could remain undetected in the current {\it HST} and ALMA observations over a range of redshifts from $z=1-6$. The only way to confirm and identify this associated transition of this emission line is through deeper observations of both this and other potential transitions.

\textbf{MACSJ0416-EL03:} This candidate is similar to A2744-EL01, in that it lies near a galaxy that is detected in 1.1\,mm continuum with S/N=4.6, MACSJ0416-ID06. This 1.1\,mm continuum source belongs to a sample of lower significance ($S/N<5$) detections selected by their close association to NIR galaxies, to be presented in \citet{Gonzalez-Lopez2017b}. Thus we favor a scenario whereby the emission line is related to the counterpart of MACSJ0416-ID06, and not to the nearest source marked by the red circle. The line and 1.1\,mm continuum association imply a detection probability of $P<0.01$. The NIR galaxy has been detected as part of the Grism Lens-Amplified Survey from Space \citep[GLASS;][]{Treu2015} with a redshift of $z=1.849$. However, at this redshift none of the bright line transitions plotted in Fig.~\ref{fig:redshift_coverage} agree as an identification for the observed emission line. Thus we additionally explore the photometric redshift range in hope of finding an identification to this line. Using the high-redshift MAGPHYS code  \citep{daCunha2008,daCunha2015,Gonzalez-Lopez2017b} we obtain a photometric redshift of $z=1.87^{+0.32}_{-0.33}$, in strong agreement with the GLASS redshift. The closest line to that redshift would be CO(7-6) at $z=1.96016\pm0.00003$. The detection of CI(2-1) would confirm the line identification, but unfortunately its expected frequency of 273.4 GHz is outside the SPW coverage.
The discrepancy between the observed frequency and the spectroscopic redshift, as well as the fact that the line width is of only $\sim32\kms$ suggest that this line is not reliable. Therefore we consider this identification as tentative.

\textbf{MACSJ1149-EL03:} We assume this line has a counterpart galaxy as marked in Fig. \ref{fig:line_reliable}. The photometric redshift of this object is $z=2.94^{+0.56}_{-0.55}$ using MAGPHYS and $z=0.28^{+0.05}_{-0.03}$ using \textit{Hyperz}. For the high redshift solution, we enter the regime of CO emission lines with $J>8$, which have been observed in only extreme "hot" systems (e.g., powerful SMGs and AGN) at high redshift. However, the multiwavelength data do not support such a counterpart. For the low redshift solution, the line would be CO(3-2) giving a redshift of $z=0.36031\pm0.00003$. The observed SFR would be $\sim0.01\ts\msol\,{\rm yr}^{-1}$, implying an expected flux for CO(3-2) of $F\sim0.002$ Jy\kms. This also appears inconsistent with the observed line properties. Thus, no solution appears viable.
 
\textbf{A370-EL01:} This line appears associated with a cluster galaxy ($z = 0.3599$) marked in Fig. \ref{fig:line_reliable2}. Based on Fig.~\ref{fig:redshift_coverage}, the best line identification corresponds to CO(3-2), implying a redshift of $z_{\rm CO}=0.35962\pm0.00001$ and a velocity shift of only $\sim60\kms$ with respect to the optical redshift (i.e., in good agreement). We consider this line identification to be secure.

\textbf{A370-EL04:} This line appears spatially coincident with the counterpart galaxy marked in Fig. \ref{fig:line_reliable3}. The photometric redshift for the galaxy is $z_{\rm ph}=0.40_{-0.23}^{+0.55}$. This is consistent with the cluster redshift ($z=0.375$) although the large uncertainties easily allow $z\la1.0$. The closest lines shown in Fig.~\ref{fig:redshift_coverage} are CO(3-2), which would put the galaxy at $z=0.27640\pm0.00001$, in the foreground of the cluster, or CO(4-3), which would put the line at $z\approx0.7$. At $z=0.27640\pm0.00001$ we obtain a ${\rm SFR}=0.01^{+0.02}_{-0.0005}\ts\msol\,{\rm yr}^{-1}$, which is fairly low to explain the observed line flux. As such, the line appears real, but since no secure identification can be made we consider it tentative.

\textbf{AS1063-EL01:} The line appears associated with a cluster galaxy ($z_{\rm GLASS} = 0.350$) marked in Fig. \ref{fig:line_reliable3}. Based on Fig.~\ref{fig:redshift_coverage}, the best line identification corresponds to CO(3-2) at a measured redshift of $z_{\rm CO}=0.35080\pm0.00001$, showing a velocity shift of only $\sim180\kms$ with respect to the optical redshift (in good agreement within the uncertainty of the GLASS redshift). We consider this line identification to be secure.

\textbf{AS1063-EL03:} The line appears robustly coincident with galaxy marked in Fig. \ref{fig:line_reliable3}. The photometric redshift for the source is $z_{\rm ph}=2.91^{+0.57}_{-0.53}$. The closest line would correspond to CO(9-8) at $z=3.0695\pm0.0002$, in good agreement with the photometric redshift. The second option is that the line corresponds to CO(8-7) at $z=2.6177\pm0.001$, which is still accommodated within the error bars of the photometric redshift. We adopt CO(9-8) since it lies closer to the best-fit photometric redshift. We estimate ${\rm SFR}=2.6^{+1.2}_{-1.0}\ts\msol\,{\rm yr}^{-1}$ using MAGPHYS. We do not attempt to estimate the SFR from the line flux, since the line emission is likely to be dominated by our assumption about the CO excitation. We consider this line identification to be tentative. 

\textbf{AS1063-EL08:} As can be seen in Fig.~\ref{fig:line_reliable4}, photometry for the nearest counterpart galaxy is plagued by scattered light from the cluster members. This leaves photometric redshift estimates highly uncertain.  Thus we consider the line identification as only tentative.

In summary, A370-EL01 and AS1063-EL01 are classified as secure identifications with CO(3-2). A2744-EL01, MACSJ0416-EL03, A370-EL04 and AS1063-EL03 are classified with a tentative identification with CO emission lines. And MACSJ0416-EL01, MACSJ1149-EL03 and AS1063-EL08 do not have any clear identifications. 
We remark that seven out of the nine reliable line candidates were selected mainly by their association to NIR sources. Out of those, five return tentative or secure line identifications given their spectroscopic or photometric redshifts. This consistency is reassuring and strengthens the case that such associations to NIR sources are real. The best case is AS1063-EL01, where the close association to a NIR galaxy and the availability of a spectroscopic redshift combined were able to fully confirm a line candidate detected with low-significance.

\subsection{Emission line widths}\label{EL_width}

We now want to check whether the observed line widths are consistent with the observed galaxy counterparts. The narrowest emission lines that have been robustly observed at high redshift are on the order of ${\rm FWHM} \sim50 \kms$ \citep{Carilli2013,Decarli2014b,Decarli2016,Walter2016}.  More than half of our emission line candidates have line widths near or lower than this value, implying that they may not be real. On the other hand, we need to take into account that our observations are more sensitive to narrower emission lines than previous observations, because of the improved ALMA sensitivity coupled with the observation of strong-lensing fields. 

On the other hand, the method used to find emission line candidates is likely biased toward selecting narrow emission lines. Any method that selects emission line candidates based only on their observed signal-to-noise will show such a bias toward narrow emission lines, independent of the technique used to estimate the FDR. Narrow emission lines at a given signal-to-noise are more likely to be produced by noise than broader ones, this happens because narrow emission lines allow for a search over a larger number of independent elements (spectral channels) compared to broader lines. Because of this, when searching for low signal-to-noise line candidates, noise contamination will readily produce false lines dominated by narrow emission lines rather the broad ones.

In conclusion, we acknowledge that the narrow emission line candidates selected have a higher probability of being false, based on the nature of the search itself and the low number of robustly detected narrow emission lines at high redshift in the literature. Imposing a reasonable velocity width cut at ${\rm FWHM} \geq50 \kms$ would reduce the number of reliable line candidates from nine to seven. However, we retain such narrow emission line candidates since they could represent the exploration of a new parameter space opened by the ALMA sensitivity and lensing fields observations.

\subsection{Magnification values}

\begin{table}
\caption[]{Magnification values for the reliable emission line candidates with tentative redshift identification. 
\label{tab:magnifications}}
\centering
\begin{tabular}{cccc}
\hline     
\noalign{\smallskip}
ID & {Assumed Redshift} & {$\mu_{\rm parametric}$} & {$\mu_{\rm all}$} \\
\hline
\noalign{\smallskip}
A2744-EL01       & $z=2.6197$    & $1.9^{+0.6}_{-0.4}$   & $1.9^{+0.6}_{-0.4}$   \\
MACSJ0416-EL03    & $z=1.9601$    & $1.5^{+0.1}_{-0.1}$   & $1.5^{+0.1}_{-0.3}$   \\
AS1063-EL03       & $z=3.0695$    & $1.6^{+0.2}_{-0.2}$   & $1.6^{+0.3}_{-0.4}$   \\
\hline
\end{tabular}
\tablefoot{\\
Models used for $\mu_{\rm parametric}$ in A2744: CATSv3, CATSv3.1, Sharonv3, Zitrin-NFWv3, Zitrin-LTM-Gaussv3, GLAFICv3\\
Models used for $\mu_{\rm parametric}$ in MACSJ0416: CATSv3, CATSv3.1, Sharonv3, Zitrin-NFWv3, Zitrin-LTM-Gaussv3, GLAFICv3\\
Models used for $\mu_{\rm parametric}$ in AS1063: CATSv1, Sharonv1, Sharonv2, Zitrin-NFWv1, Zitrin-LTMv1 and Zitrin-LTM-Gaussv1\\
Models used for $\mu_{\rm all}$ in A2744: CATSv3, CATSv3.1, Sharonv3, Zitrin-NFWv3, Zitrin-LTM-Gaussv3, GLAFICv3, Williamsv3\\
Models used for $\mu_{\rm all}$ in MACSJ0416: CATSv3, CATSv3.1, Sharonv3, Zitrin-NFWv3, Zitrin-LTM-Gaussv3, GLAFICv3, Williamsv3, Williamsv3.1, Diegov3\\
Models used for $\mu_{\rm all}$ in AS1063: CATSv1, Sharonv1, Sharonv2, Zitrin-NFWv1, Zitrin-LTMv1, Zitrin-LTM-Gaussv1, Williamsv1 and Mertenv1\\}
\tablebib{(1)~\citet{Bradac2005}; (2) \citet{Bradac2009}; (3) \citet{Diego2015}; (4) \citet{Jauzac2015a};
(5) \citet{Jauzac2015b}; (6) \citet{Jauzac2016}; (7) \citet{Johnson2014}; (8) \citet{Kawamata2016};
(9) \citet{Liesenborgs2006}; (10) \citet{Merten2009}; (11) \citet{Merten2011}; (12) \citet{Sebesta2015}; (13) \citet{Williams2014}; (14) \citet{Zitrin2009}; (15) \citet{Zitrin2013}.
}
\end{table}

We use the tentative redshift identifications for the emission line candidates behind the clusters to more precisely estimate their magnification values. We estimate the magnification values as in \cite{Coe2015}, by taking the 16th, 50th and 84th percentiles of the magnification values (representing the central $\approx68\%$, or $1\sigma$ range, of the magnification distribution) estimated from both a range of parametric mass models and from all available models in FF archives.
For A2744 and MACSJ0416 we only use the v3 or newer models, since these account for the most recent published redshifts and multiple image identifications. When no v3 models are available (such as for MACSJ1149), we use the latest available for each modeler. For AS1063 we use the only two models available: Zitrin-NFWv1 and Zitrin-LTM-Gaussv1. 

In Table~\ref{tab:magnifications} we present the derived magnification values for the sources with tentative redshifts using the magnification webtool\footnote{https://archive.stsci.edu/prepds/frontier/lensmodels/webtool/magnif.html}. All of the secure or tentative sources are consistent with having $\mu\la2$, similar to what was found for the ALMA-FF 1.1\,mm continuum detected sources \citep[GL17;][]{Laporte2017}.

\subsection{Molecular gas}
For the emission line candidates with the most reliable identifications, A2744-EL01, MACSJ0416-EL03, A370-EL01 and AS1063-EL01, we estimate the amount of molecular mass inferred from the corresponding line detections. We do not include AS1063-EL03, because the CO(9-8) line is not easy to model. In the first two cases the lines are tentatively identified with high J values (CO(7-6) and CO(8-7)), while the latter correspond to CO(3-2).
Our estimate of the molecular mass will be limited by two associated uncertainties, the conversion factor ($\alpha_{\rm CO}$), which relates the CO(1-0) luminosity to molecular gas, and the CO excitation level of the galaxies.
The conversion factor depends strongly on the properties of the galaxy and whether they are in a normal star-forming or starburst phase. We use the value found for local Ultra Luminous Red Galaxies (ULIRGs) of $\alpha_{\rm CO}= 0.8\pm0.8\ts\ts\msol / ({\rm K}\kms{\rm s}^{-1}{\rm pc}^2)$ for dusty star-forming galaxies (DSFGs) at high redshift \citep{Downes1998} and $\alpha_{\rm CO}= 3.6\pm0.8\ts\ts\msol / ({\rm K}\kms{\rm s}^{-1}{\rm pc}^2)$ for low redshift / cluster sources \citep{Daddi2010} given their low estimated SFRs. 

The CO excitation ladder depends on the temperature and density of the molecular gas. Without more information about the emitting galaxies, it is highly uncertain to estimate the excitation level from CO(8-7) and CO(7-6) to CO(1-0). We rely here on the values determined for SMGs of ${\rm L'}_{CO(8-7)}/{\rm L'}_{CO(1-0)}={\rm L'}_{CO(7-6)}/{\rm L'}_{CO(1-0)}=R_{71}\sim0.39$. For CO(3-2), we use a value of $R_{31}\sim0.56$. 
For A2744-EL01, MACSJ0416-EL03, A370-EL01, and AS1063-EL01, we obtain  intrinsic molecular masss of 
${\rm M}_{\rm H2}=(2.4\pm0.4)\times10^{10}\times\left(\frac{\alpha_{\rm CO}}{0.8}\right)\left(\frac{R_{81}}{0.39}\right)\left(\frac{1.9}{\mu}\right)\msol$, ${\rm M}_{\rm H2}=(2.4\pm0.3)\times10^{9}\times\left(\frac{\alpha_{\rm CO}}{3.6}\right)\left(\frac{R_{71}}{0.39}\right)\left(\frac{1.5}{\mu}\right)\msol$, ${\rm M}_{\rm H2}=(4.2\pm0.7)\times10^{9}\times\left(\frac{\alpha_{\rm CO}}{3.6}\right)\left(\frac{R_{31}}{0.56}\right)\msol$ and 
${\rm M}_{\rm H2}=(2.0\pm0.6)\times10^{9}\times\left(\frac{\alpha_{\rm CO}}{3.6}\right)\left(\frac{R_{31}}{0.56}\right)\msol$, respectively.

The molecular mass values calculated for the four galaxies above can be used to estimate the expected continuum flux densities adopting relation from \citet{Scoville2016}. For A2744-EL01, the expected continuum level is $\sim$0.4 mJy; this should have been detected in the $S/N\geq5$ analysis presented in GL17. The associated continuum source A2744-ID01 has a flux density of $1.570\pm0.073$ mJy, which is in reasonable agreement (although with a puzzling spatial offset). 
For the other three emission line sources, the expected continuum level is $<$0.1 mJy, which is of the same order as the rms values of the continuum images. Such low values expected for the continuum emission would explain the lack of continuum counterparts for the majority of emission line candidates shown in Figure \ref{fig:line_reliable}. In the case of MACSJ0416-EL03, we actually detect continuum at low significance; this 'enhanced' continuum is presumably within the dispersion of the estimate.

We can then estimate the depletion timescales over which the star-formation would use up the available molecular gas, $t_{\rm dep}={\rm M}_{\rm H2}/{\rm SFR}$. We find depletion times of $t_{\rm dep}\sim0.1$ Gyr and $t_{\rm dep}\sim0.02$ Gyr for A2744-EL01 and MACSJ0416-EL03 respectively. Alternatively, for the cluster galaxies A370-EL01 and AS1063-EL01, we find depletion times of $t_{\rm dep}\sim0.9$ Gyr and $t_{\rm dep}\sim0.07$ Gyr, respectively. For A370-EL01 we use a ${\rm SFR}=4.6\,\msol\,{\rm yr}^{-1}$ and ${\rm SFR}=28\,\msol\,{\rm yr}^{-1}$ for AS1063 (estimated from H$\alpha$ observations, see below). 

At $z\sim2$, starburst galaxies show depletion timescales in the range $\sim0.01-0.1$ Gyr, while main-sequence (MS) galaxies are expected to have $\sim0.3-0.6$ Gyr \citep{Saintonge2013,Aravena2016}. 
MACSJ0416-EL03 is in the middle of the range of depletion timescales for starburts galaxies, supporting the usage of $\alpha_{\rm CO}= 0.8$ estimated for starburst-like DSFGs at $z>1$. A2744-EL01 is in the top end of the depletion timescale range for starburst galaxies. If we use $\alpha_{\rm CO}= 3.6$, as estimated for color selected galaxies at $z\sim1.5$ \citep{Daddi2010}, we would obtain a molecular mass of ${\rm M}_{\rm H2}\sim1.1\times10^{11}\times\left(\frac{\alpha_{\rm CO}}{3.6}\right)\left(\frac{R_{81}}{0.39}\right)\left(\frac{1.9}{\mu}\right)\msol$ and a depletion time of $t_{\rm dep}\sim0.6$ Gyr, in the middle of the range for MS at $z\sim2$.
We conclude that, assuming our correct identification of the emission line candidate, MACSJ0416-EL03 would correspond to a starburst galaxy while A2744-EL01 would be similar to a MS galaxy, both at $z\sim2$.

Likewise, for A370-EL01 and AS1063-EL01, the first galaxy shows a depletion time in the range of MS galaxies while the latter shows depletion times comparable to starburst galaxies. For the case of A370-EL01, even using $\alpha_{\rm CO}= 0.8$ would give a depletion time near MS galaxies. \citet{Vulcani2016} presented H$\alpha$ observations for several galaxies of the FFs, from which the host galaxy of A370-EL01 is classified as a galaxy undergoing a major merger, with highly concentrated H$\alpha$ emission. The host galaxy of AS1063-EL01 is classified as a so-called 'jellyfish' galaxy with signs of ram pressure stripping, as the H$\alpha$ emission is asymmetric, typical of jellyfish galaxies falling onto galaxy clusters. The gas in such galaxies appears partially stripped, with substantial H$\alpha$ emission showing offset from the optical emission; in this case the offset is $\sim1.1$ kpc. It is important to note that the stripping direction is the same as the offset observed between the nuclear optical emission and the CO(3-2) emission detected by ALMA (see Fig. \ref{fig:line_reliable3}). It has been observed that the denser molecular gas is not perturbed as much as the atomic hydrogen gas in the infalling galaxies \citep{Ebeling2014,Vollmer2001}. Our observations would confirm this scenario, where the hydrogen gas has been almost completely removed but the denser molecular gas is still near the center. \citep{Vulcani2016} also suggested a merger signature for AS1063-EL01, again supporting the starbursting phase observed. 

\subsection{Comparison with blind line scan surveys}

We compare our results with those obtained in ASPECS in ALMA bands 3 + 6 \citep{Walter2016}. While the goal of our observations was to obtain a large area continuum image at 1.1\,mm, ASPECS aimed to obtain spectral scan observations covering a large frequency coverage (212-272 GHz) over the smaller HUDF area ($\approx1.5$ square arcminutes). In ASPECS, 11 emission lines are blindly detected, with the faintest emission line flux there being roughly equal to the faintest emission line discovered in this study, with $\approx0.22$ Jy km s$^{-1}$. 

The emission line detection rate in ASPECS is $0.13\pm0.04$ GHz$^{-1}$ square arcmin$^{-1}$ with the errors given by Poisson statistics. Applying the same restriction of $P_{S/N}\leq0.6$ (eight sources) as for ASPECS, we obtain a detection rate of $0.05^{+0.02}_{-0.02}$ GHz$^{-1}$ square arcmin$^{-1}$ in the ALMA-FFs. This substantially lower detection rate in the ALMA-FFs could be a product of the different observational design (continuum vs. line scan), cosmic variance or due to differences in how FDR is estimated. Only when we take the whole list of 26 lines candidates (assuming $P_{S/N}\leq0.99$) do we obtain a detection rate of $0.15^{+0.03}_{-0.03}$ GHz$^{-1}$ square arcmin$^{-1}$, more in line with the ASPECS results, although not directly comparable. 

\section{Summary and conclusion}
We have presented here an exhaustive emission line search through the band 6 ALMA  observations of five Frontier Fields. 

We searched for emission lines in the spectral cubes generated from the ALMA observations, recovering 26 emission line candidates with $P_{S/N}\le0.99$ (i.e., $S/N$$>$5.6). Two of these are considered secure based on their low probability of being produced by noise ($P_{S/N}\leq0.05$). We also identified line candidates which are located in close proximity to galaxies detected in {\it HST} NIR images and estimate the probability of finding such separations by chance. From this, we identified nine reliable emission lines candidates that have a low probability ($P=P_{S/N}P_{sep}\leq0.05$) of being false based on their  $S/N$ and association to a NIR galaxy. 

We used the {\it HST} photometry to estimate the photometric redshift of the counterpart galaxies and/or spectroscopic redshifts from the literature, allowing us to identify likely molecular line transitions for reliable line candidates. For six out of the nine line candidates, we find secure or tentative line transition identifications based on the redshift and observed line frequencies. For these, we further compared the observed line fluxes with those estimated from the observed SFRs and redshifts to weed out inconsistent identifications.

From the reliable CO lines identification, we estimate molecular gas masses for the counterpart galaxies in the range of ${\rm M}_{\rm H2}=(0.2-2.4)\times10^{10}\msol$. These imply depletion times in the range $\sim0.1-0.9$ Gyr. Because of this, we conclude that A2744-EL01 and AS1063-EL01 have depletion times consistent with starburst galaxies, while MACSJ0416-EL03 and A370-EL01 are more consistent with main sequence galaxies. 
We also note that AS1063-EL01 is associated with a jellyfish galaxy that shows signatures of stripping by ram pressure, as traced by offset H$\alpha$. The direction and level of offset between the NIR continuum, H$\alpha$ and CO(3-2) emission are consistent with the scenario where the less dense atomic hydrogen is stripped first/easier than the more dense molecular gas, as the CO emission arises from a region closer to the nucleus of the galaxy. 

For the emission lines still considered as tentative, spectroscopic redshifts for the NIR galaxies could help reveal whether the lines are real and correspond to expected bright emission lines. Alternatively, confirmation of other molecular transitions would help constrain the identification possibilities. 

We have shown that the combination of large area ALMA continuum observations and deep optical and NIR images can be successfully used to search for the detection of serendipitous emission lines. However, the narrow frequency coverage afforded by a single ALMA frequency setup clearly limits the detectability of secure bright emission lines to specific ranges of redshifts. Broader frequency coverage is required to assemble more complete samples, as well as improve the ability to confirm tentative line detections via the detection of multiple line transitions.

\begin{appendix}
\section{Emission line candidates}
Here we present the spectra and collapsed maps of the emission line candidates presented in Table \ref{tab:ELC_detection_properties}.
\begin{figure*}[!htbp]
\includegraphics[width=0.4\textwidth]{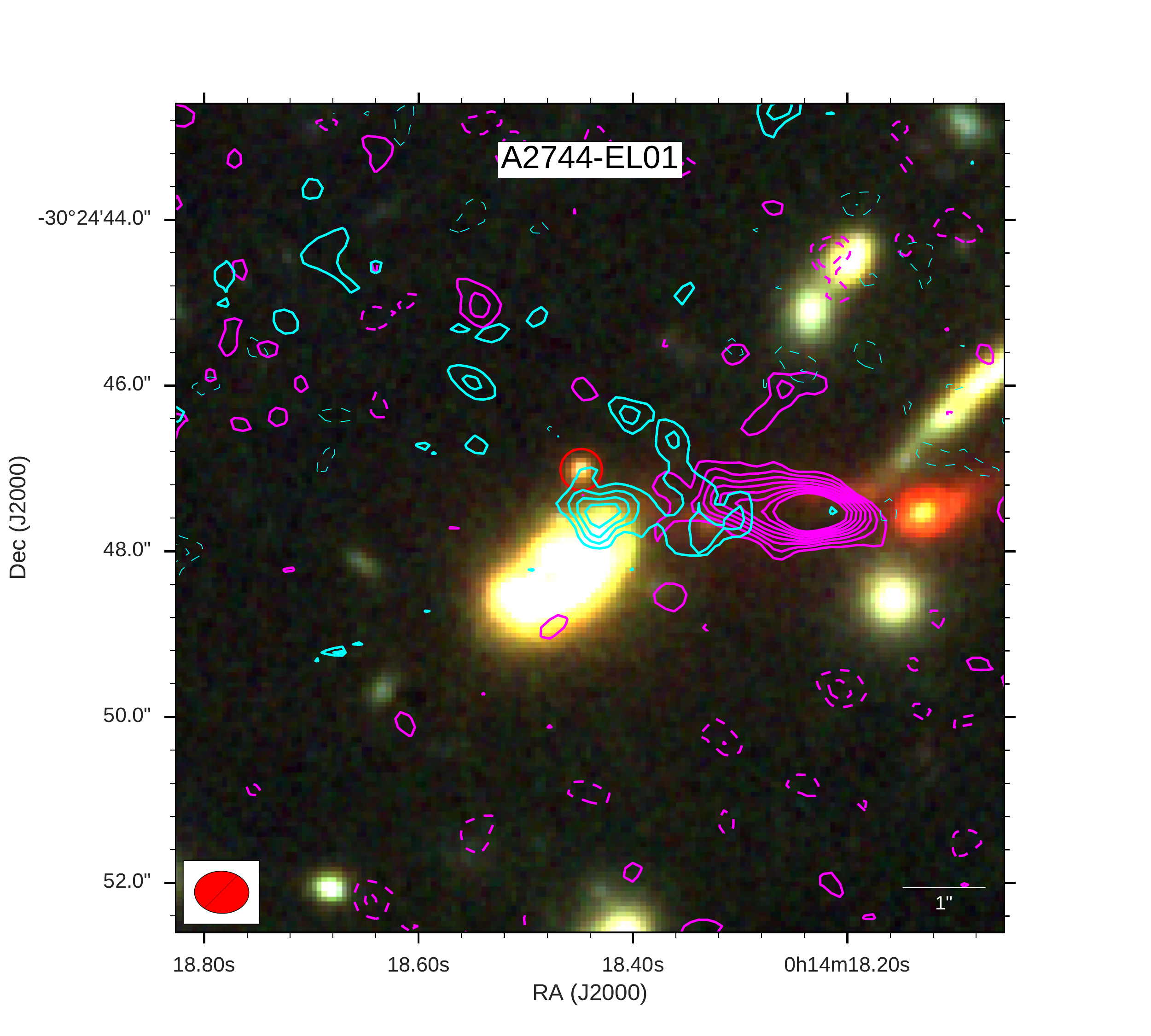}
\includegraphics[width=0.6\textwidth]{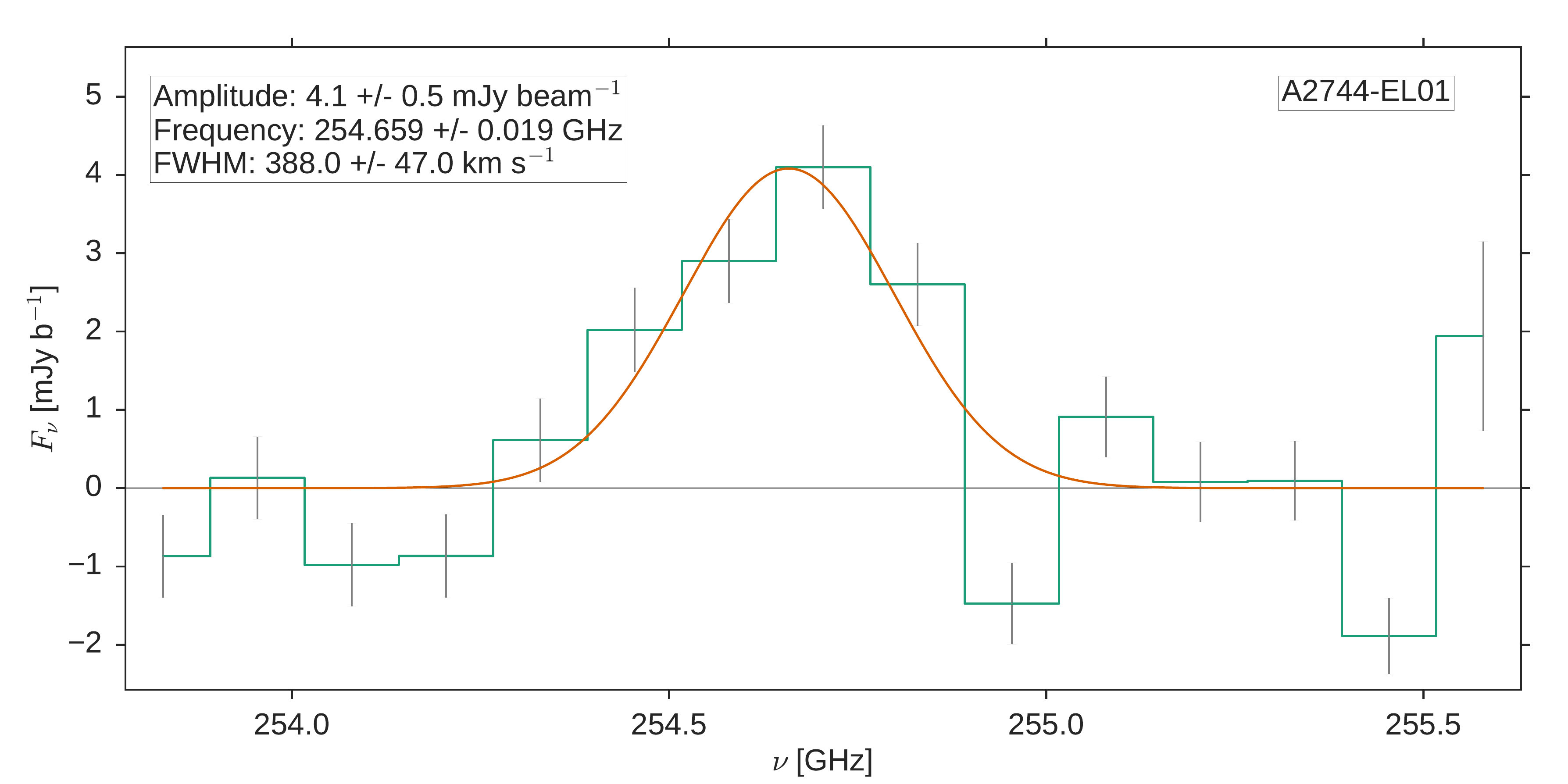}
\includegraphics[width=0.4\textwidth]{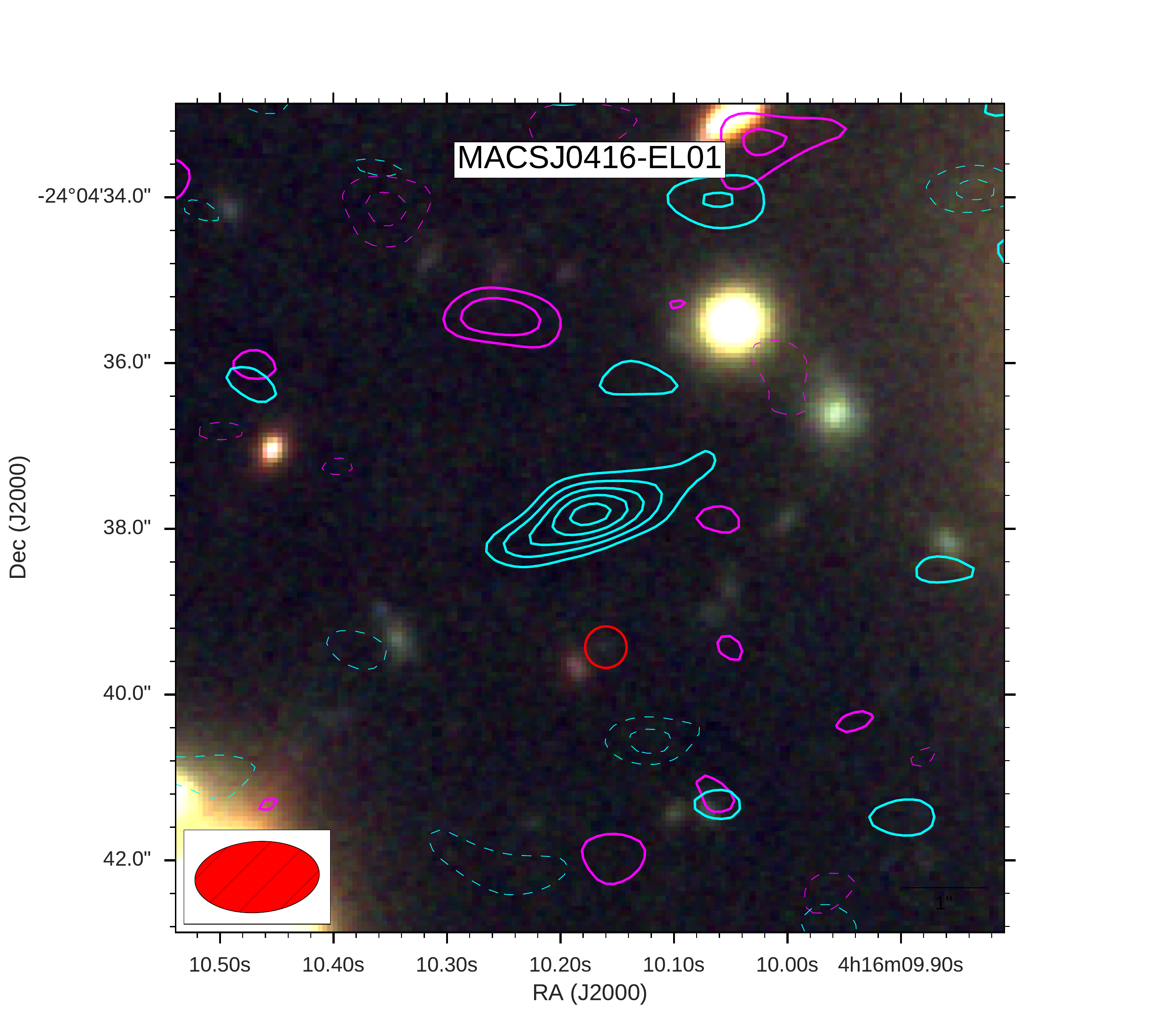}
\includegraphics[width=0.6\textwidth]{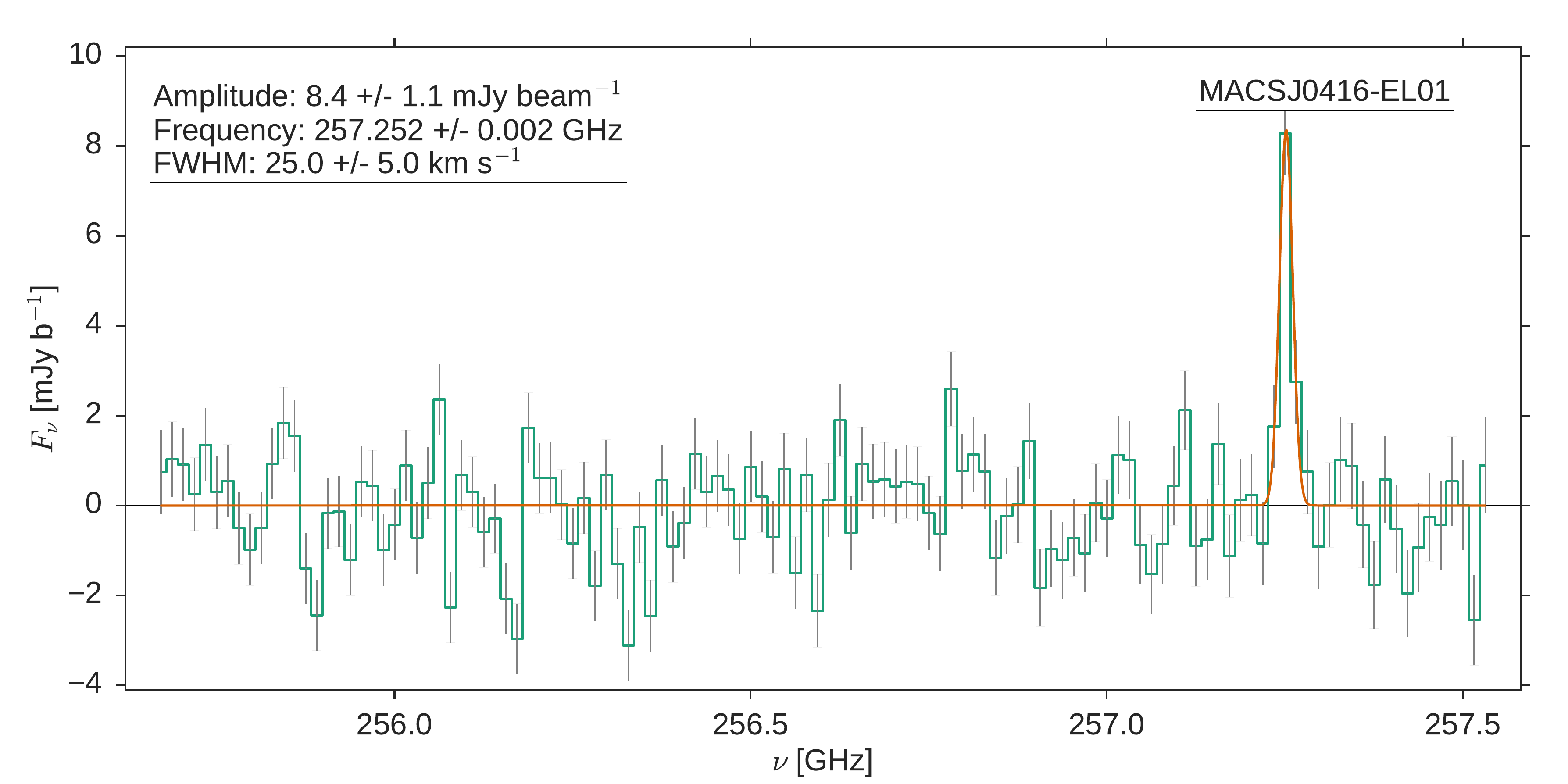}
\includegraphics[width=0.4\textwidth]{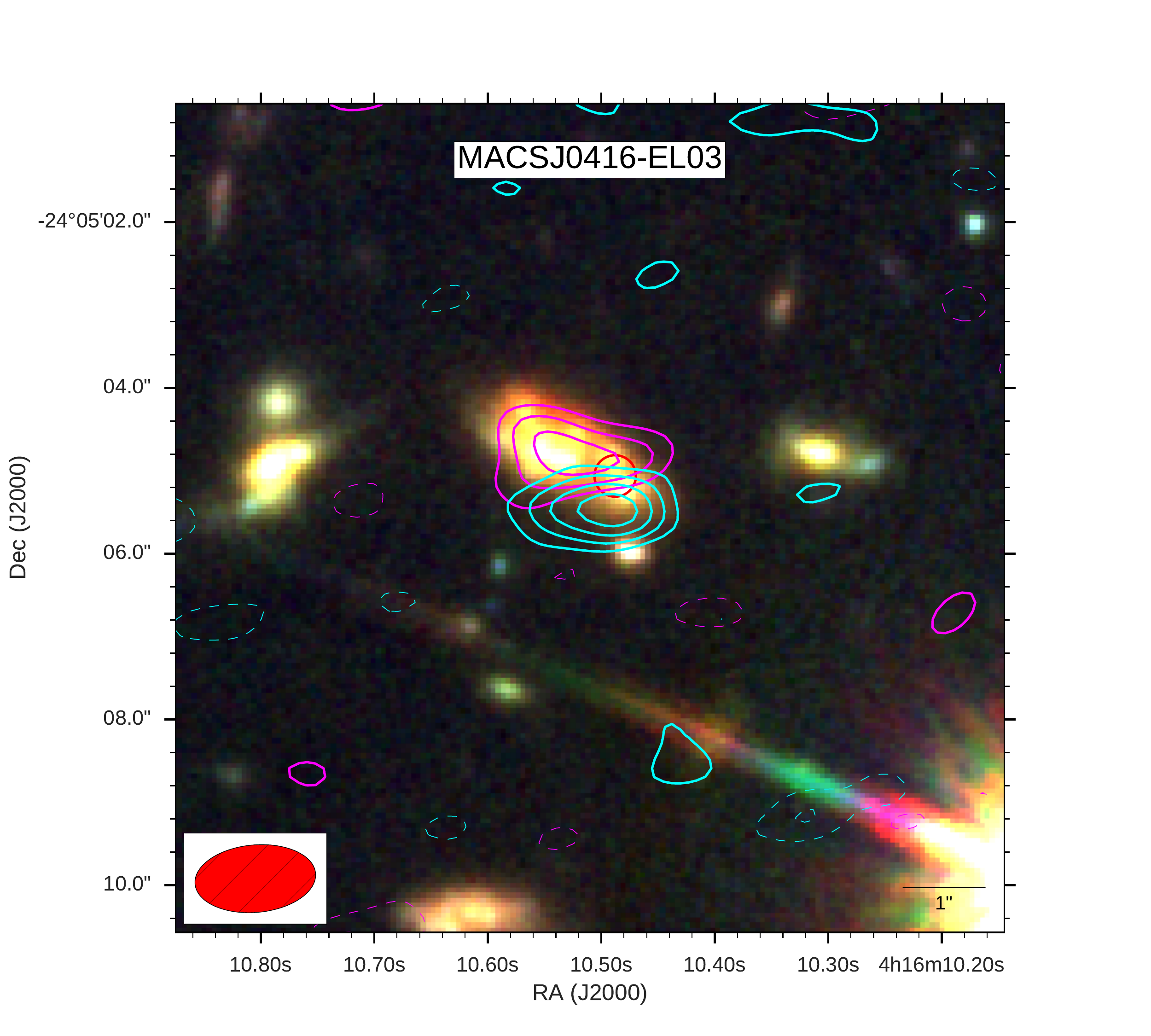}
\includegraphics[width=0.6\textwidth]{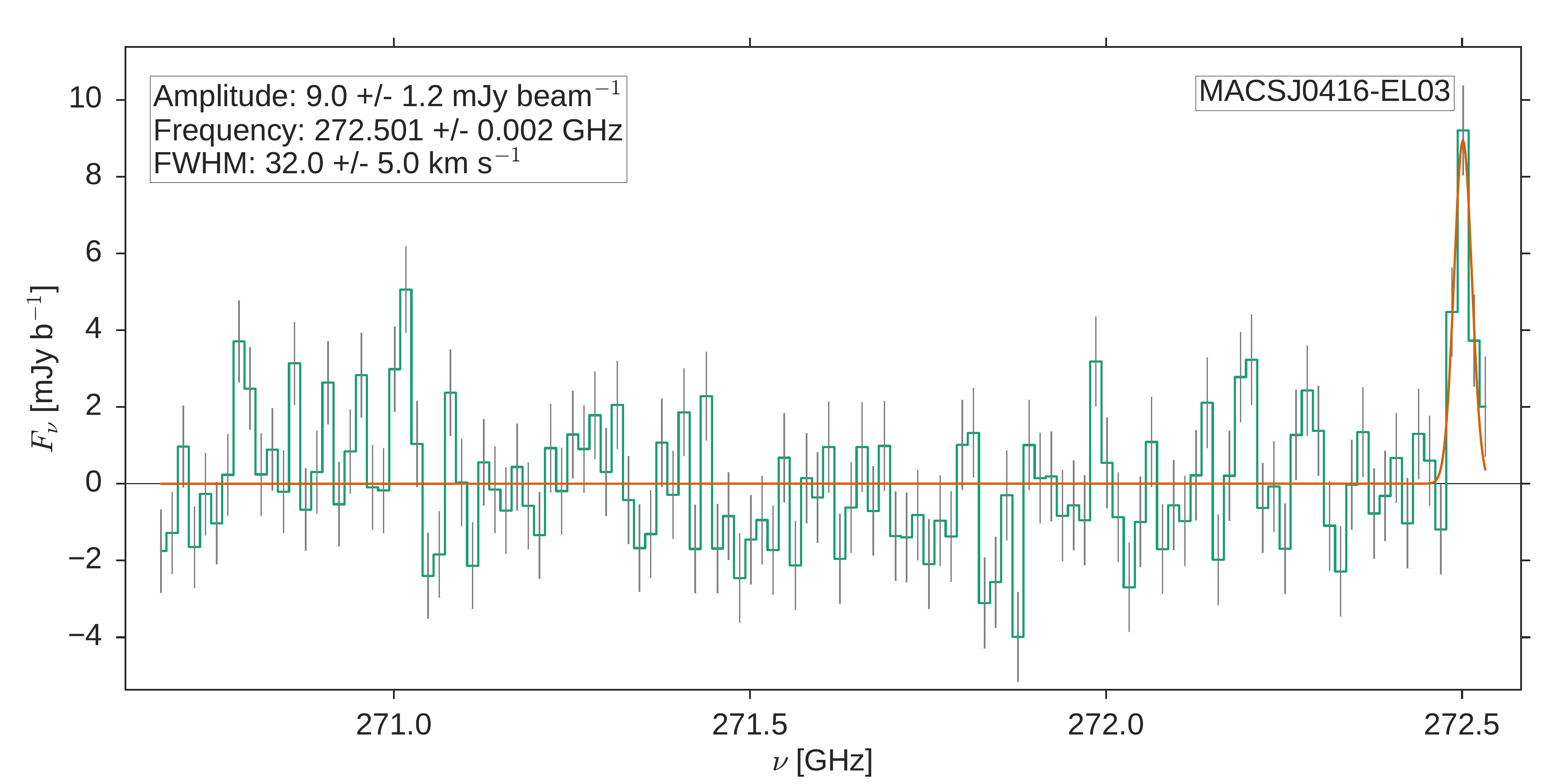}
\caption{Reliable candidates. Left: {\it HST} color image cutouts for the ALMA-FFs emission line candidates. Here bands $F435W+F060W$ correspond to blue, $F810W+F105W$ to green and $F125W+F140W+F160W$ to red, respectively. The cyan contours correspond to $S/N$ levels of the ALMA emission line candidates in $1\sigma$ steps starting at $\pm2$ times the primary beam corrected noise level at the position of the source. The continuum 1.1\,mm emission is plotted in magenta in steps starting at $\pm2\sigma$ up to $9\sigma$.
In the bottom left corner, we show the ALMA synthesized beam. Red circles mark the position of the nearest galaxy used to estimate $P_{sep}$.
Right: Spectra of the emission line candidates. We plot the observed spectrum extracted in the peak for each line in green with $1\sigma$ error bars and the best fit Gaussian function in orange. In each panel we show the complete frequency coverage (1.875 MHz) of the SPW where the emission line candidate is detected. The spectral resolution for each spectrum was chosen to best highlight the detection in each case. A2744-EL01 is identified tentatively as CO(8-7) at $z=2.61975\pm0.0003$, MASCJ0416-EL01 has no identification and MACSJ0416-EL03 is identified tentatively as CO(7-6) at $z=1.96016\pm0.00003$.
\label{fig:line_reliable}}
\end{figure*}

\begin{figure*}[!htbp]
\includegraphics[width=0.4\textwidth]{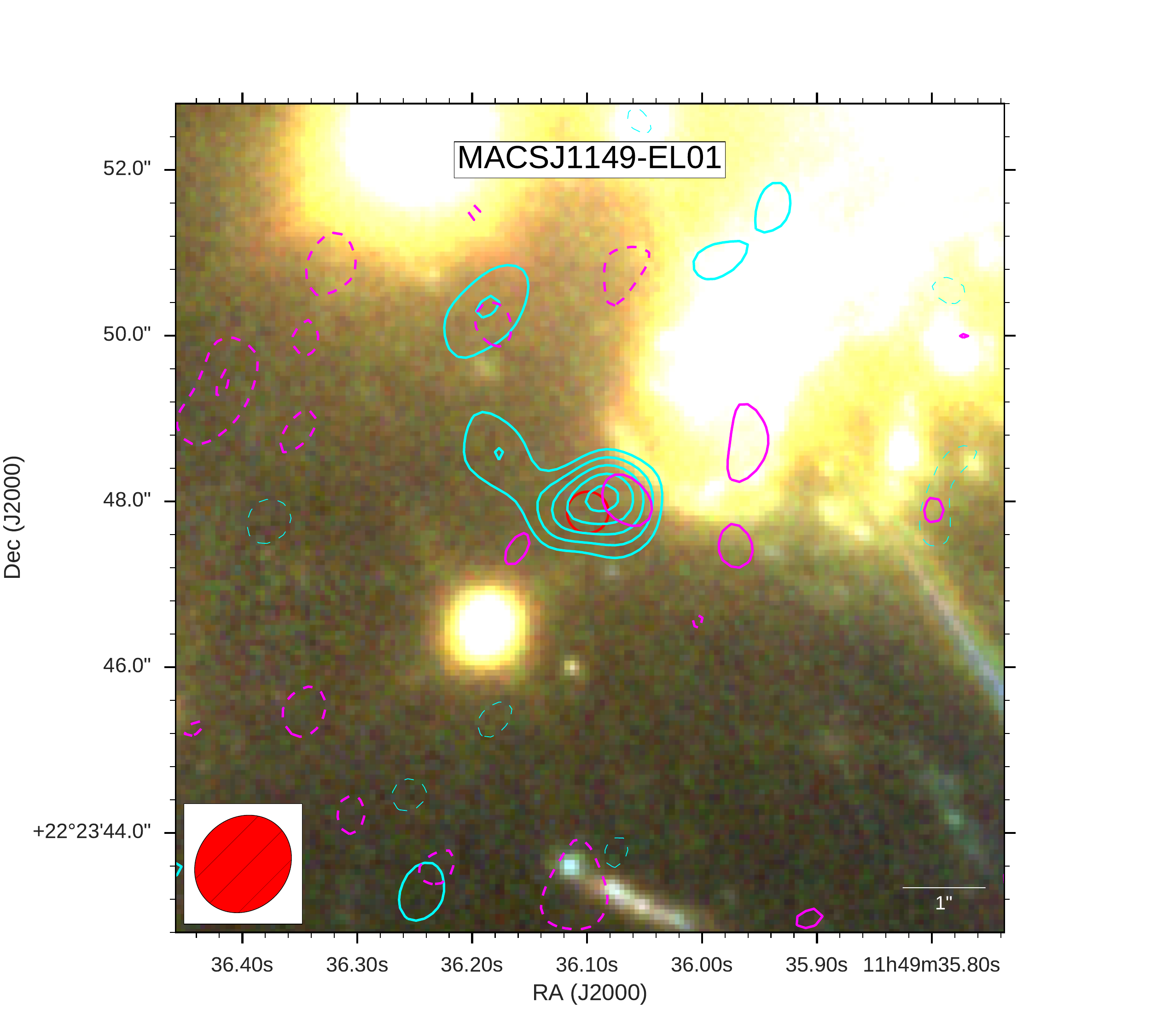}
\includegraphics[width=0.6\textwidth]{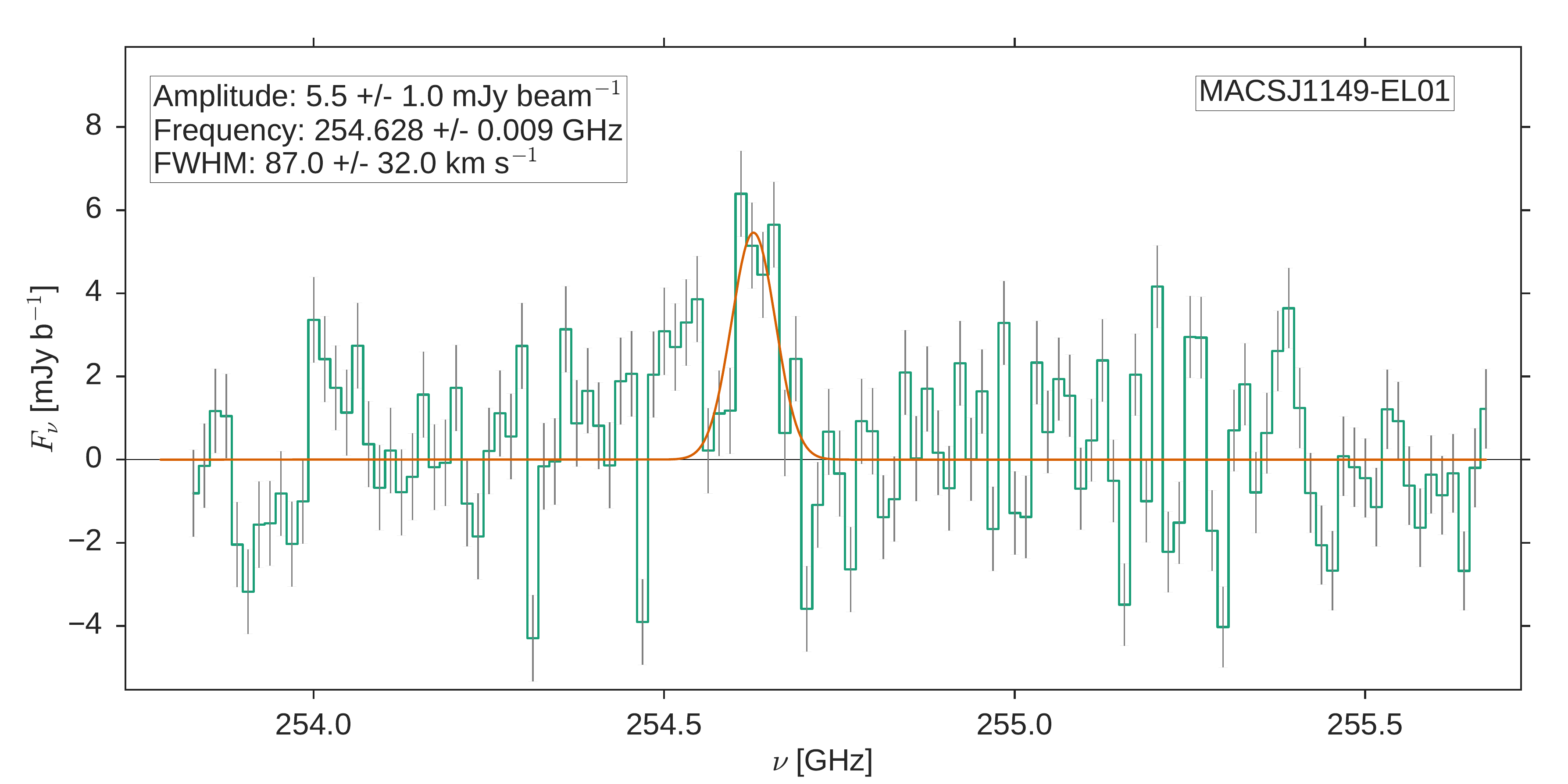}
\includegraphics[width=0.4\textwidth]{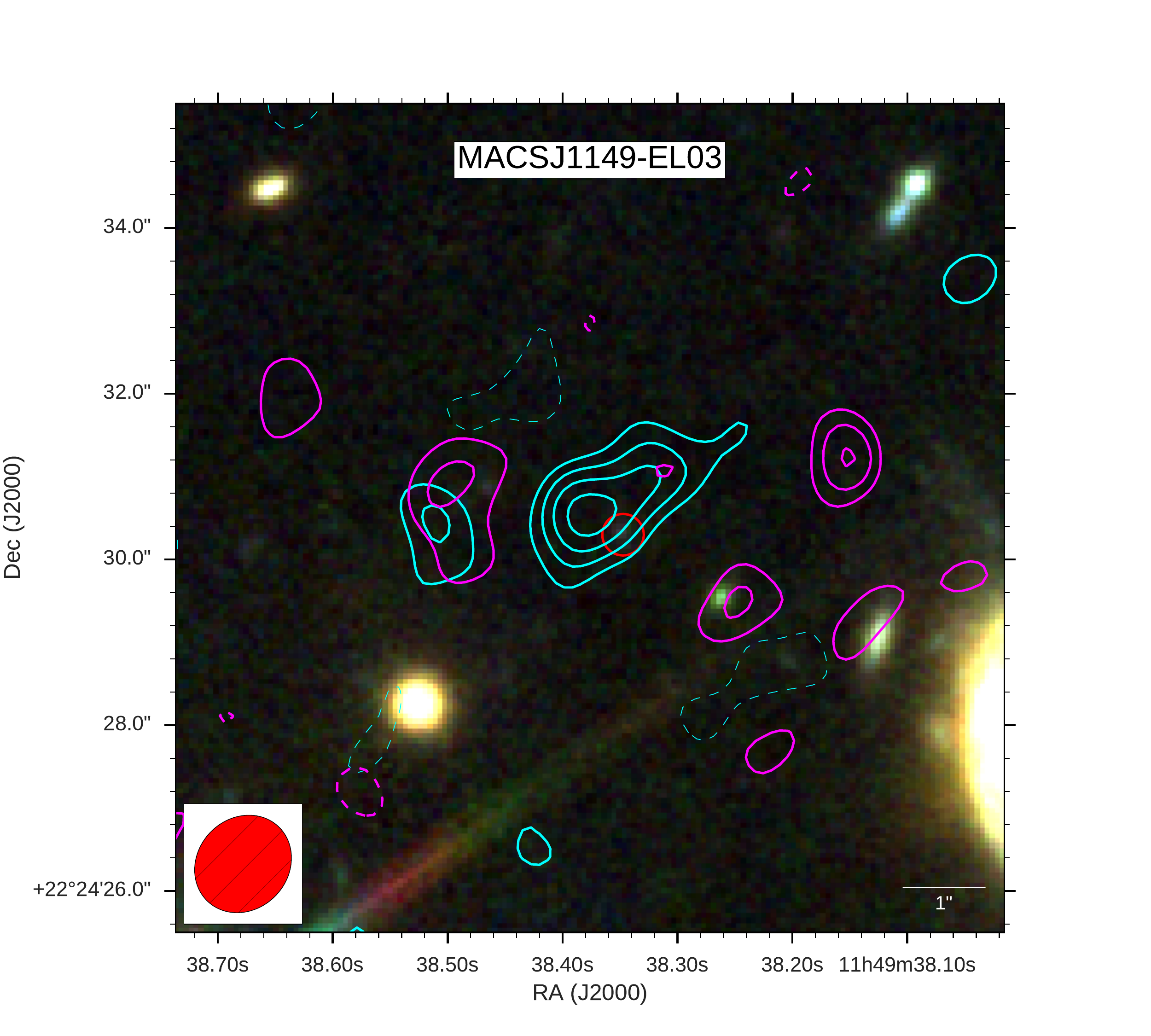}
\includegraphics[width=0.6\textwidth]{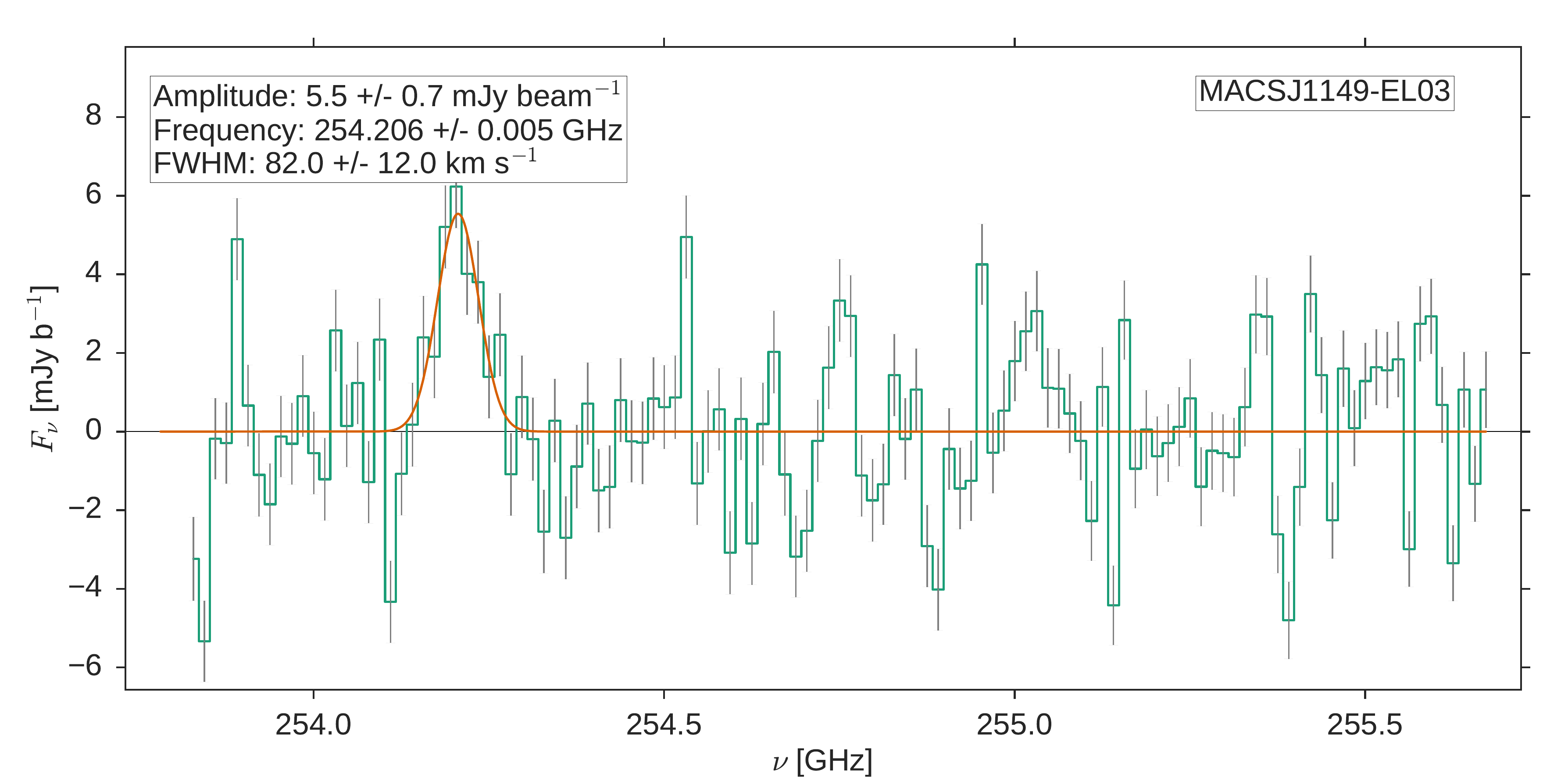}
\includegraphics[width=0.4\textwidth]{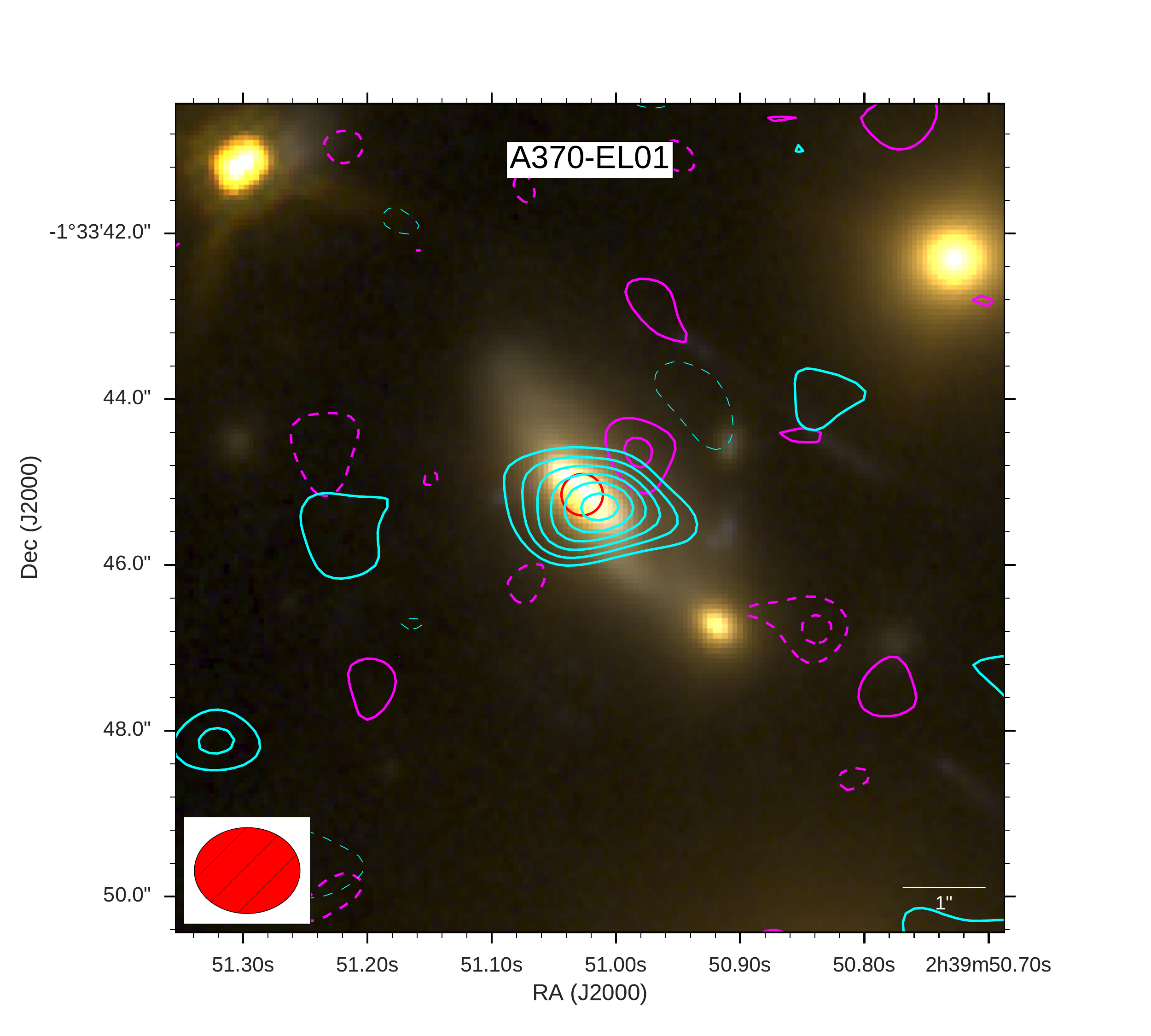}
\includegraphics[width=0.6\textwidth]{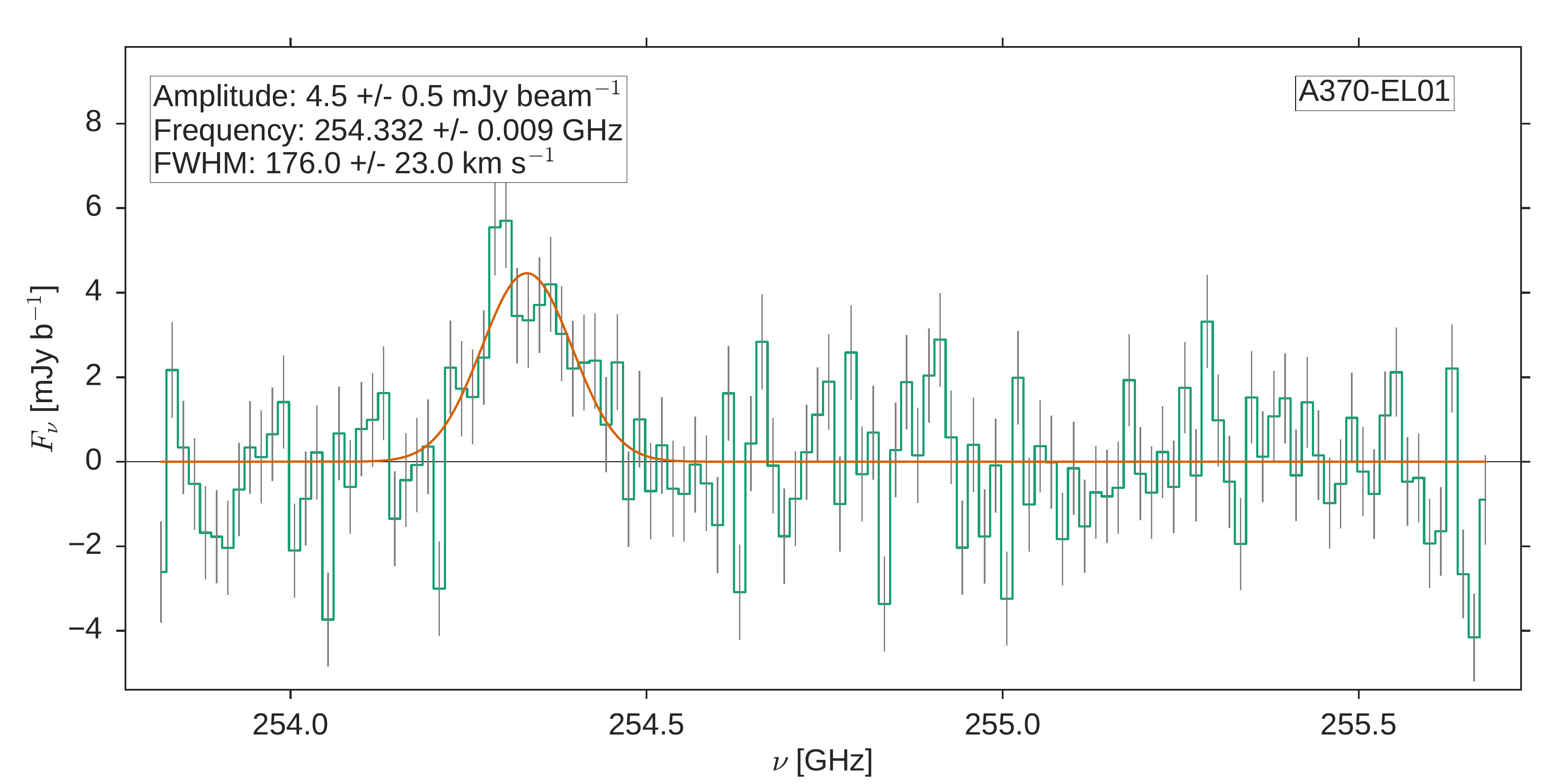}
\caption{Continuation of Figure~\ref{fig:line_reliable}. MACSJ1149-EL01 and MACSJ1149-EL03 have no identification while A370-EL01 is securely identified as CO(3-2) at $z=0.35962\pm0.00001$. The color scale for A370-EL01 has been modified to show the features of the counterpart galaxy.\label{fig:line_reliable2}}
\end{figure*}

\begin{figure*}[!htbp]
\includegraphics[width=0.4\textwidth]{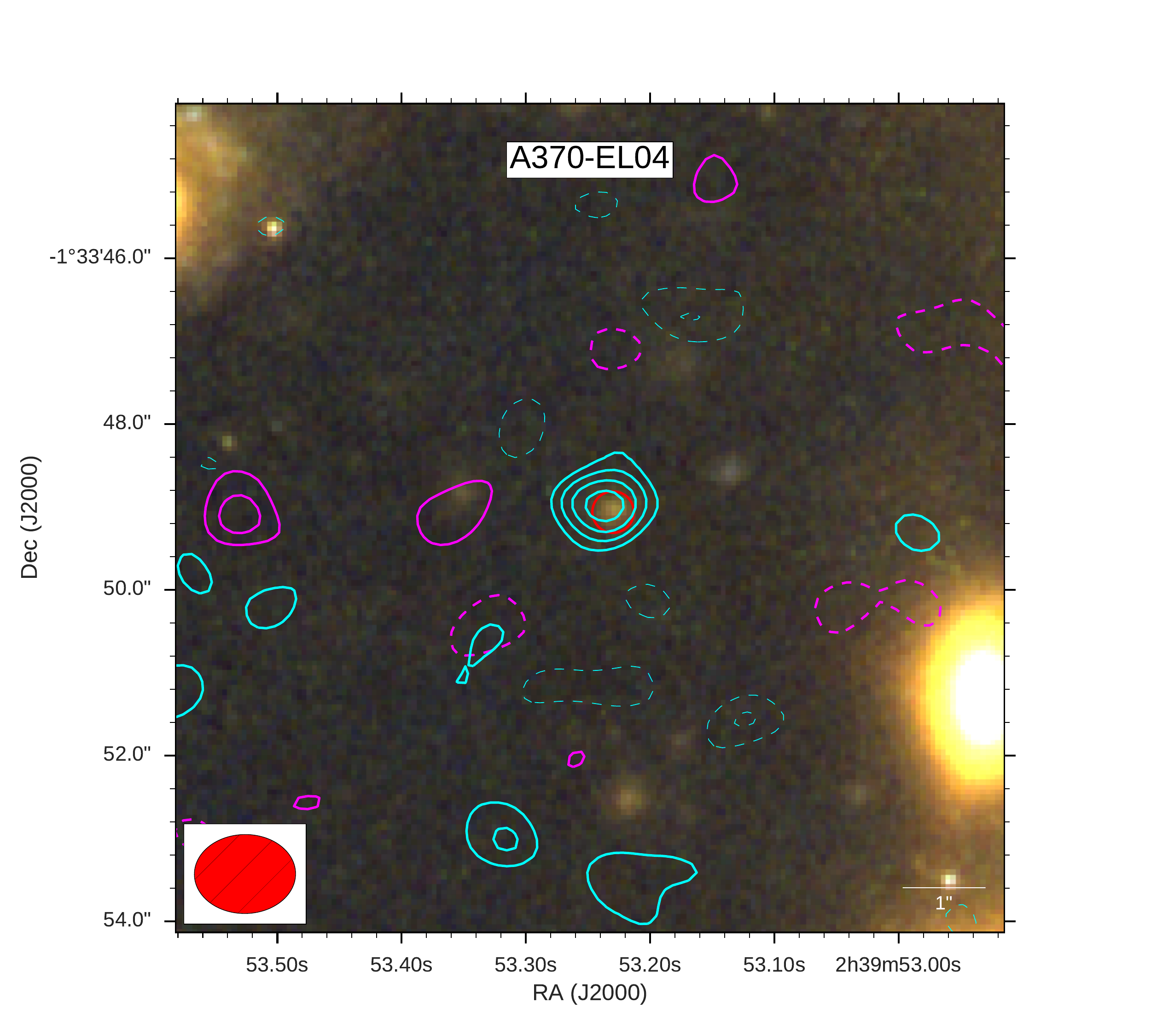}
\includegraphics[width=0.6\textwidth]{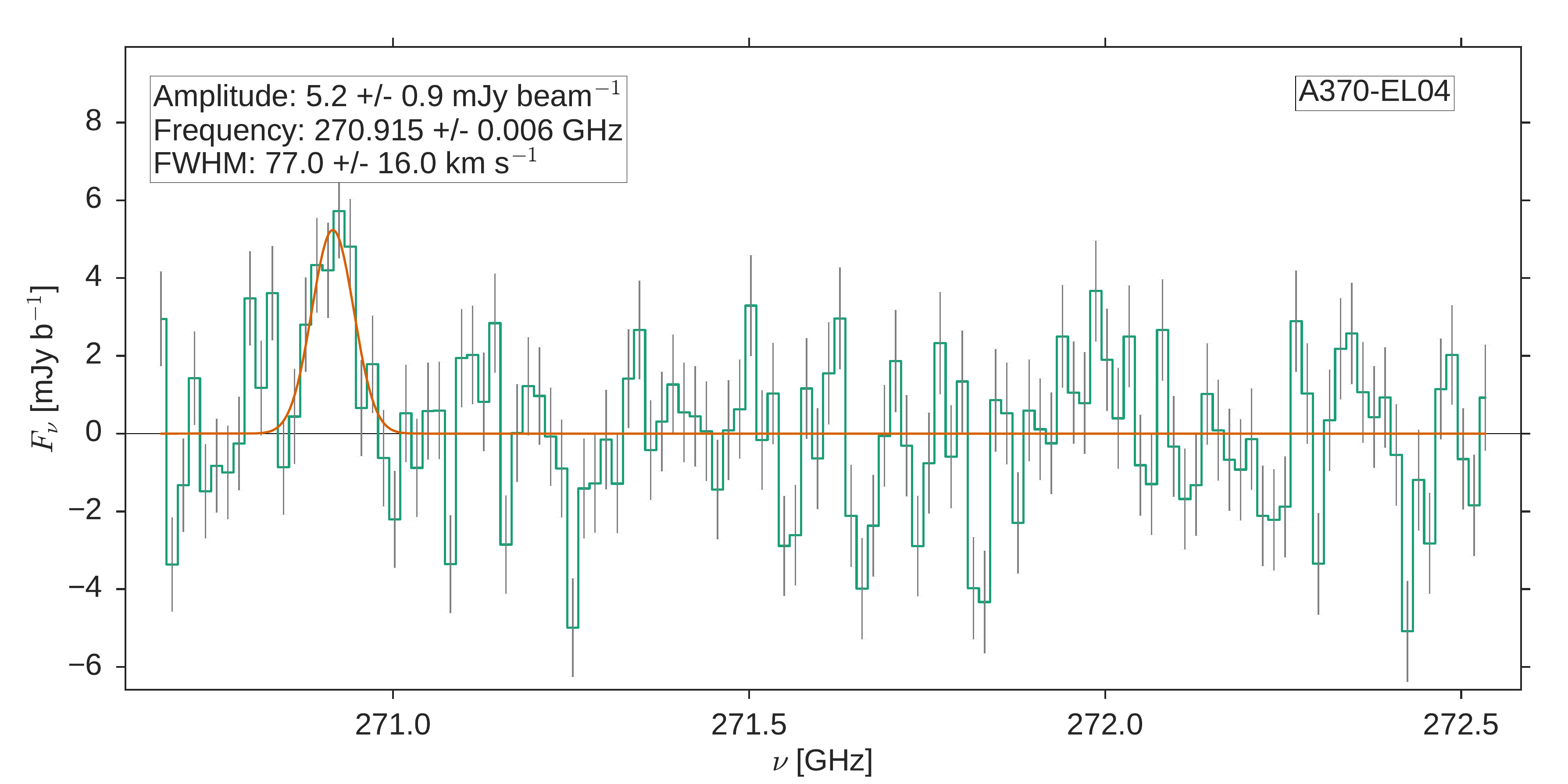}
\includegraphics[width=0.4\textwidth]{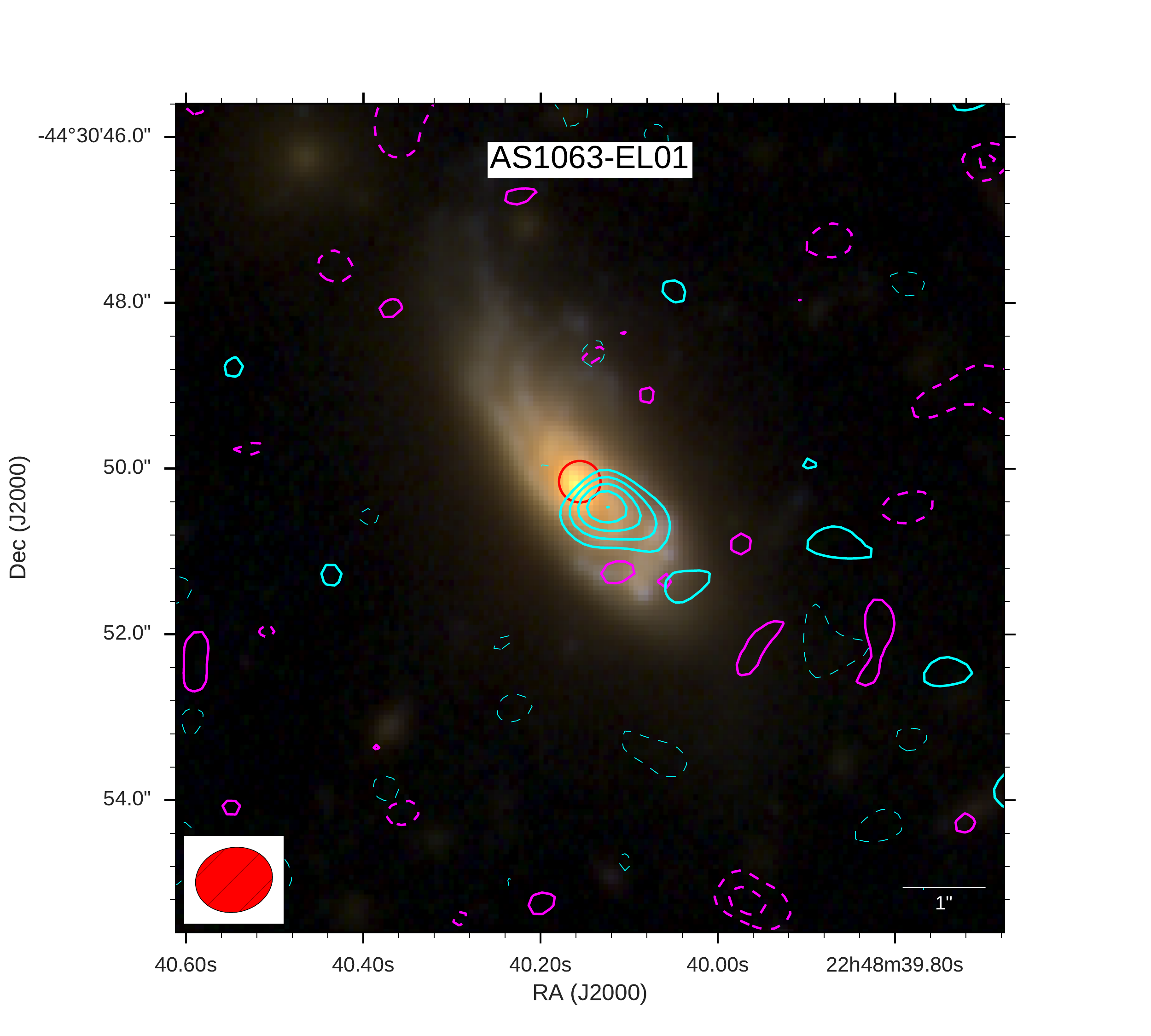}
\includegraphics[width=0.6\textwidth]{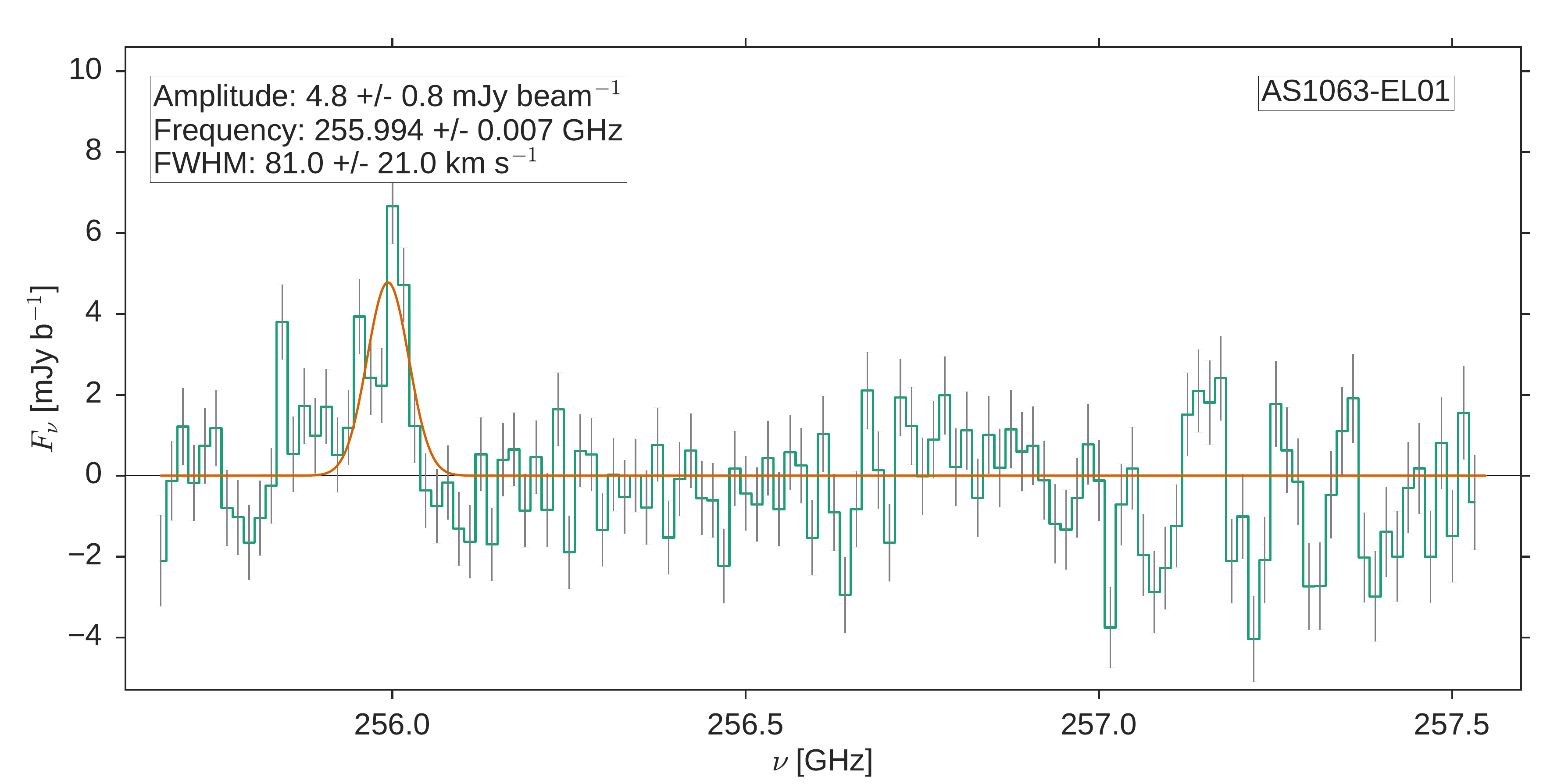}
\includegraphics[width=0.4\textwidth]{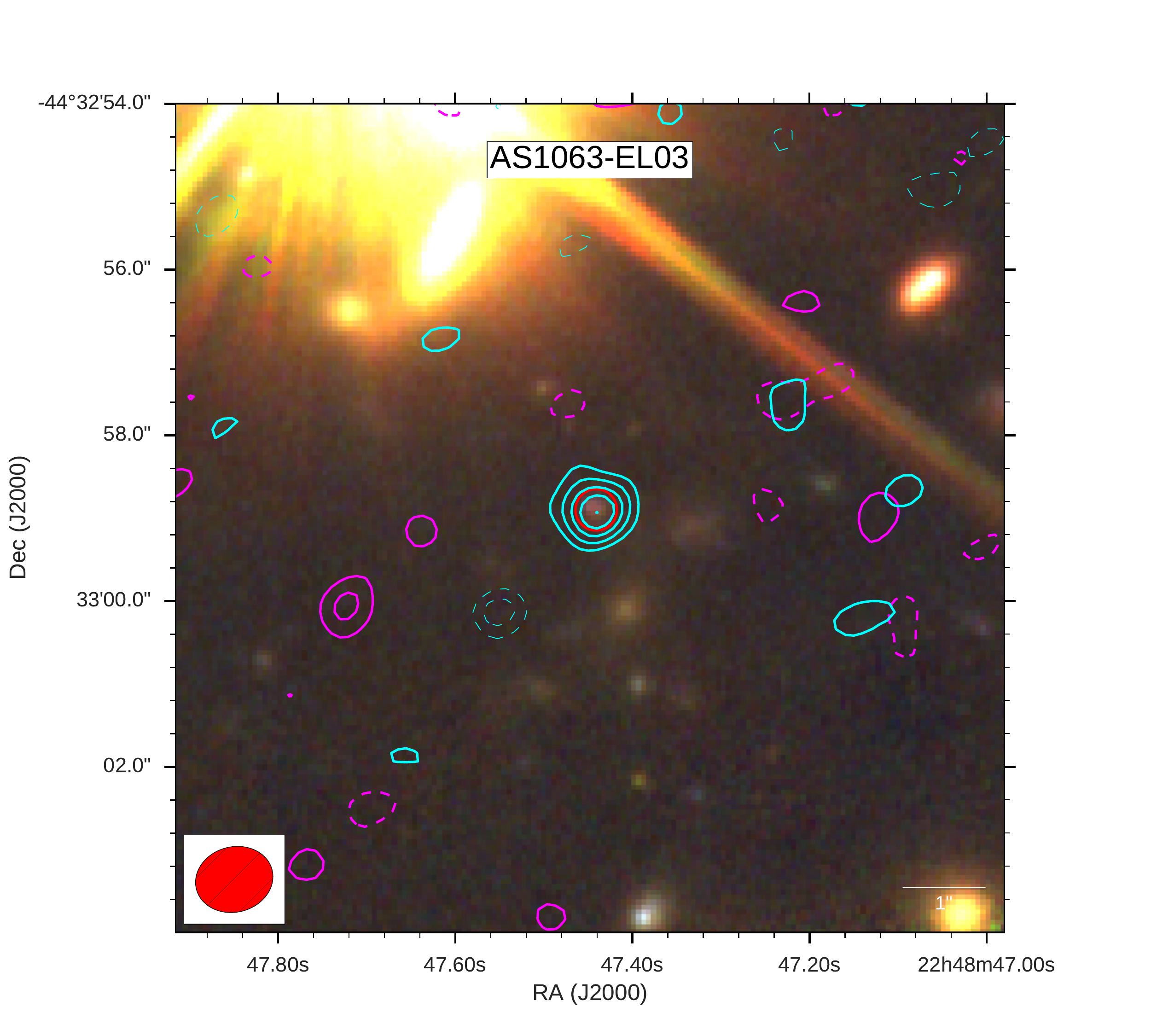}
\includegraphics[width=0.6\textwidth]{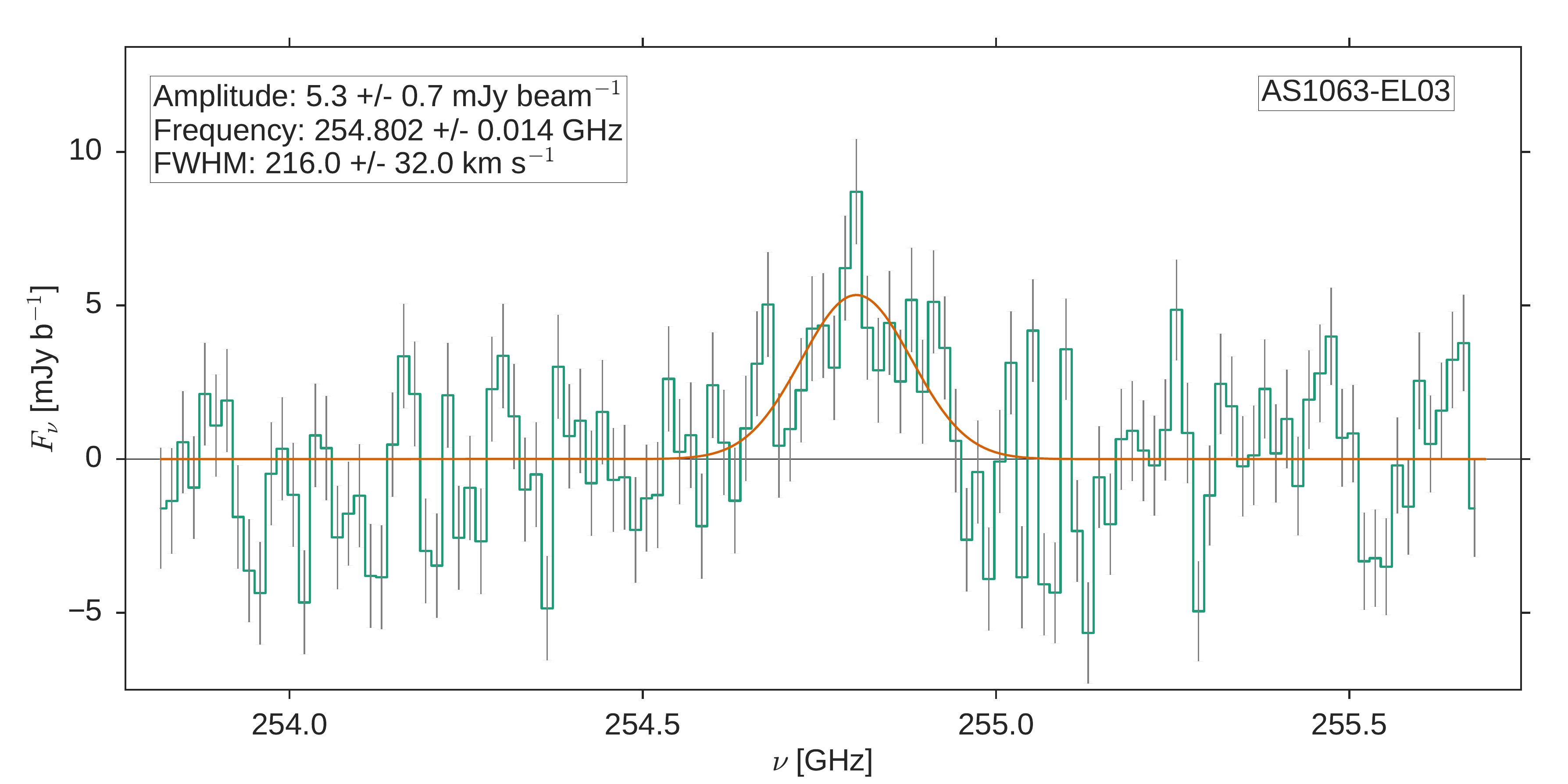}
\caption{Continuation of Figure~\ref{fig:line_reliable}. A370-EL04 is tentatively identified  as CO(3-2) at $z=0.27640\pm0.00001$, AS1063-EL01 is securely identified as CO(3-2) at $z=0.35080\pm0.00001$ while AS1063-EL03 is tentatively identified as CO(9-8) at $z=3.0695\pm0.0002$. The color scale for AS1063-EL01 has been modified to show the features of the counterpart galaxy.\label{fig:line_reliable3}}
\end{figure*}

\begin{figure*}[!htbp]
\includegraphics[width=0.4\textwidth]{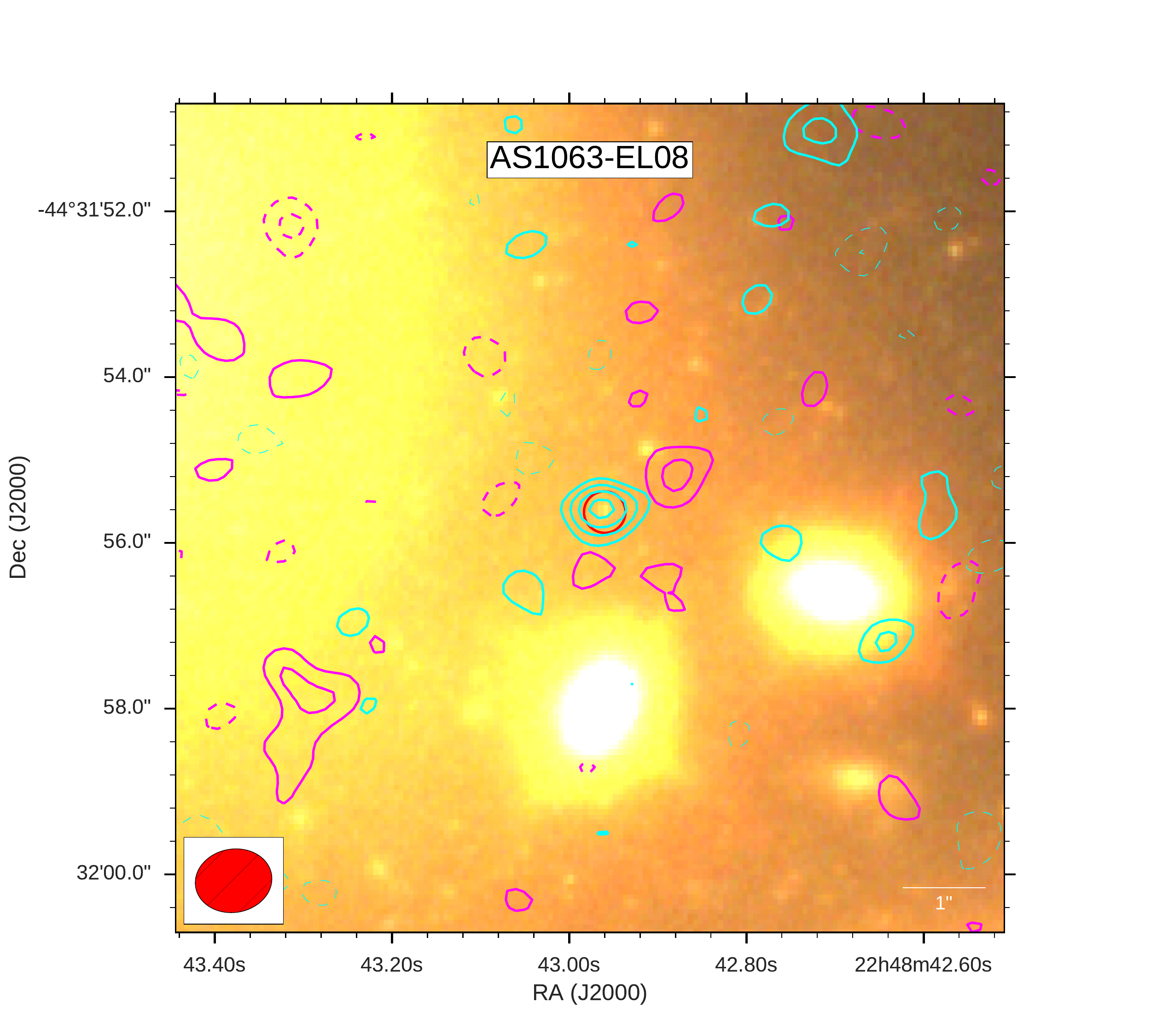}
\includegraphics[width=0.6\textwidth]{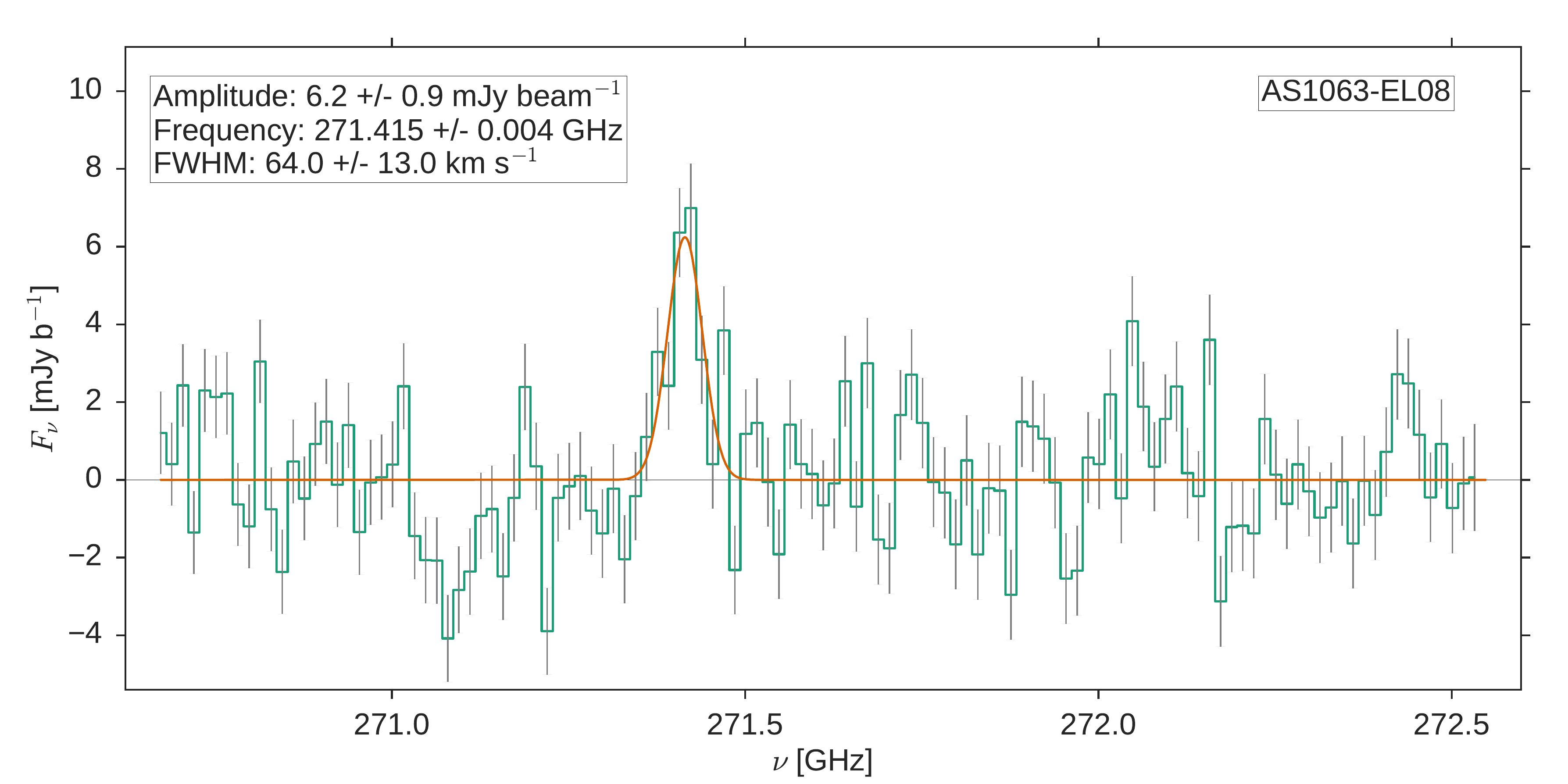}
\caption{Continuation of Figure~\ref{fig:line_reliable}. AS1063-EL08 has no identification.\label{fig:line_reliable4}}
\end{figure*}

\begin{figure*}[!htbp]
\includegraphics[width=0.4\textwidth]{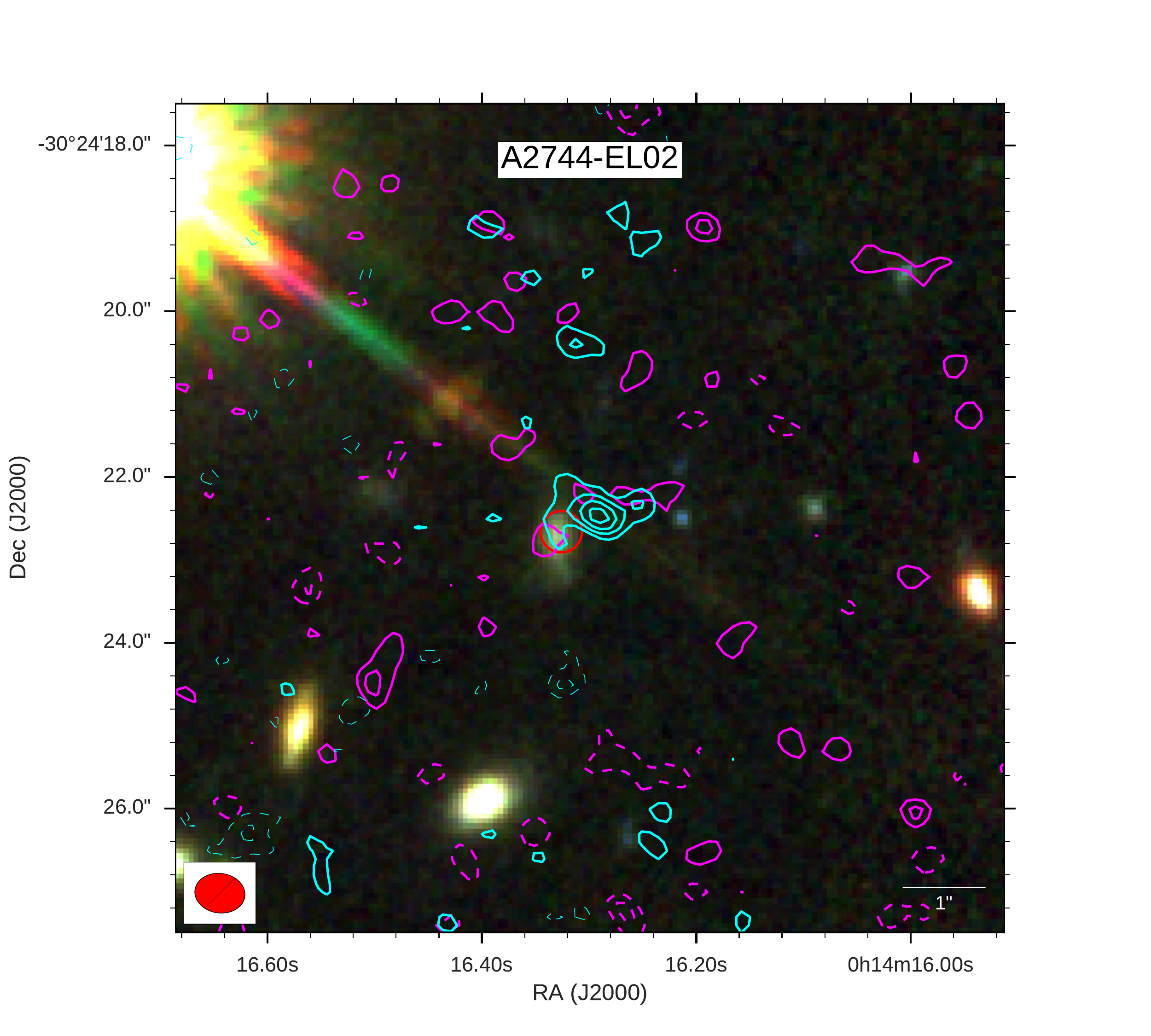}
\includegraphics[width=0.6\textwidth]{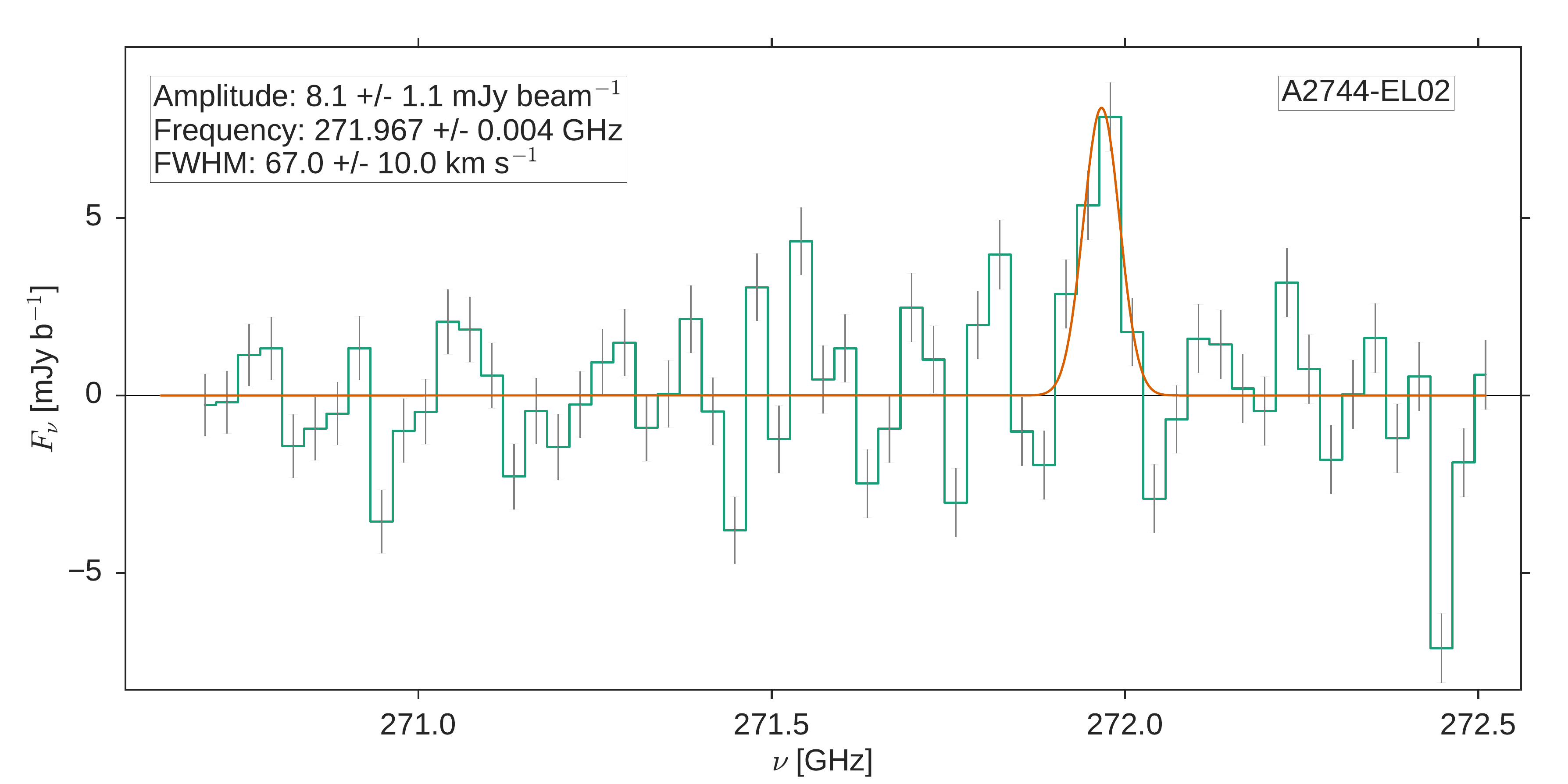}
\includegraphics[width=0.4\textwidth]{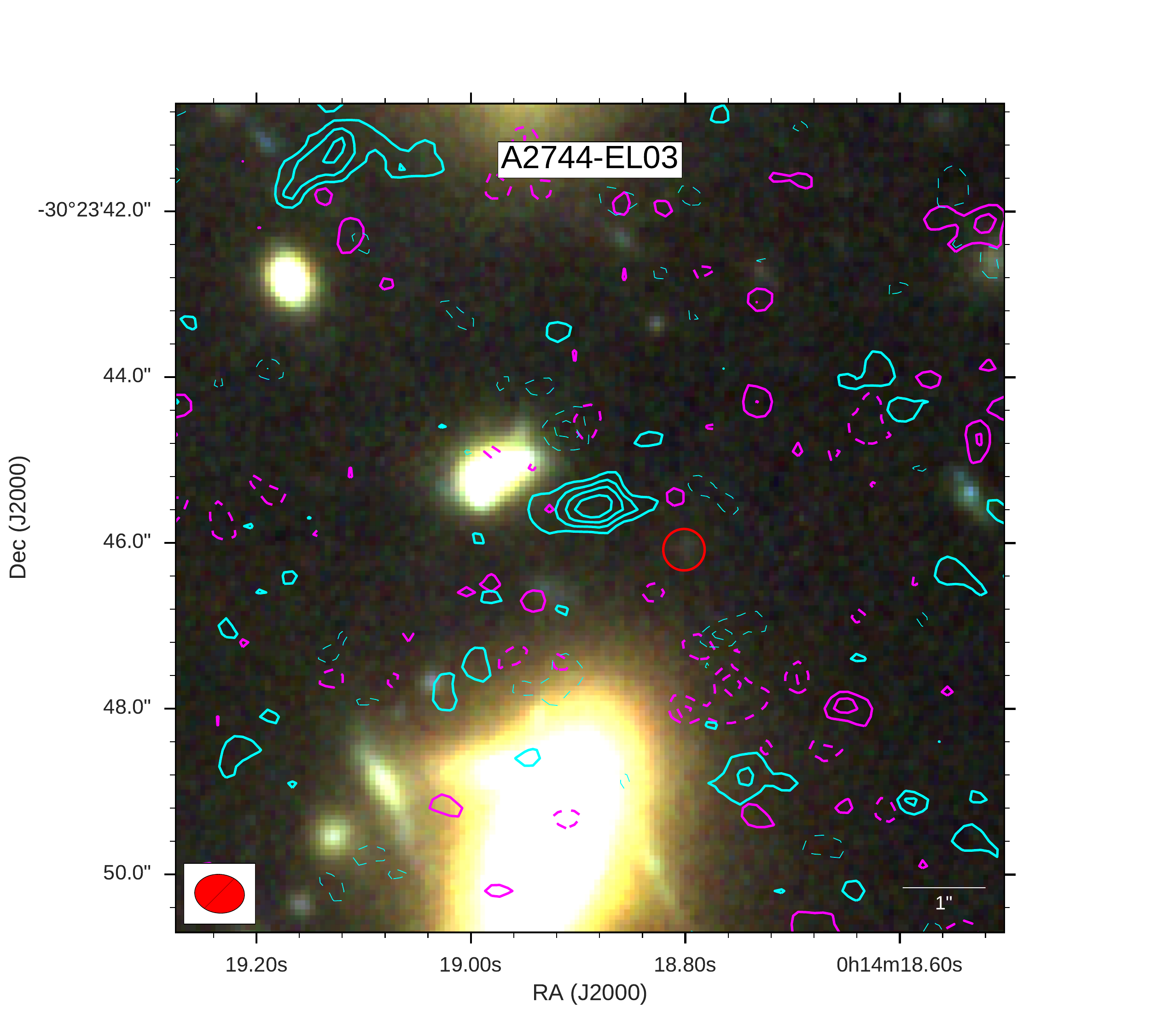}
\includegraphics[width=0.6\textwidth]{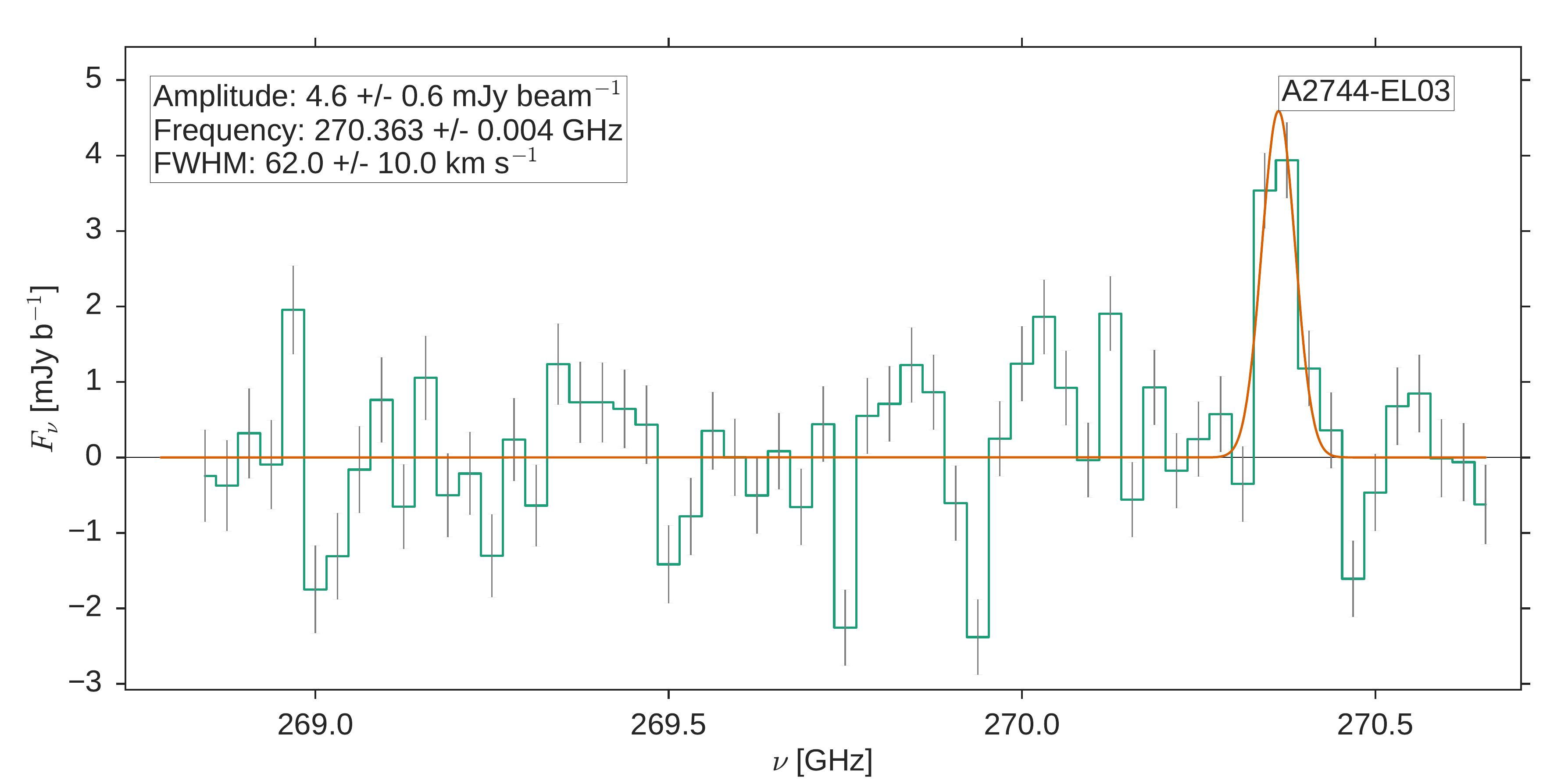}
\includegraphics[width=0.4\textwidth]{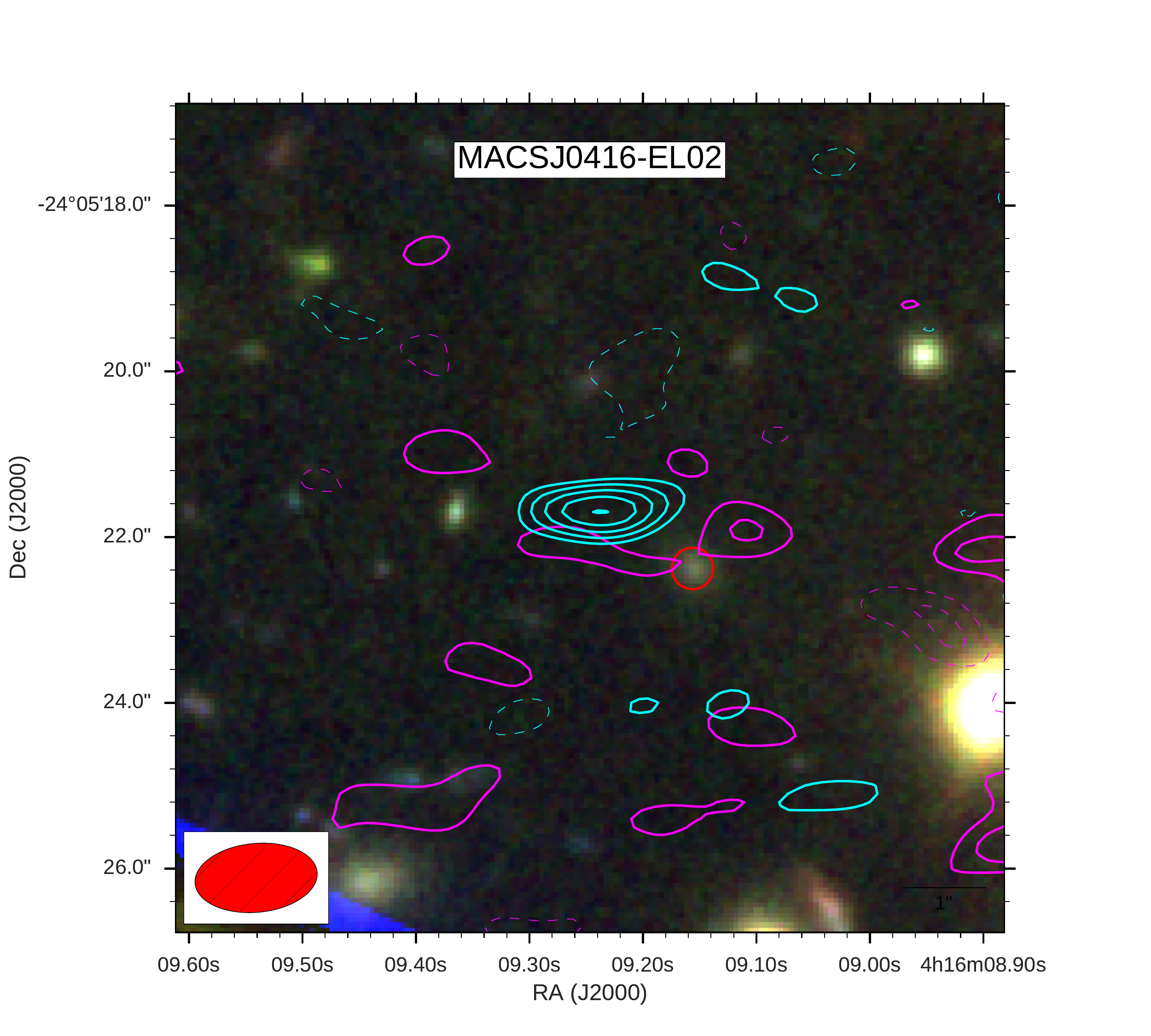}
\includegraphics[width=0.6\textwidth]{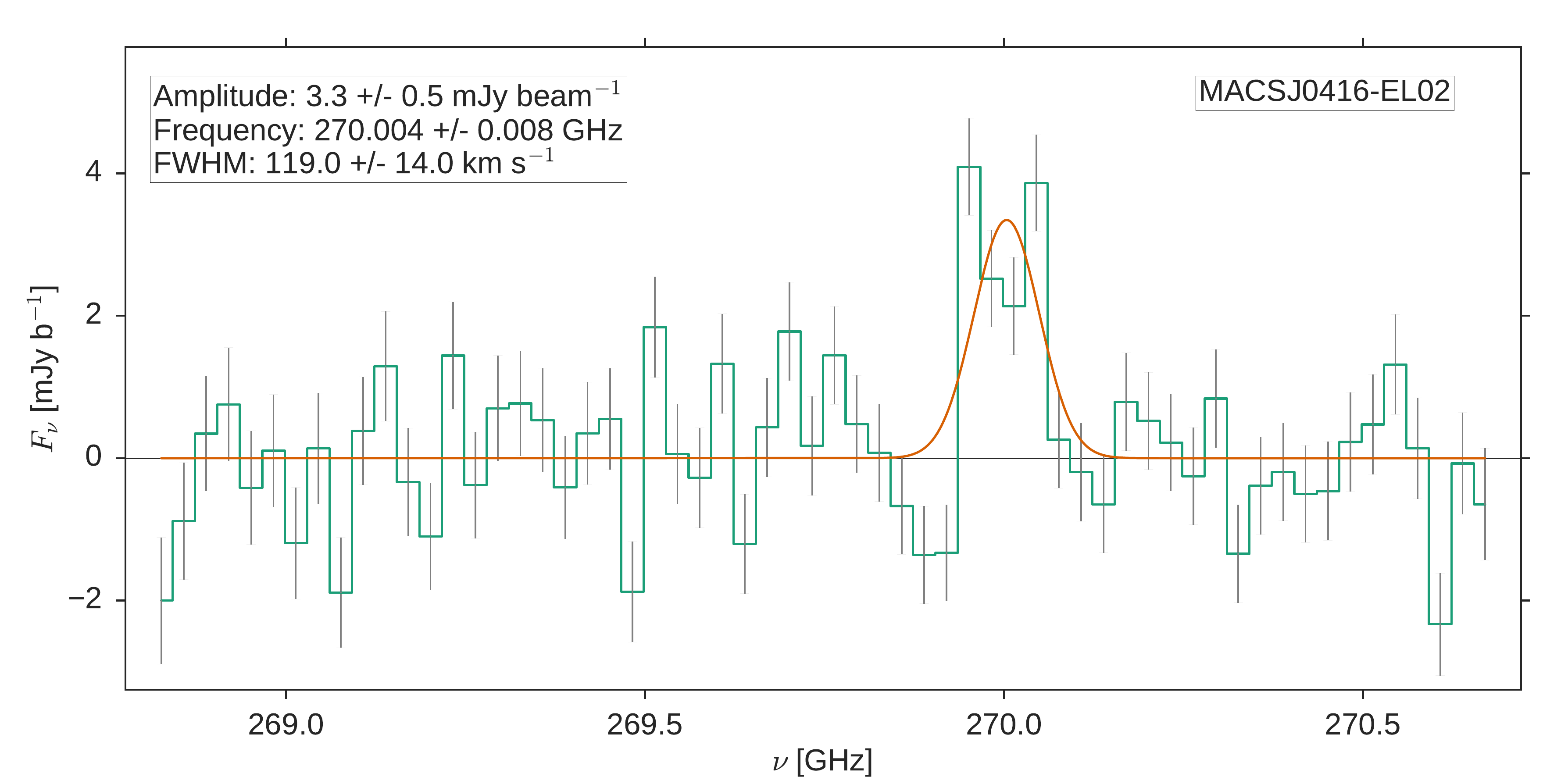}
\caption{The same as Figure \ref{fig:line_reliable} but for the rejected line candidates.
\label{fig:line_candidate}}
\end{figure*}

\begin{figure*}[!htbp]
\includegraphics[width=0.4\textwidth]{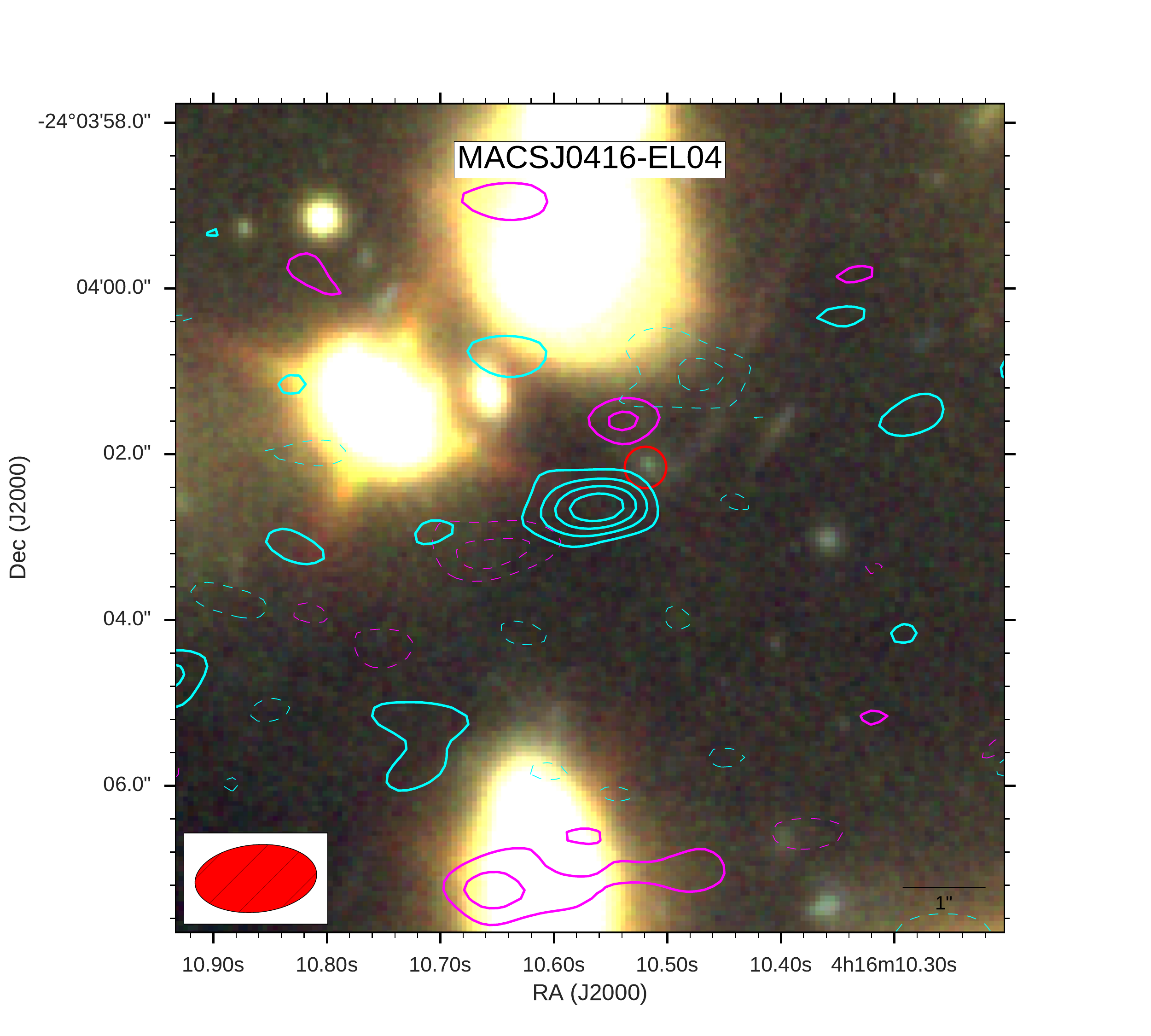}
\includegraphics[width=0.6\textwidth]{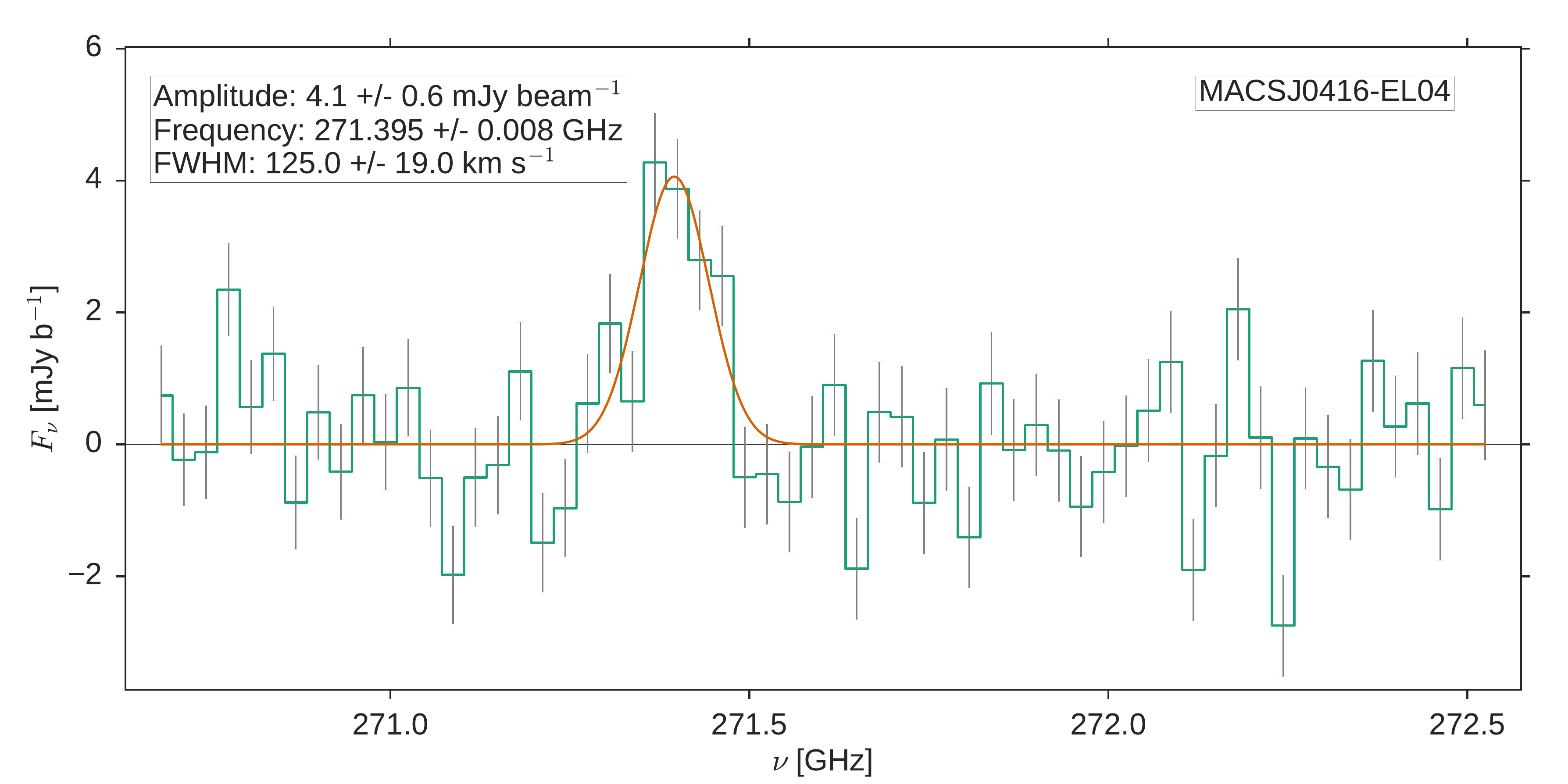}
\includegraphics[width=0.4\textwidth]{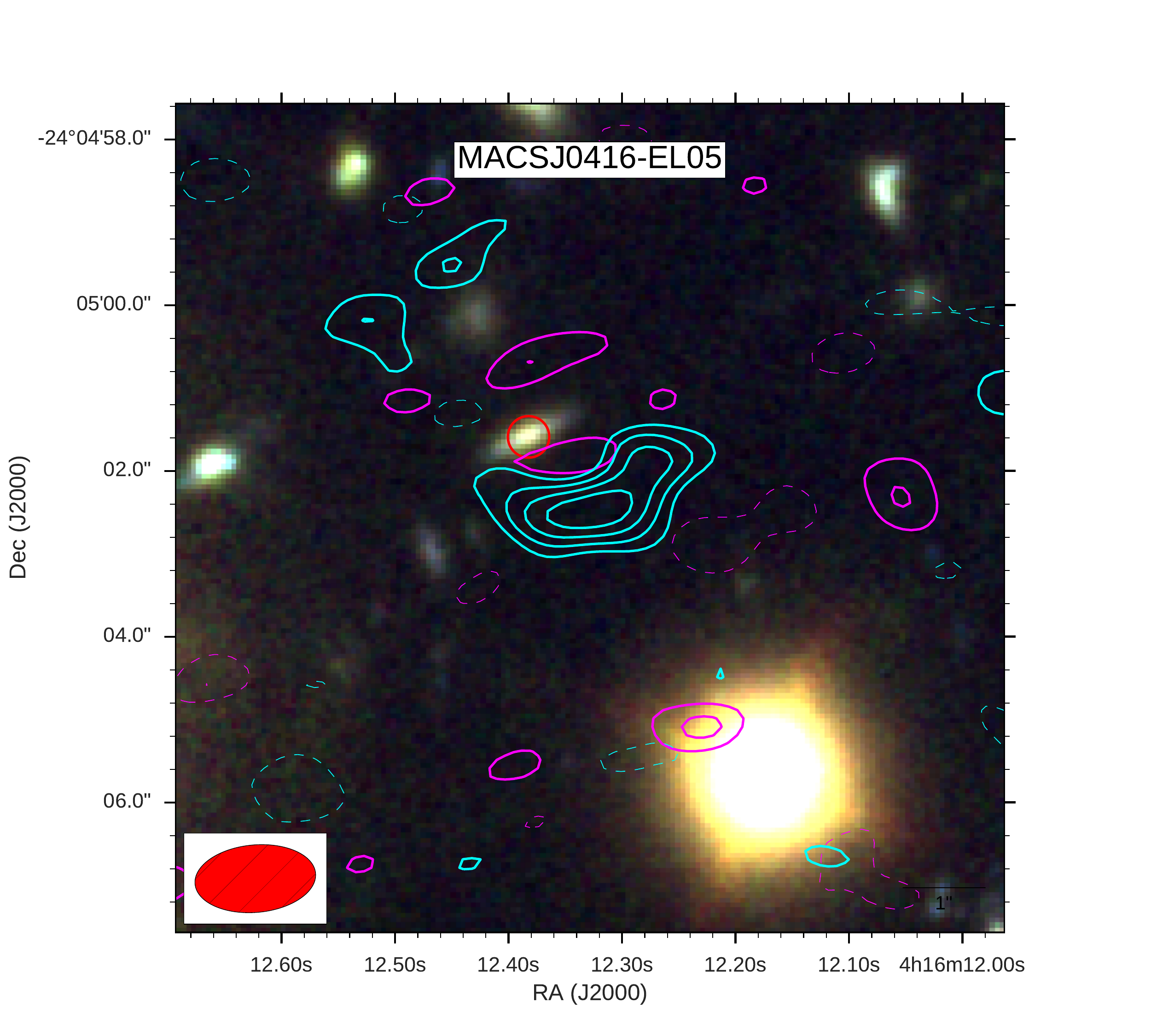}
\includegraphics[width=0.6\textwidth]{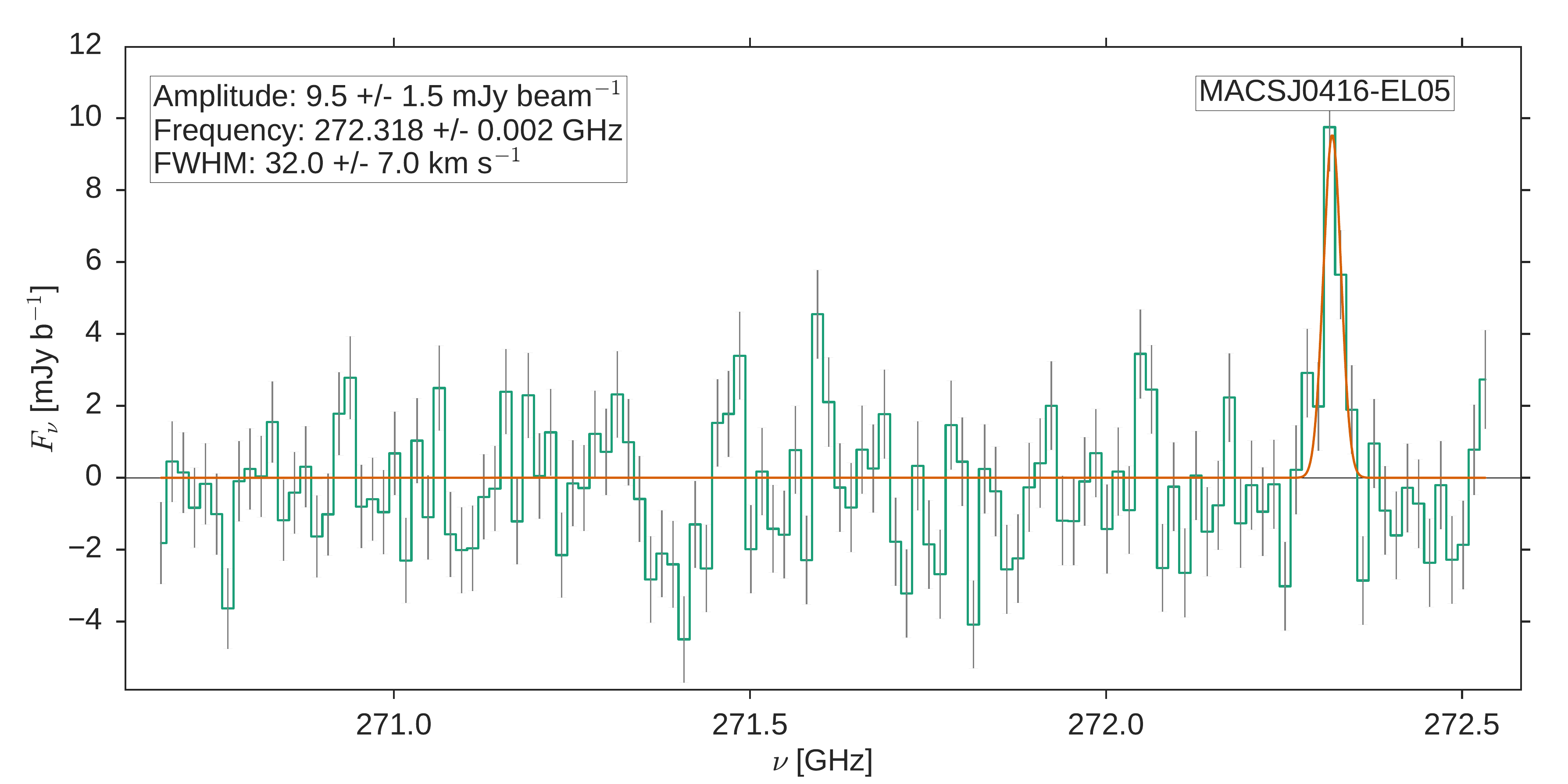}
\includegraphics[width=0.4\textwidth]{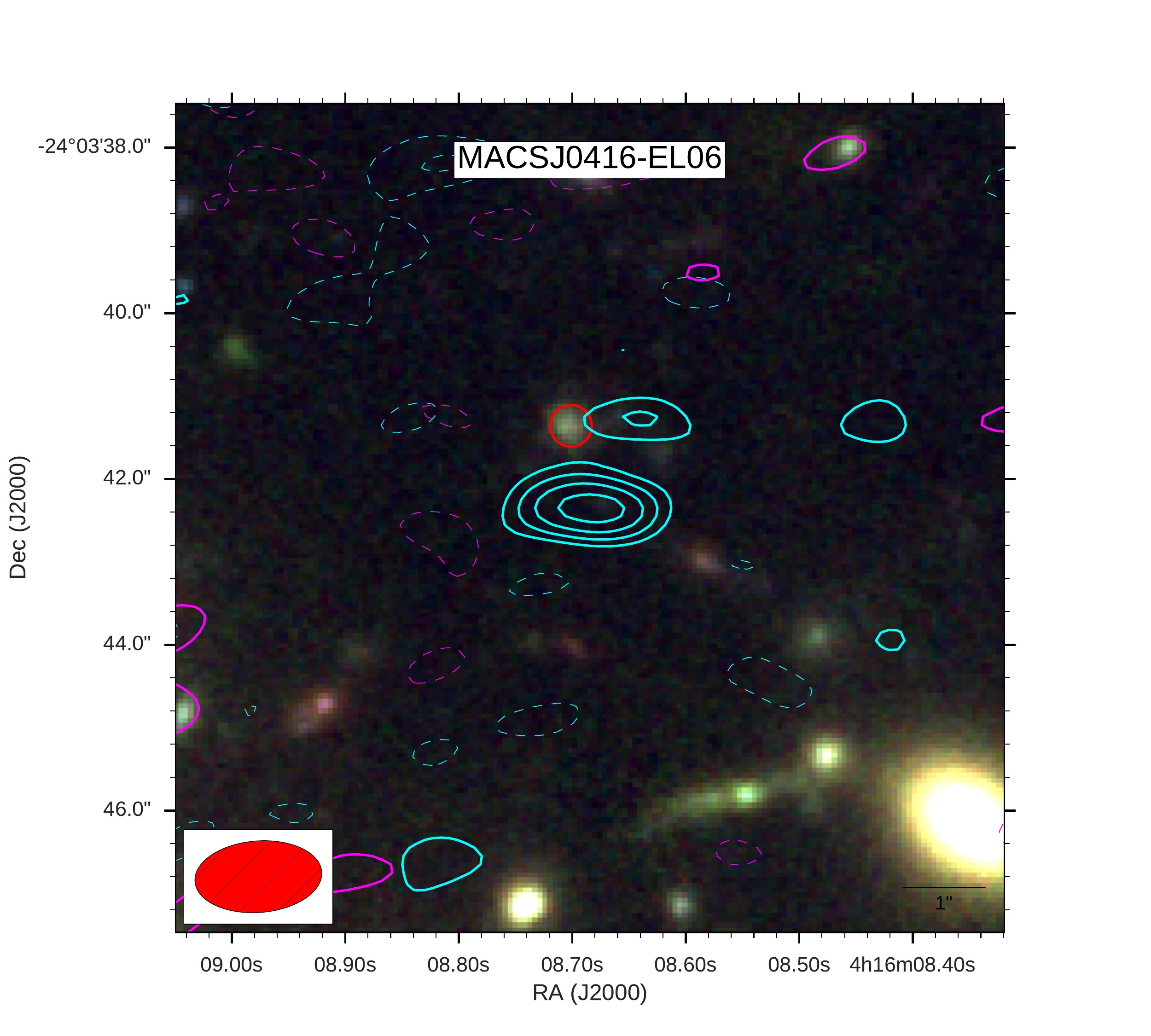}
\includegraphics[width=0.6\textwidth]{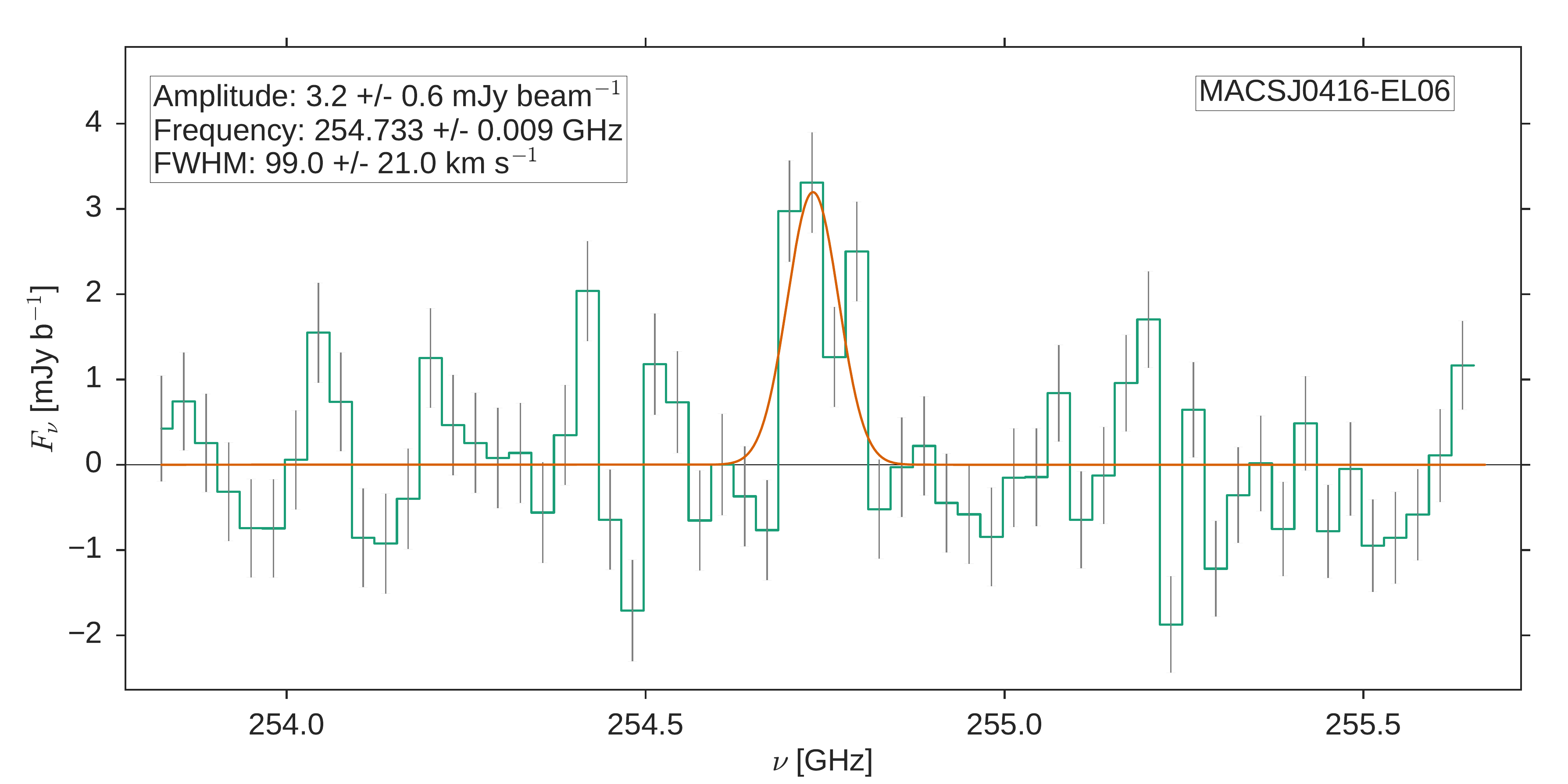}
\caption{Continuation of Figure~\ref{fig:line_candidate}\label{fig:line_candidate2}}
\end{figure*}

\begin{figure*}[!htbp]
\includegraphics[width=0.4\textwidth]{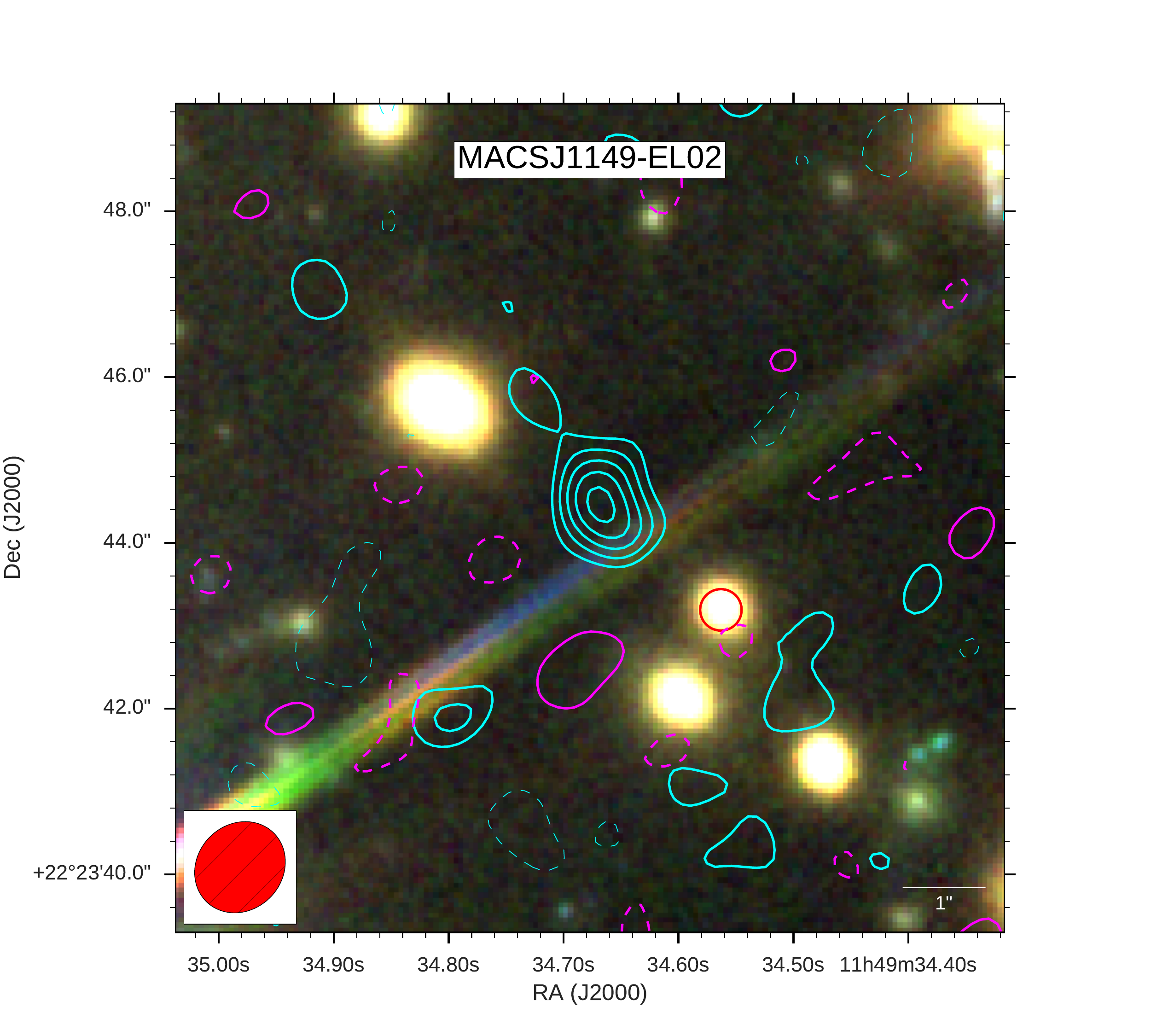}
\includegraphics[width=0.6\textwidth]{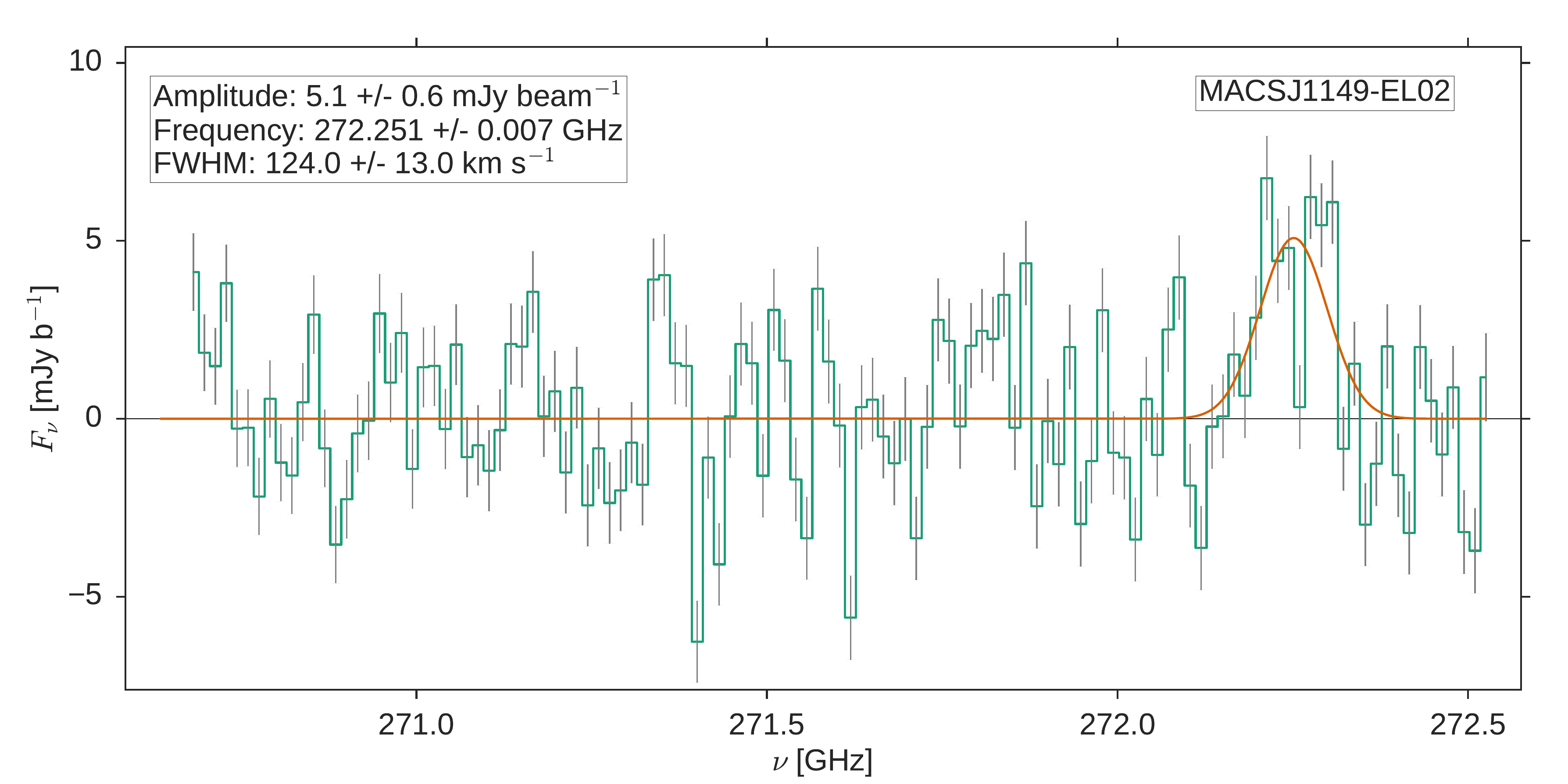}
\includegraphics[width=0.4\textwidth]{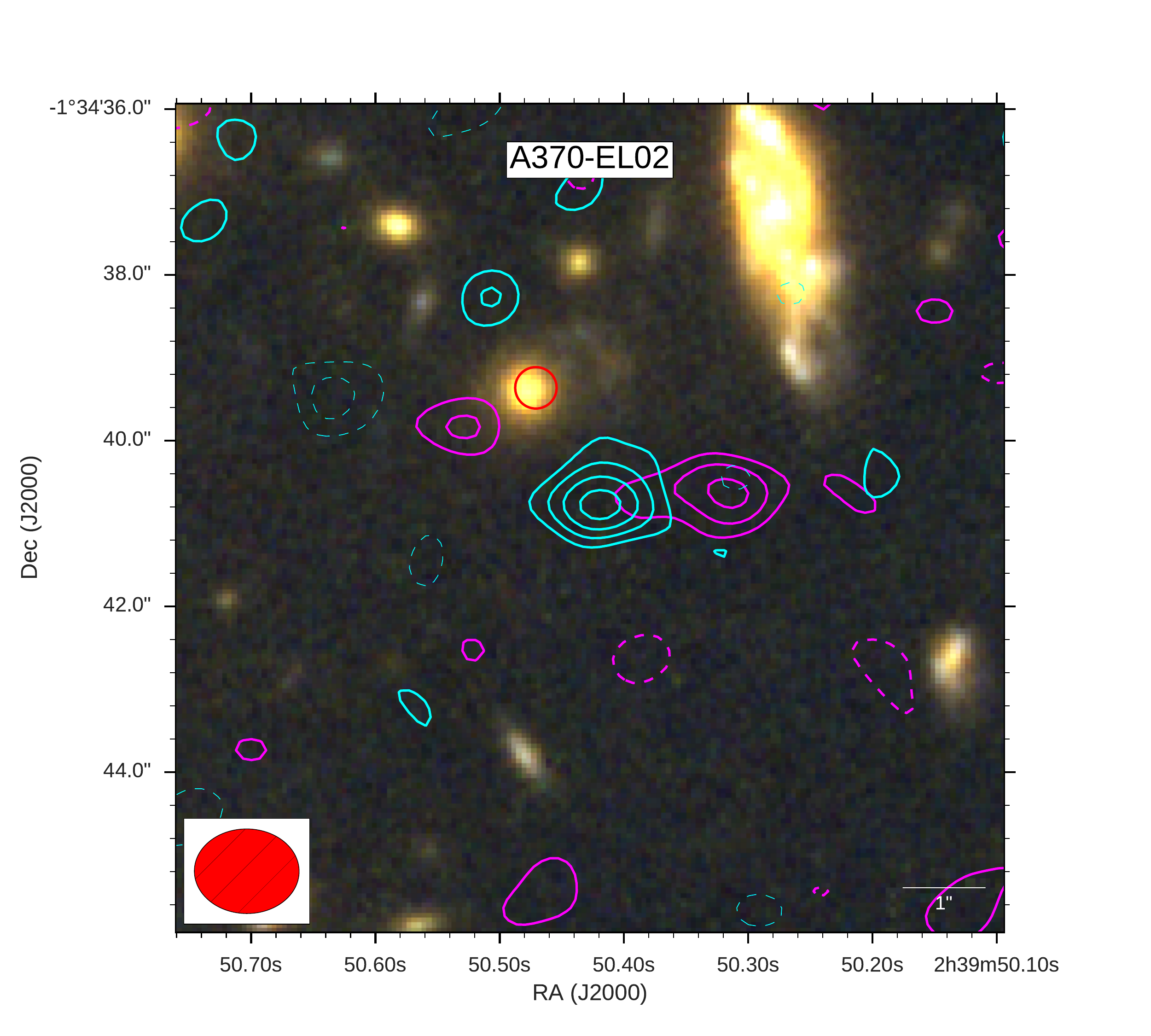}
\includegraphics[width=0.6\textwidth]{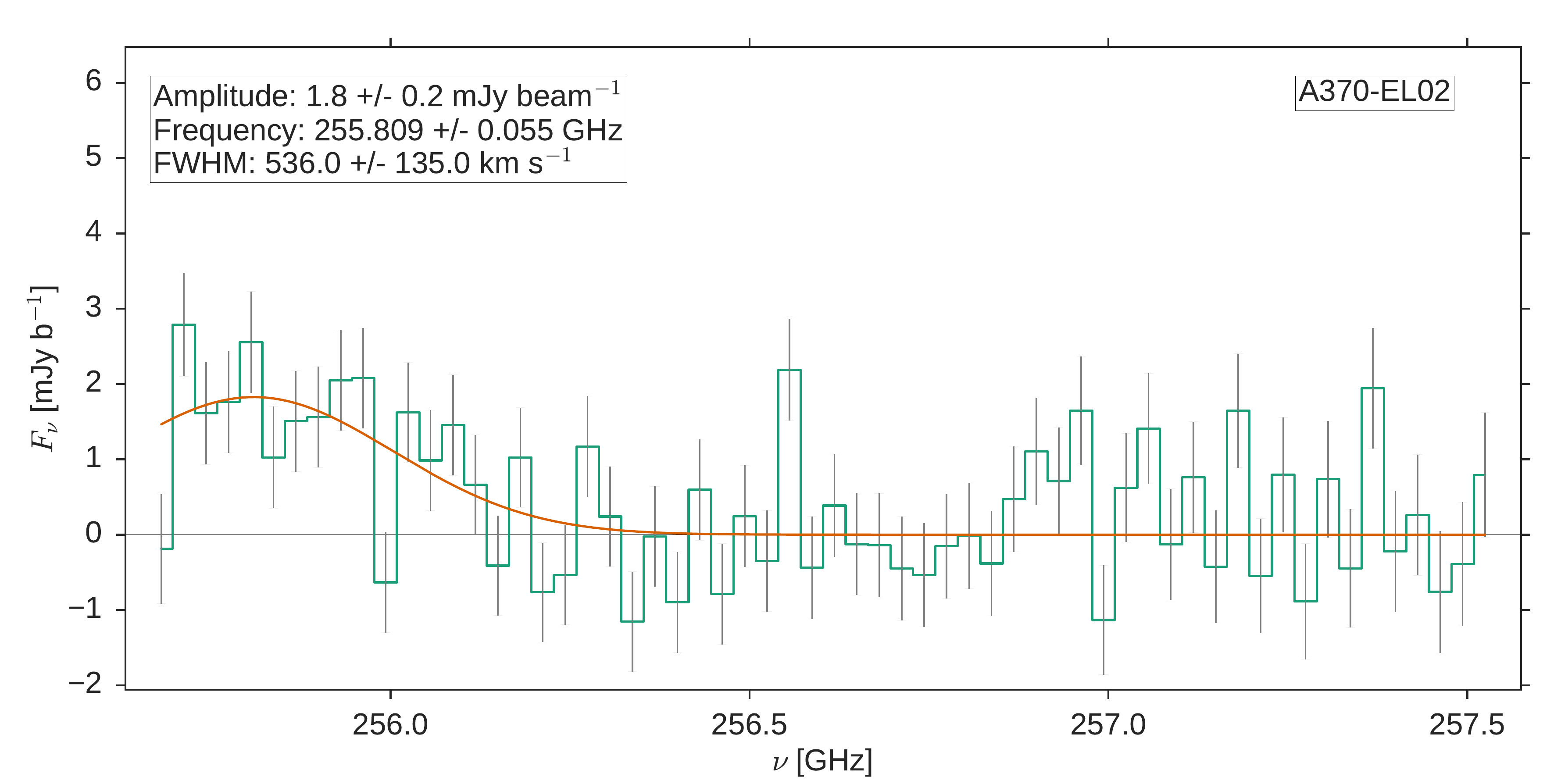}
\includegraphics[width=0.4\textwidth]{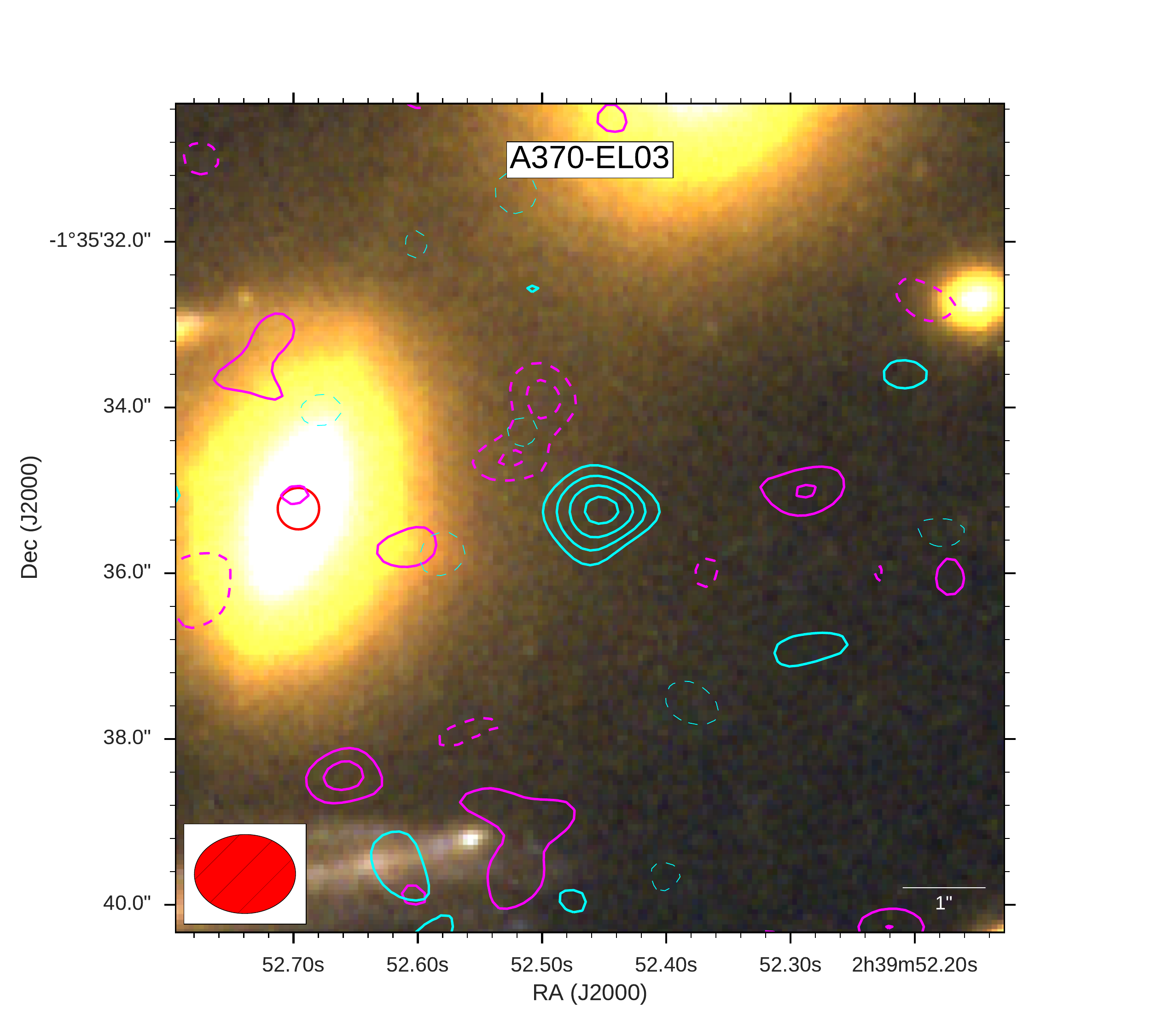}
\includegraphics[width=0.6\textwidth]{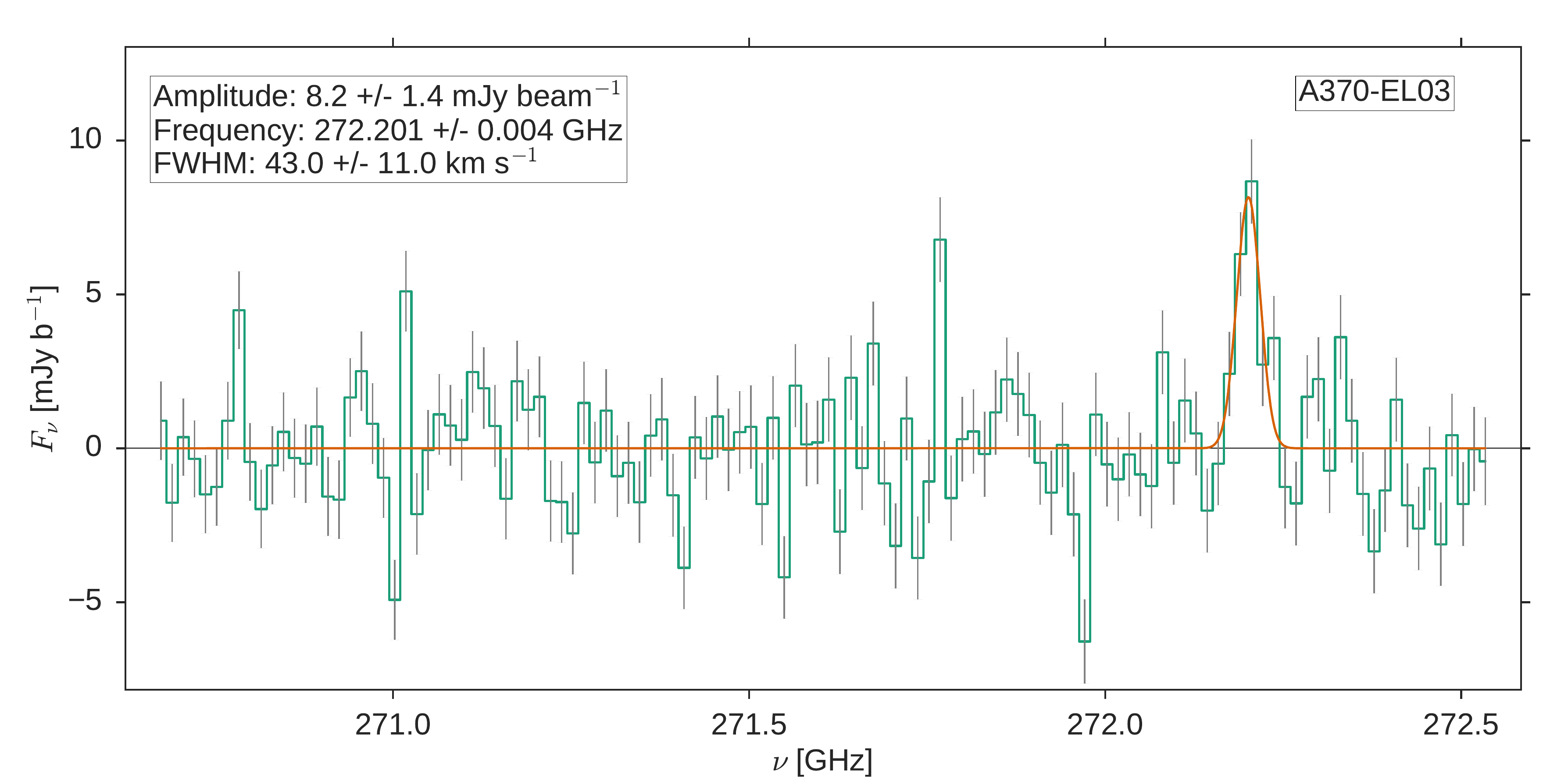}
\caption{Continuation of Figure~\ref{fig:line_candidate}\label{fig:line_candidate3}}
\end{figure*}

\begin{figure*}[!htbp]
\includegraphics[width=0.4\textwidth]{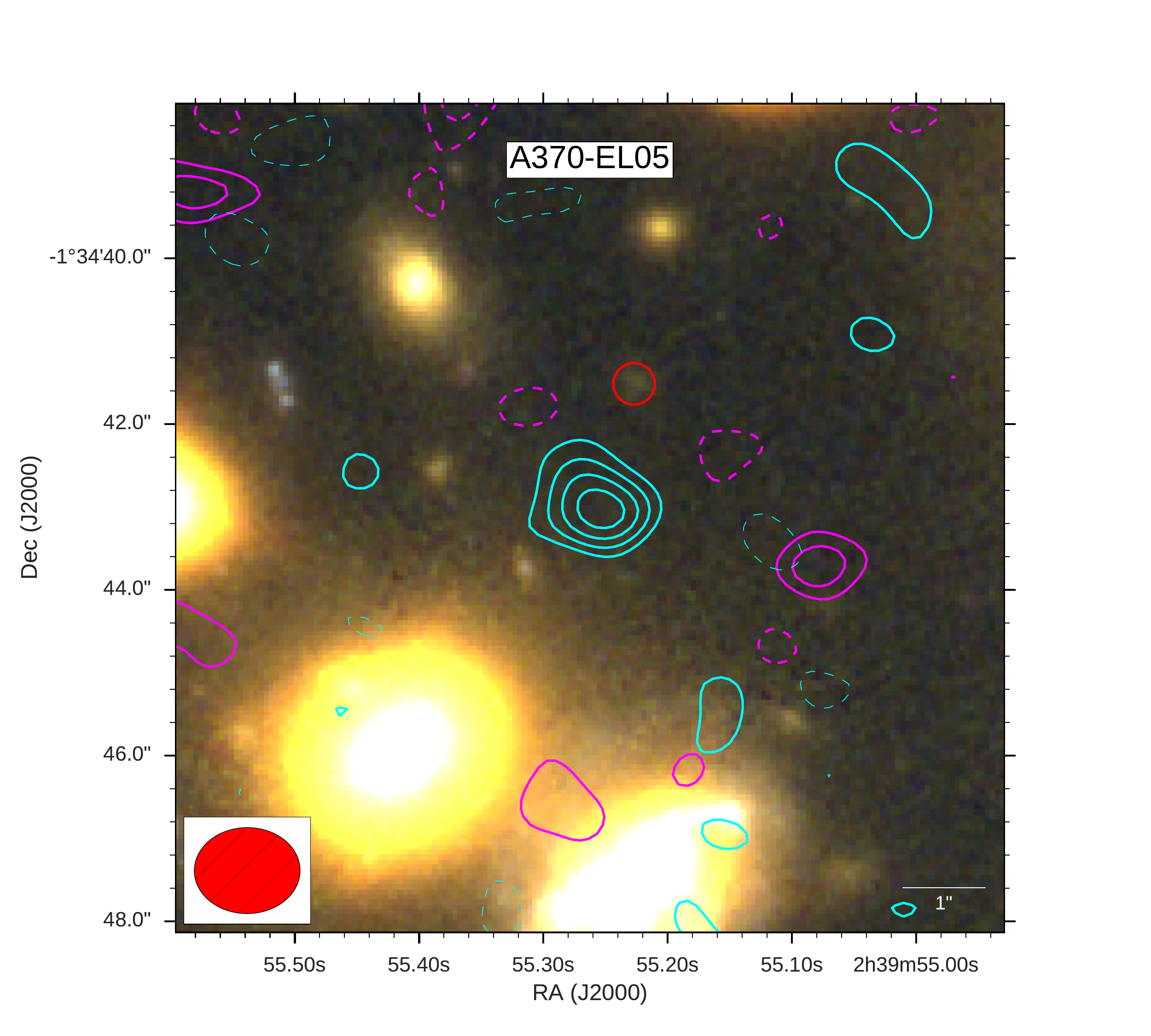}
\includegraphics[width=0.6\textwidth]{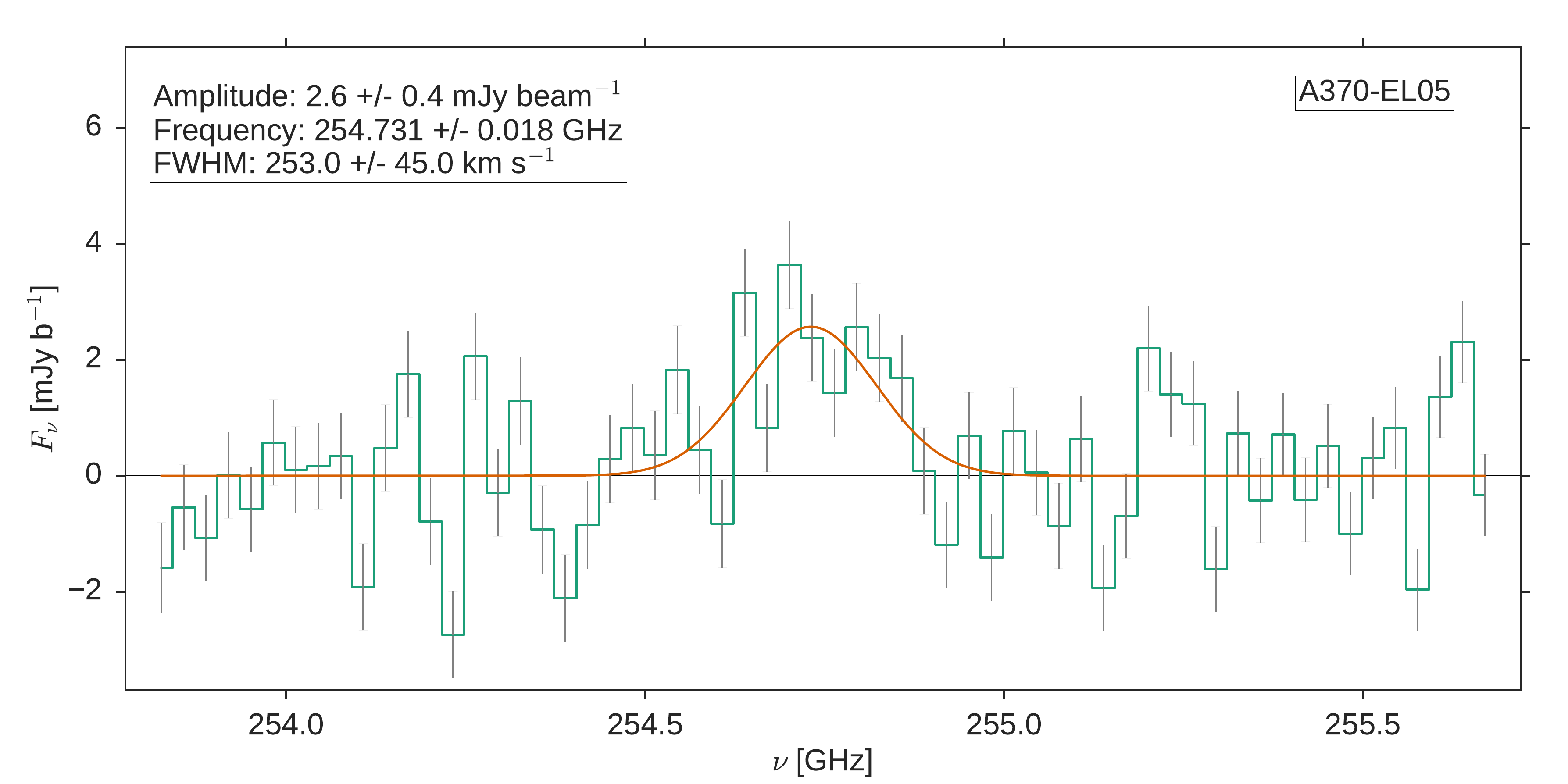}
\includegraphics[width=0.4\textwidth]{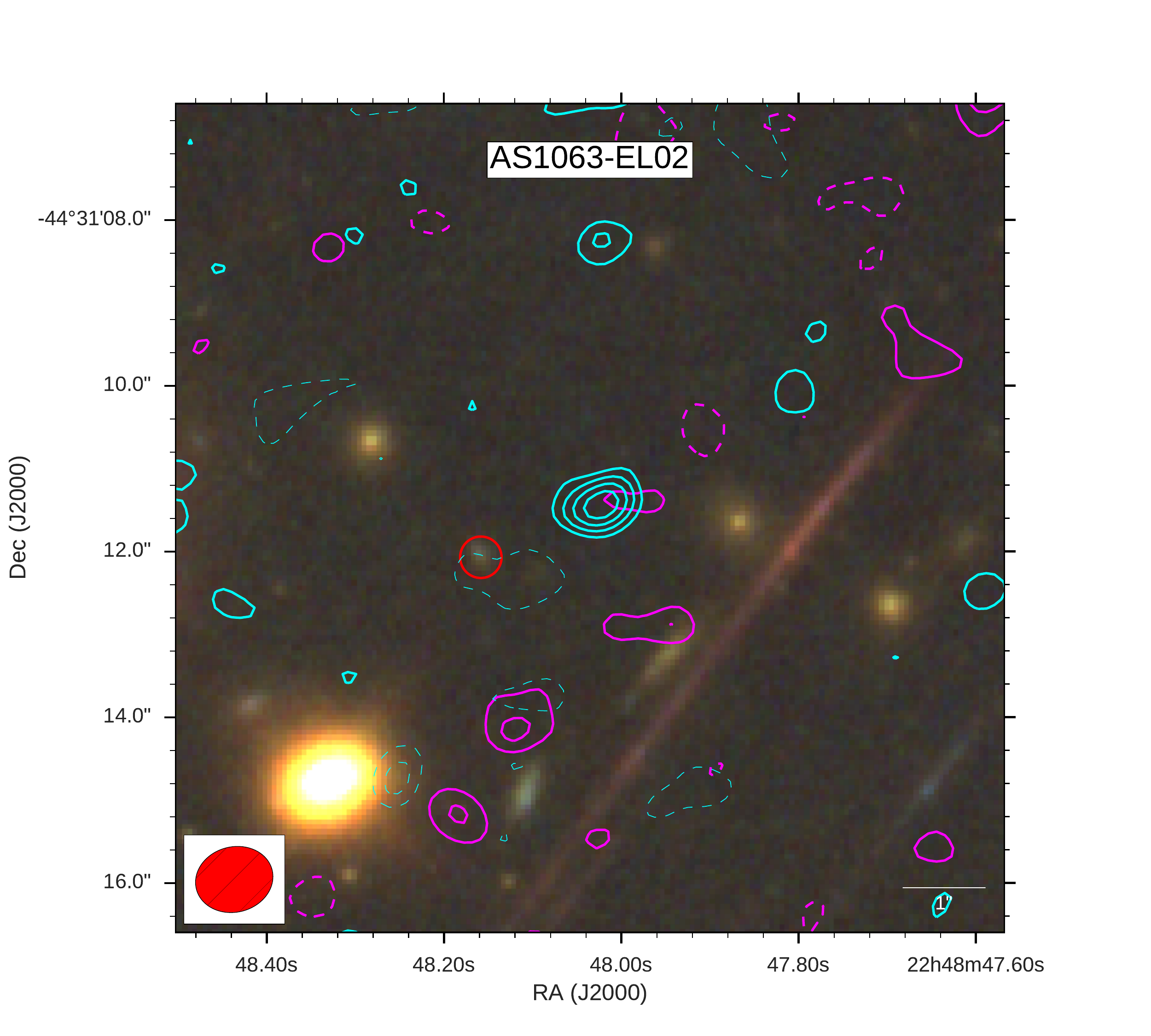}
\includegraphics[width=0.6\textwidth]{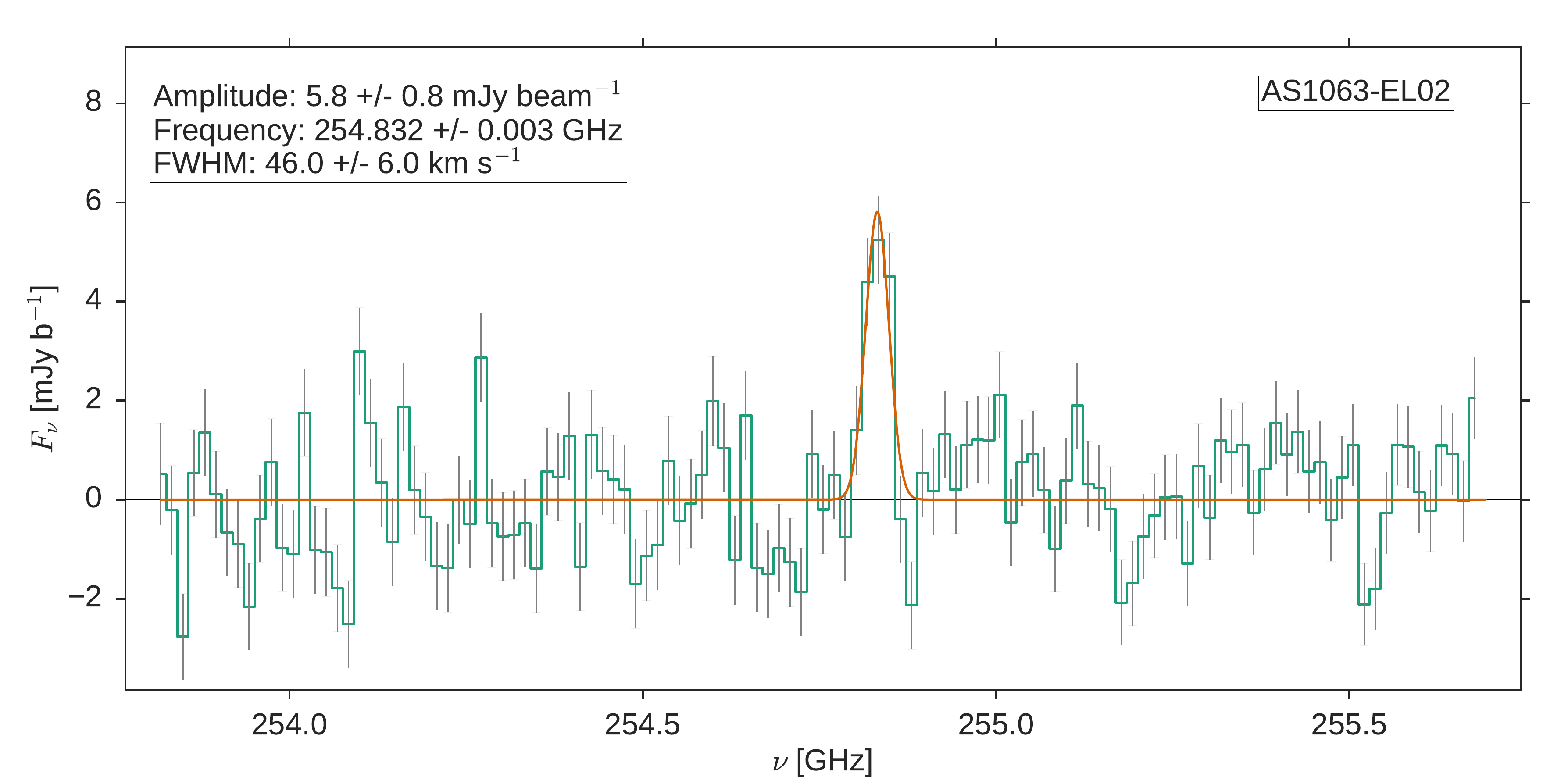}
\includegraphics[width=0.4\textwidth]{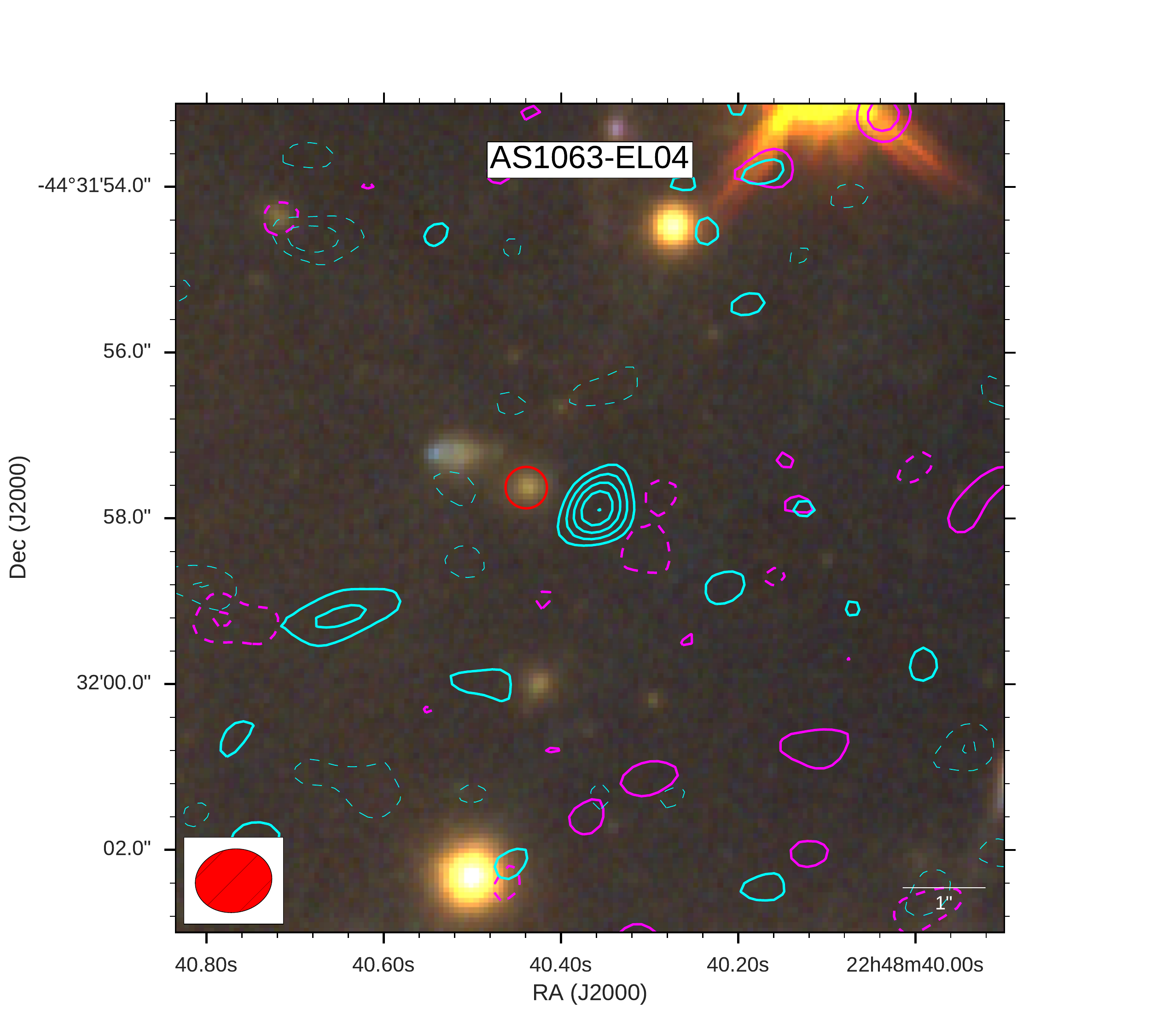}
\includegraphics[width=0.6\textwidth]{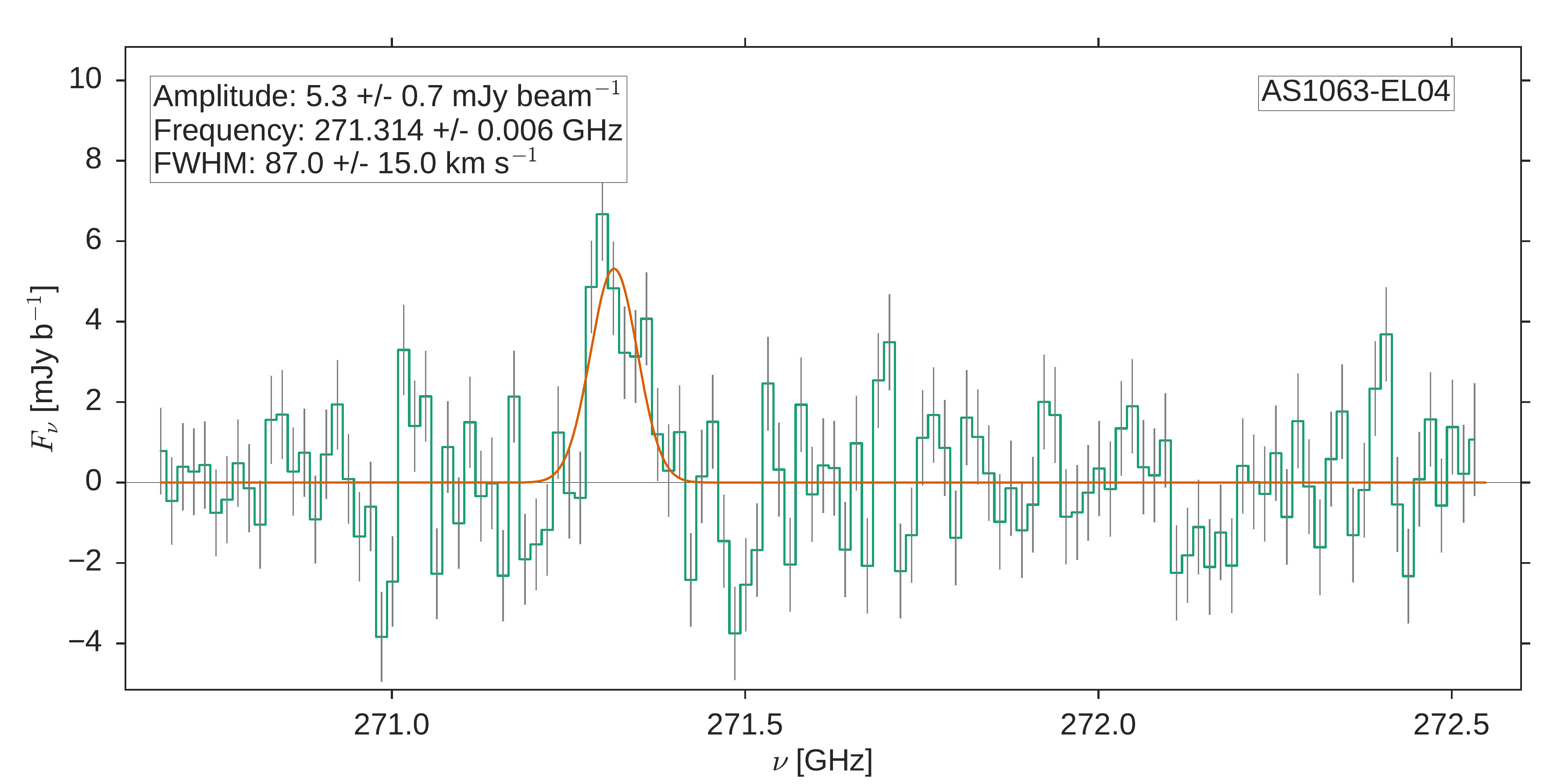}
\caption{Continuation of Figure~\ref{fig:line_candidate}\label{fig:line_candidate4}}
\end{figure*}

\begin{figure*}[!htbp]
\includegraphics[width=0.4\textwidth]{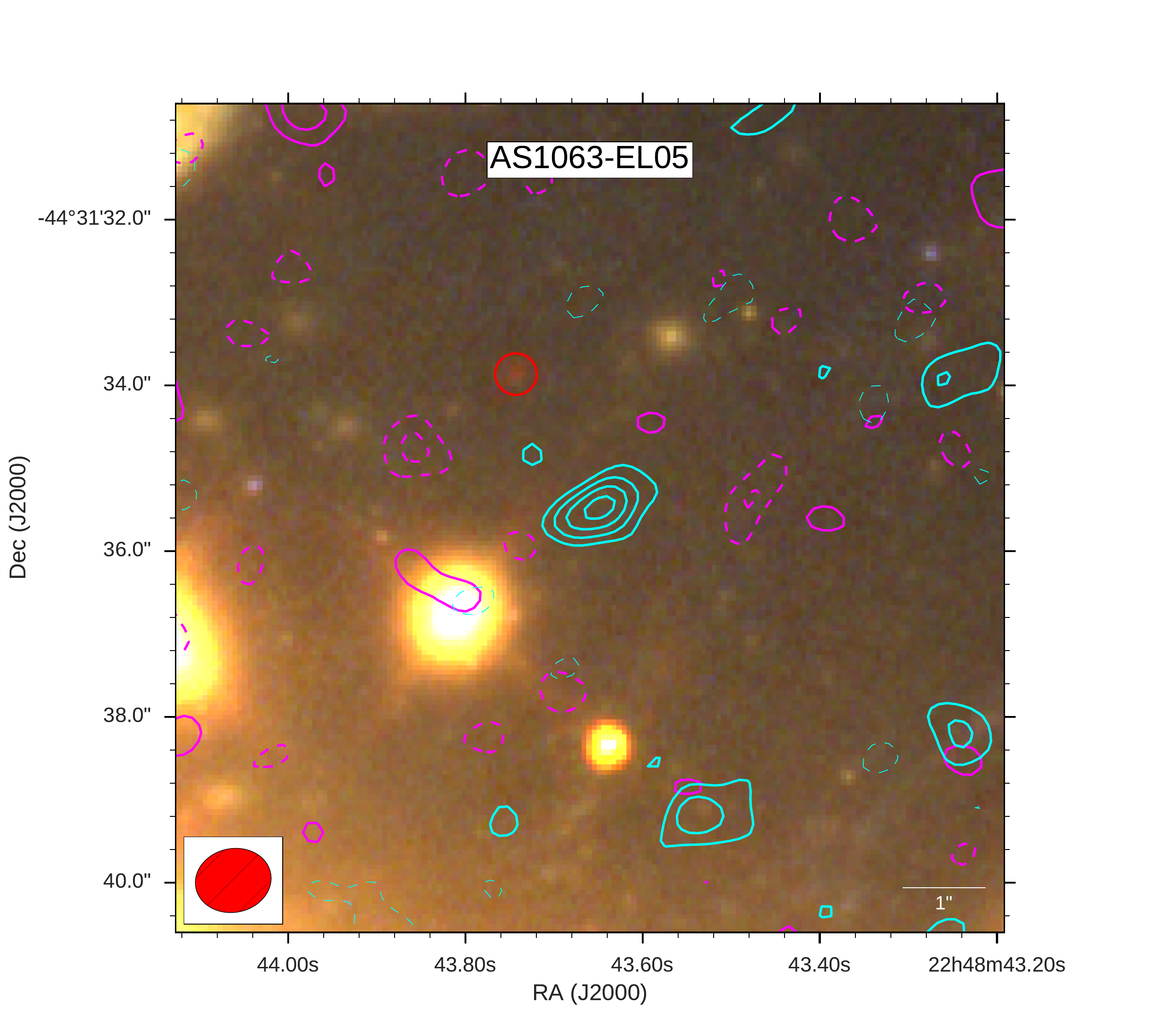}
\includegraphics[width=0.6\textwidth]{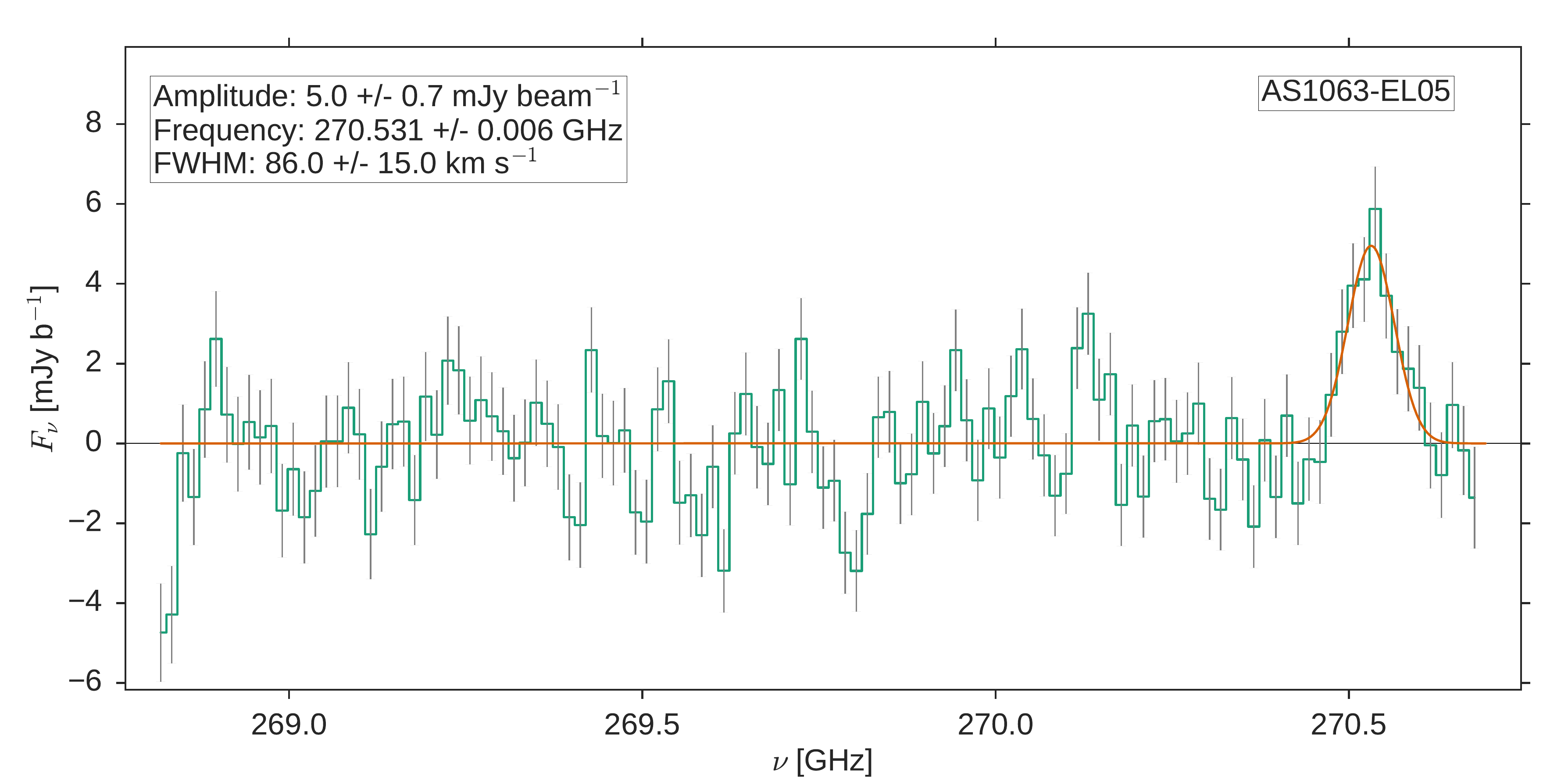}
\includegraphics[width=0.4\textwidth]{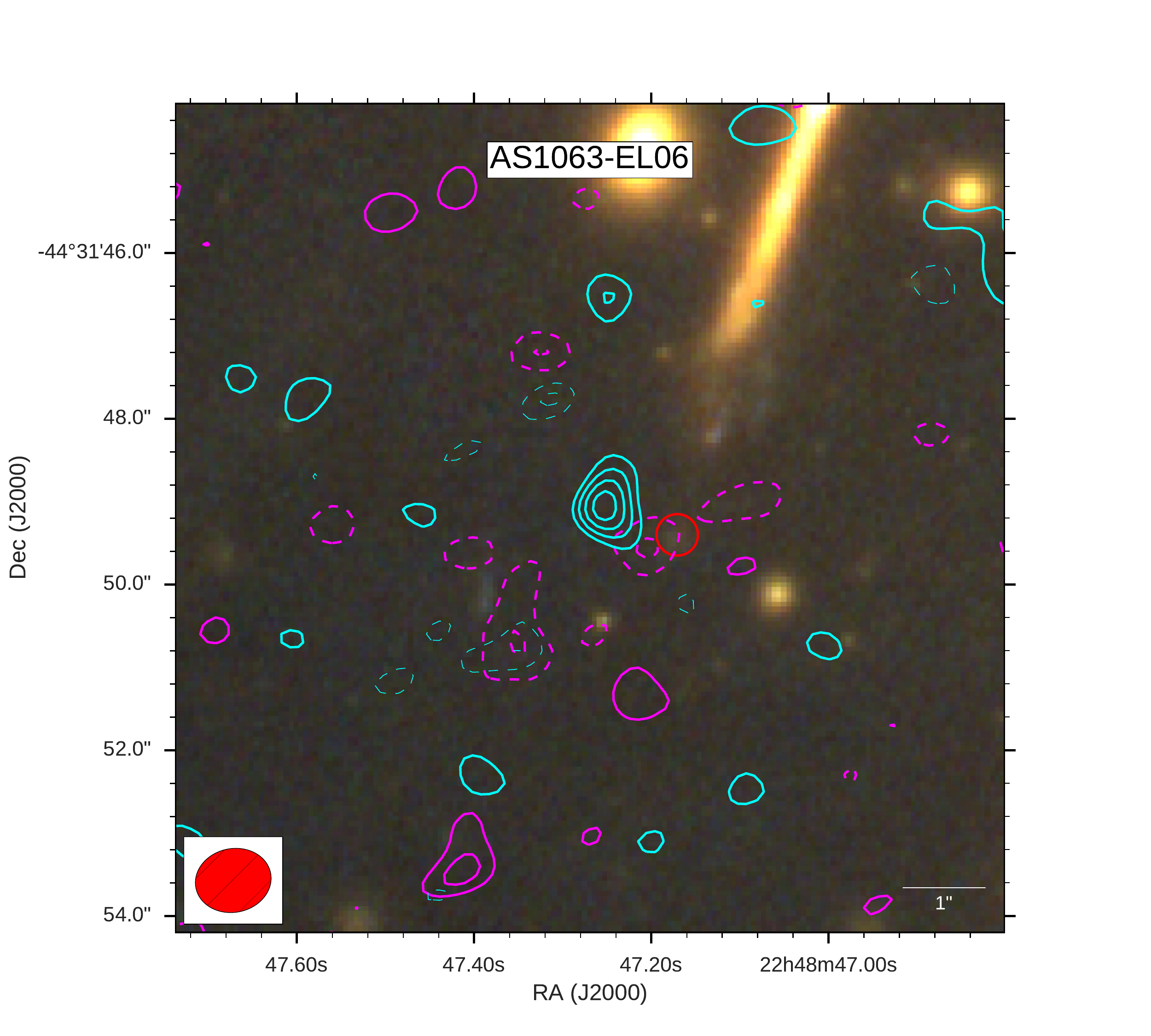}
\includegraphics[width=0.6\textwidth]{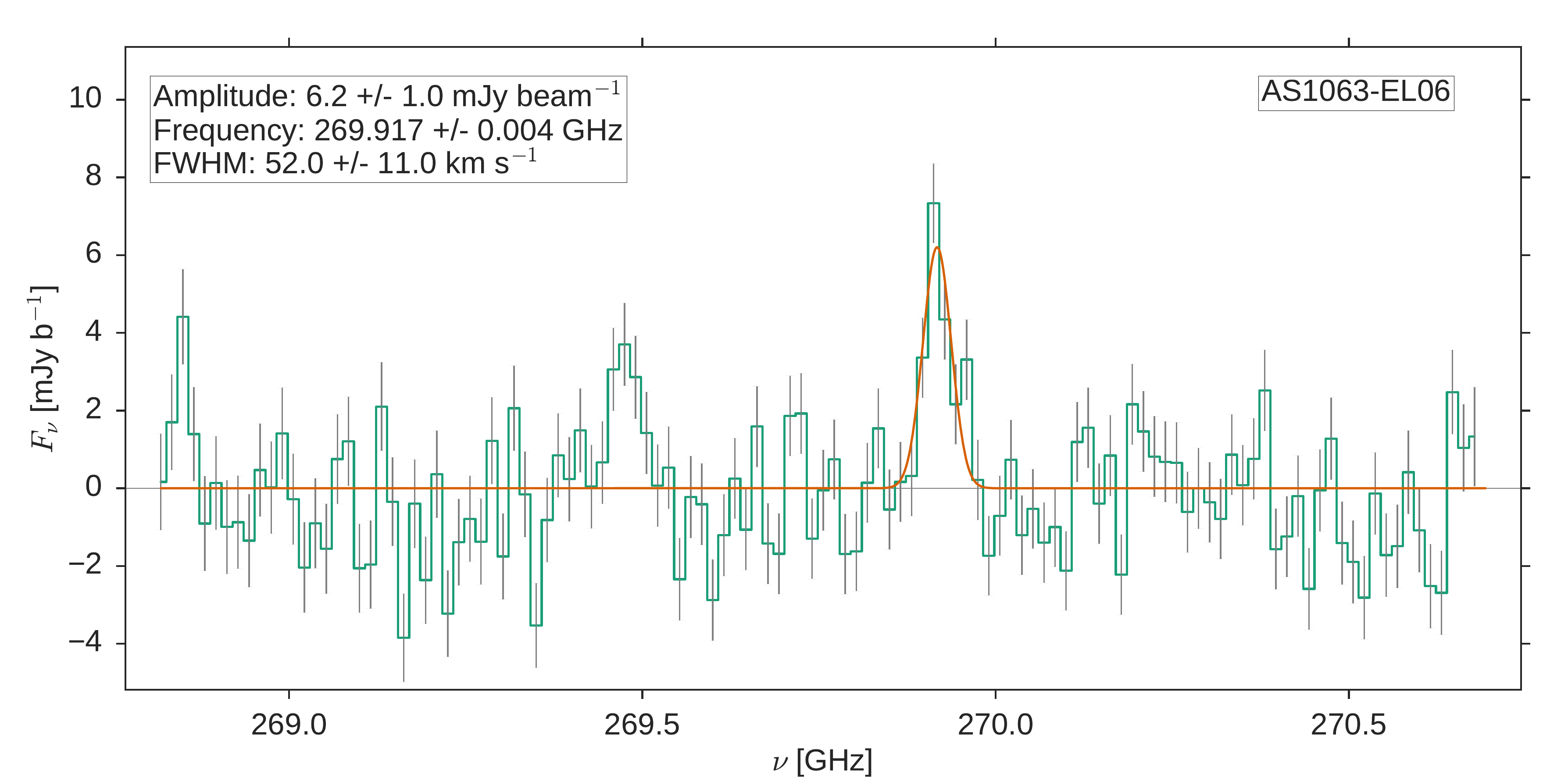}
\includegraphics[width=0.4\textwidth]{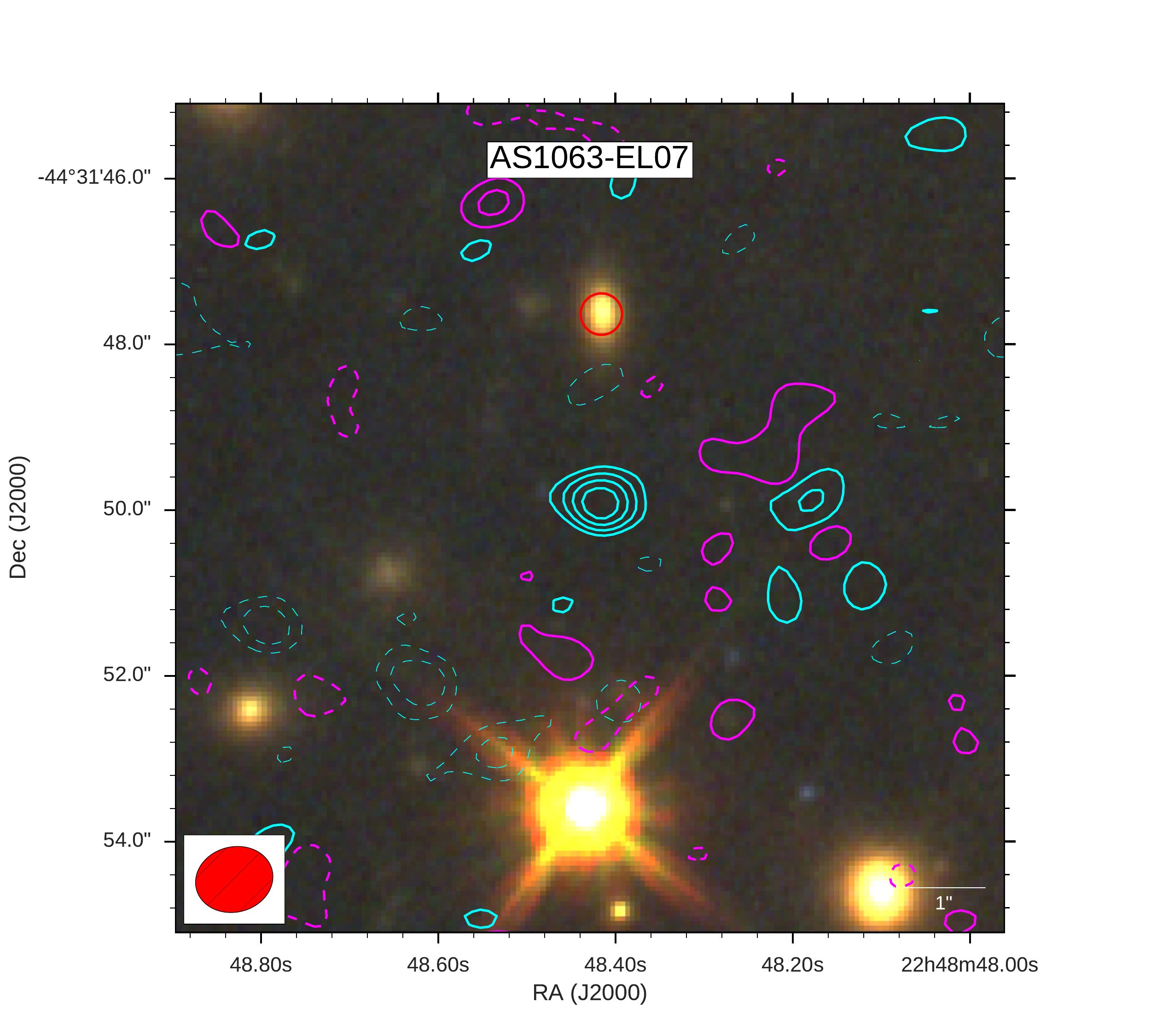}
\includegraphics[width=0.6\textwidth]{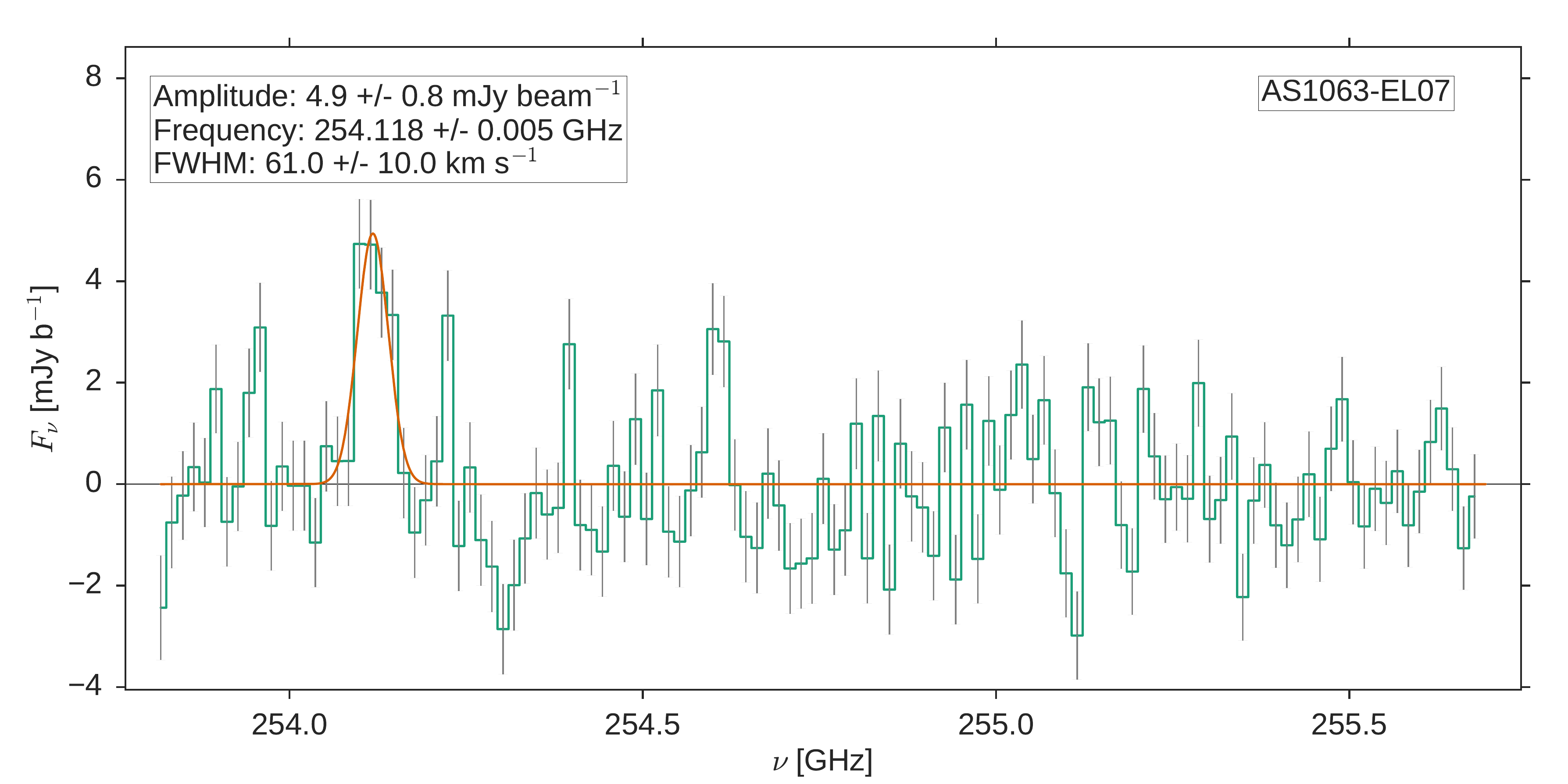}
\caption{Continuation of Figure~\ref{fig:line_candidate}\label{fig:line_candidate5}}
\end{figure*}

\begin{figure*}[!htbp]
\includegraphics[width=0.4\textwidth]{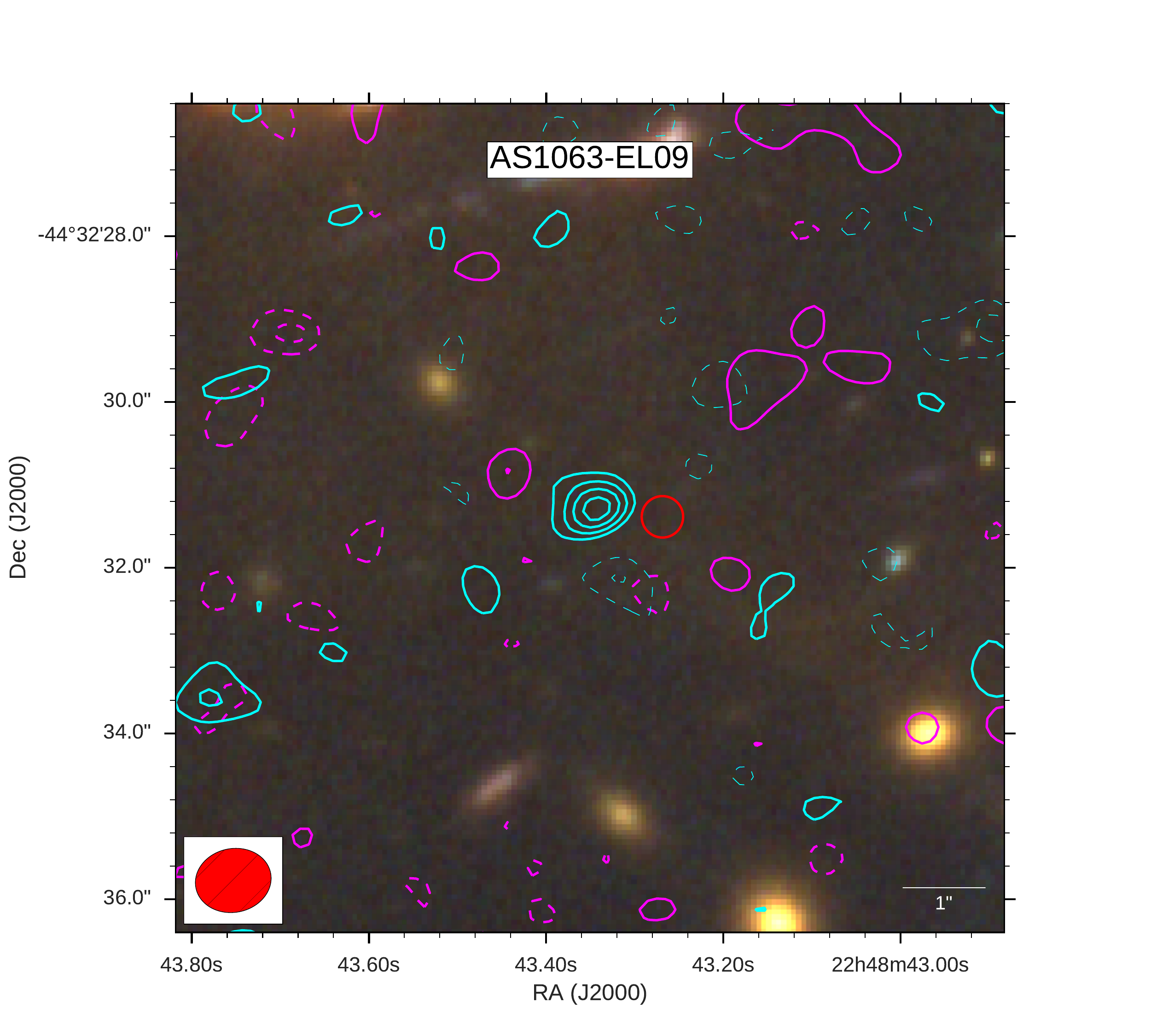}
\includegraphics[width=0.6\textwidth]{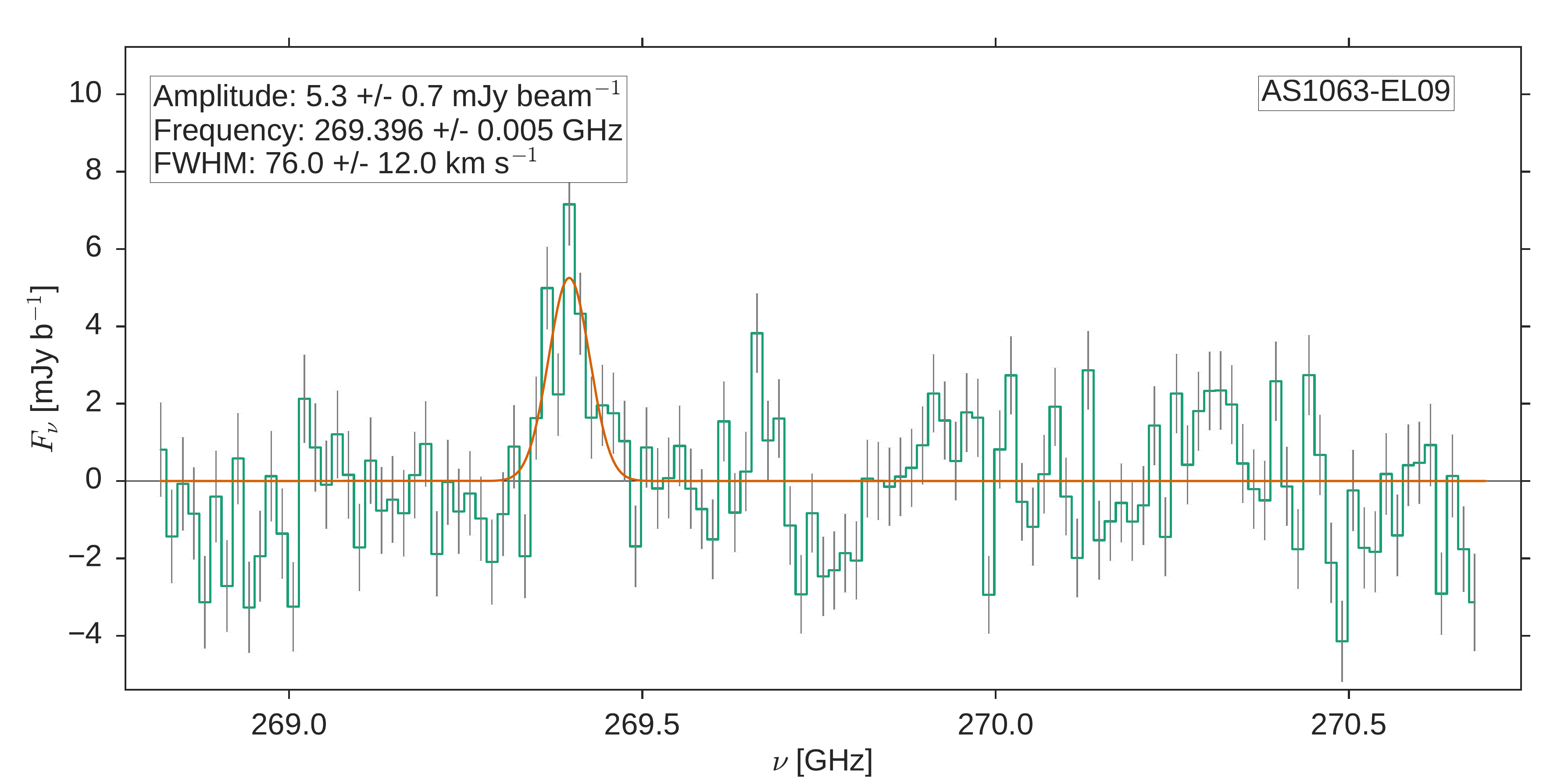}
\caption{Continuation of Figure~\ref{fig:line_candidate}\label{fig:line_candidate6}}
\end{figure*}

\end{appendix}

%%%%%%%%%%%%%%%%%%%%%%%%%%%%%%%%%%%%%%%%%%%%%%%%%%%%%%%%%%%%%%%%%%%%%%%%%%%%%%%%%
\begin{acknowledgements}
This paper makes use of the following ALMA data: ADS/JAO.ALMA\#2013.1.00999.S and \#2015.1.01425.S.
ALMA is a partnership of ESO (representing its member states), NSF (USA) and
NINS (Japan), together with NRC (Canada) and NSC and ASIAA (Taiwan), in
cooperation with the Republic of Chile. The Joint ALMA Observatory is operated
by ESO, AUI/NRAO and NAOJ.\\
% %
We acknowledge support from
CONICYT-Chile grants Basal-CATA PFB-06/2007 (JGL, FEB, RC), 
FONDECYT Regular 1141218 (JGL, FEB, RC), 
"EMBIGGEN" Anillo ACT1101 (FEB), and
the Ministry of Economy, Development, and Tourism's Millennium Science
Initiative through grant IC120009, awarded to The Millennium Institute
of Astrophysics, MAS (FEB,CRC).\\
M. Carrasco's  research is supported by the SFB-Transregio TR33 ”The Dark
Universe”.
RD gratefully acknowledges the support provided by the BASAL Center for Astrophysics and Associated Technologies (CATA), and by
FONDECYT grant N. 1130528.
PT acknowledge support from ANILLO ACT1417.
A.M.M.A. acknowledges support from FONDECYT grant 3160776.
NL acknowledges financial support from European Research Council Advanced Grant FP7/669253.
M.A. acknowledges partial support from FONDECYT through grant 1140099
This work utilizes gravitational lensing models produced by PIs Bradač, Natarajan \& Kneib (CATS), Merten \& Zitrin, Sharon, and Williams, and the GLAFIC and Diego groups. This lens modeling was partially funded by the HST Frontier Fields program conducted by STScI. STScI is operated by the Association of Universities for Research in Astronomy, Inc. under NASA contract NAS 5-26555. The lens models were obtained from the Mikulski Archive for Space Telescopes (MAST).
\end{acknowledgements}
%%%%%%%%%%%%%%%%%%%%%%%%%%%%%%%%%%%%%%%%%%%%%%%%%%%%%%%%%%%%%%%%%%%%%%%%%%%%%%%%%

%-------------------------------------------------------------------

\end{document}